\documentclass[11pt]{article}
\pdfoutput=1 
\usepackage{jheppub}
\usepackage{graphicx}

\usepackage[T1]{fontenc}
\usepackage{xcolor}
\usepackage{caption}
\usepackage{amsmath}
\usepackage{amsfonts}
\usepackage{mathrsfs}
\usepackage{comment}
\usepackage{subcaption,longtable,stmaryrd}
\usepackage{bigints}
\usepackage{multicol}
\usepackage{tikz,lipsum,lmodern}
\usepackage{dsfont}
\usepackage[most]{tcolorbox}
\usepackage{blindtext}
\usepackage{adjustbox}
\usepackage{tikz}
\usepackage{graphicx}
\usepackage{booktabs}
\usepackage{float}
\usepackage{upgreek}
\usetikzlibrary{shapes, arrows.meta, positioning}

\tikzset{
  box/.style={
    rectangle,
    draw=black,  % Make the box visible
    thick,
    text width=12cm,
    minimum height=1.5cm,
    align=center,
    font=\large,
    rounded corners,
  },
  arrow/.style={thick, -{Latex[length=3mm]}},
}

\newcommand{\bea}{\begin{eqnarray}}
\newcommand{\eea}{\end{eqnarray}}

\newcommand*\circled[1]{\tikz[baseline=(char.base)]{
  \node[shape=circle,draw,inner sep=1pt] (char) {#1};}}

\title{On $\sqrt{T\overline{T}}$ deformed pathways: CFT to CCFT}
\author[1,2]{Aritra Banerjee,}
\author[3,4]{Pulastya Parekh,}
\author[5,1]{and Robin Raj }
\affiliation[1]{Birla Institute of Technology and Science, Pilani Campus, Pilani, Rajasthan 333031, India.}
\affiliation[2]{Asia Pacific Center for Theoretical Physics, Postech, Pohang 37673, Korea.}
\affiliation[3]{Centro de Estudios Cient\'{i}ficos (CECs), Av. Arturo Prat 514, Valdivia, Chile.}
\affiliation[4]{Facultad de Ingenier\'{i}a, Arquitectura y Dise\~{n}o, Universidad San Sebasti\'{a}n, sede Valdivia,
General Lagos 1163, Valdivia 5110693, Chile.}
\affiliation[5]{Departamento de Ciencias Fisicas, Universidad Andres Bello, Av. Republica 252, Santiago,Chile}

\emailAdd{aritra.banerjee@pilani.bits-pilani.ac.in}
\emailAdd{parekh@cecs.cl}
\emailAdd{r.raj@uandresbello.edu}

\abstract{We discuss the marginal $\sqrt{T\overline{T}}$ deformation of massless scalar field theories in two dimensions from a dynamical perspective. The operator flow equations for such deformations induce a particular Legendre Transformation between flowed Lagrangians and flowed Hamiltonians. The marginal deformation does not change the conformal symmetries of the theory, until some special points in the moduli space are reached, and the relativistic conformal algebra smoothly changes to the Carrollian conformal (equivalently BMS) one. We investigate this change of symmetry from both configuration space and phase space point of view, while keeping the notion of Legendre Transformation unchanged during the flow. By expanding the actions, in the extreme limits of the flow parameter, we recover the usual ``Electric'' Carroll theory and further uncover a novel ``Magnetic'' counterpart. We discuss the intriguing geometric understanding of such dynamical maps for the deformed theories, and also provide a concrete example for the same from a deformed string theory in flat space. }
\begin{document}
\maketitle

\section{Introduction}
Studying deformation of quantum theories is absolutely essential if one wants to understand natural phenomena at different energy scales as a consequence of Renormalisation Group flows (RG) \cite{Wilson:1974mb}. This allows us to employ effective theories for particular phenomena using a controlled way based on the fixed points and associated symmetry properties of the system. One such powerful way, in the space of Quantum Field Theories (QFTs), to define new theories from seed ones, is via judicious deformations via local quantum operators \cite{Zamolodchikov:1986gt}.
For two dimensional QFTs, among a zoo of such operators, we can construct a particularly interesting \textit{irrelevant} one from the stress-energy tensor: 
\begin{align}
O_{T \bar T}=\frac{1}{2}\left( T^{\mu \nu} T_{\mu \nu}- T_\mu{ }^\mu T_\nu{ }^\nu\right).    
\label{TbarT}\end{align}
This operator is referred to as the $T\overline T$ operator in the literature \cite{Smirnov:2016lqw,Cavaglia:2016oda}, is exactly solvable and tractable, and has further been studied for almost a decade. The expectation value of $T\overline T$ can be factorised (Zamolodchikov factorisation \cite{Zamolodchikov:2004ce}), which ensures the $T\overline T$ operator is UV-finite and local, which in turn is remarkable for an irrelevant deformation. For more details, one should look at the comprehensive reviews \cite{Jiang:2019epa, He:2025ppz} and references therein.
\medskip

Such generic deformations produce a class of theories away from the fixed point that are exactly solvable, augmented by the deformed Lagrangian satisfying a well defined flow equation \cite{Bonelli:2018kik,Smirnov:2016lqw}. The presence of infinite conserved commuting charges also ensures that the classical integrability is retained for the deformed theory \cite{Smirnov:2016lqw}. However, added to a Conformal Field Theory (CFT), this breaks scale-invariance. Further, the quantum spectrum of this theory satisfies a Burger's hierarchy \cite{Datta_2018,Dubovsky_2018,Aharony_2019}, but the classical evolution can be thought of as a special field redefinition \cite{Conti:2018tca}, making the observables in this theory especially tractable. These properties distinguish 
$T\overline T$ as exceptional among irrelevant deformations, which typically lack such control. For more interesting properties of $T\overline T$ deformed theories that relate to Holography, the interested reader can look up \cite{Chakraborty_2018,Asrat:2017tzd,Giveon:2017nie} and follow-ups. This list is in no way exhaustive.
\medskip

A related deformation could be produced by simply taking a square root of the $T\overline T$ operator at the classical level \cite{Rodriguez:2021tcz,Ferko:2022cix}. Added to a CFT, this mass dimension-2 operator works perfectly as a \textit{marginal} deformation that makes sure scale invariance is untouched. This further induces a novel class of integrable theories \cite{Borsato:2022tmu}.
However, due to the non-polynomial nature of the 
$\sqrt{T\overline T}$
  operator, standard point-splitting regularisation fails, and Zamolodchikov factorisation cannot be directly applied. This obstructs a straightforward quantum definition of such a deformed theory, and the epithet `operator' could turn out to be a misnomer. Some attempts towards such a definition was made in \cite{Ebert:2024zwv,Hadasz:2024pew}\footnote{In fact, in \cite{He:2025fdz} the density of states at large energy and high temperature of these theories have been described to have a modified Cardy-like behaviour.}. Given a proper quantum definition, one hopes to use such a deformation to introduce new families of CFTs with smoothly connected spectra. But for this work, we will focus on the classical aspect only. 
  \medskip
  
   After the initial introduction,
$\sqrt{T\overline T}$ deformations have been applied to many physical cases, including two dimensional bosonic field theories \cite{Ferko:2022cix,Ebert:2024zwv,Babaei-Aghbolagh:2022leo,Conti:2022egv}, higher spin theories \cite{Bielli:2024ach,Bielli:2025uiv}, p-form gauge theories
\cite{Babaei-Aghbolagh:2020kjg,Babaei-Aghbolagh:2022itg,Ferko:2024zth,Kuzenko:2025jgk,Hutomo:2025dfx}, coupled dynamical systems \cite{Garcia:2022wad}, and different gravity models 
\cite{Li:2025lpa,Adami:2025pqr,Babaei-Aghbolagh:2024hti,Babaei-Aghbolagh:2025tim,Tsolakidis:2024wut}. What is more powerful in this construction is to realise that for the well known Modified Maxwell (ModMax) theories \cite{Bandos:2020jsw}, the four dimensional duality-invariant conformal electromagnetic actions are nothing but $\sqrt{T\overline T}$ deformation of Maxwell theory \cite{Babaei-Aghbolagh:2022uij,Chen:2024vho}. This deformation has found widespread use in construction and mapping of integrable models due to the deep rooted connection to the Lax formalism \cite{Conti:2018jho,Sakamoto:2025hwi,Babaei-Aghbolagh:2025hlm}, which makes sure if the seed theories are integrable, the infinite tower of conserved charges persist for the deformed case as well \cite{Borsato:2022tmu}. Such deformations have also appeared in the holographic context, for example in \cite{Ran:2025xas,Tian:2024vln}, and from a modified boundary condition point of view as well \cite{Ebert:2023tih}. 
 \medskip

All these diverse advances make sure concrete studies involving the $\sqrt{T\overline T}$ operators are absolutely important to investigate.
 However, our interest in this class of deformations stem from something far more interesting; that is the flow of the spacetime symmetry algebra, focusing on the two dimensional case. For general values of the deformation parameters, of course we should have the conformal symmetries unbroken in this case. But, for certain values in the moduli space, the deformed generators run into a singularity and the classical relativistic conformal algebra (Virasoro $\times$ Virasoro) changes smoothly into those of conformal \textit{Carrollian} algebra  in same dimension (CCA$_2$), or equivalently the isomorphic BMS$_3$ algebra in one higher dimension \cite{Duval:2014uva,Duval:2014lpa}. Carroll conformal theories generally occur as a speed of light going to zero ($c\to 0$) contraction \cite{SenGupta:1966qer,Levy-Leblond:1965dsc,Henneaux:1979vn} of its relativistic counterpart (for example, see \cite{Bagchi:2019xfx,Henneaux:2021yzg}), and has been emerging widely as the key player in a myriad of physical phenomena\footnote{This is a very active research area with a lot of important references at place, one could instead see the recent reviews \cite{Bagchi:2025vri,Nguyen:2025zhg}, the first of which encompasses most important aspects of Carroll and Carroll conformal symmetries. The latter review mostly focuses on a Carrollian version of holography \cite{Donnay:2022aba,Bagchi:2022emh,Donnay:2022wvx,Bagchi:2023fbj}.}. The flow from CFT to a Carrollian Conformal Field Theory (CCFT) via the $\sqrt{T\overline T}$ operator, noticed for the scalar case in \cite{Rodriguez:2021tcz,Tempo:2022ndz, Bagchi:2022nvj}, offers a brilliant alternate avenue for realising these interesting symmetries. The deformation, nonlinear in nature due to the square root, could equivalently be thought of as an ``infinite boost'' on the (anti-)holomorphic coordinates in a simplified setting \cite{Bagchi:2022nvj,Bagchi:2024unl} involving current bilinears of $U(1)$ Kac-Moody type. One could also think of $\sqrt{T\overline T}$ deformed string worldsheet theories where the flow to CCFT is realised as a change in residual gauge symmetry algebra \cite{Isberg:1993av,Bagchi:2013bga,Bagchi:2015nca,Bagchi:2019cay,Bagchi:2020fpr,Bagchi:2021ban}. Given the widespread interest, clearly, this line of investigation still needs to be clarified more, especially in the full nonlinear level, and that is where the current work comes into focus.

\subsection*{In this paper:}

\begin{figure}[h!]
\centering
\begin{tikzpicture}[
    node distance=1.6cm,
    round/.style={circle,draw,minimum size=14pt,fill=gray!5},
    >={Stealth[length=2mm,width=2mm]},
    scale=1.1
]

% ========== Left CFT box ==========
\node[draw, rectangle, minimum width=1.6cm, minimum height=1cm, fill=gray!5] (cft) at (-6,0) {CFT};

% ========== Left nodes L, H ==========
\node[round] (L) at (-3.8,1.2) {$\mathcal{L}$};
\node[round] (H) at (-3.8,-1.2) {$\mathcal{H}$};

% Arrow L <-> H
\draw[<->] (H) -- node[right] {{\tiny $\text{LT}_{(0)}$}} (L);

% ========== Right CCFT box ==========
\node[draw, rectangle, minimum width=1.8cm, minimum height=1cm, fill=gray!5] (ct) at (6.5,0) {CCFT};

% ========== Right nodes C, D ==========
\node[round] (C) at (3.8,1.2) {$\mathcal{L}_\alpha^+$};
\node[round] (D) at (3.8,-1.2) {$\mathcal{H}_\alpha^-$};

% ========== Middle transition labels and dashed arrows ==========

\foreach \x/\name/\up/\down in {
   -1.5/{(\alpha_1)}/1.8/-1.84,
    0/{(\alpha_2)}/1.9/-1.85,
    1.5/{(\alpha_3)}/1.64/-1.65
}{
    \node at (\x,0) {{\tiny $\text{LT}_{\name}$}};
    \draw[dashed,<-] (\x, \up) -- (\x,0.2);
    \draw[dashed,->] (\x,-0.2) -- (\x,\down);
}

% ========== Top and bottom middle nodes ==========
\node[round] (Z) at (4.8,3.0) {$\mathcal{L}_\alpha^-$};
\node[round] (Y) at (4.8,-3.0) {$\mathcal{H}_\alpha^+$};

% Curved arrows from middle to boundary circle
\draw[->, thick] (1.8,1.6) .. controls (3.5,2) .. (Z);
\draw[->, thick] (1.8,-1.6) .. controls (3.5,-2) .. (Y);

% Curved arrows L to C and H to D
\draw[->, thick] (L) .. controls (-0.8,2.12) .. node[above] {\small $\partial_\alpha \mathcal{L}_\alpha= \mathcal{O}$} (C);
\draw[->, thick] (H) .. controls (-0.8,-2.12) .. node[below] {\small $\partial_\alpha \mathcal{H}_\alpha={\mathfrak{O}}$} (D);

% Connection between Z and Y
\draw[<-] (4.8,2.4) -- (4.6,0.2);
\draw[->]  (4.6,-0.2)--(4.8,-2.4);
\node at (4.5,0){\tiny $\text{LT}_{(\infty)}$};

% Connection between Z and Y
\draw[<-] (4,0.8) -- (4.3,0.2);
\draw[<-]  (4,-0.8)--(4.3,-0.2);

% Infinity labels
\node at (3.5,2.3){\tiny $-\infty$};
\node at (2.6,1.3){\tiny $+\infty$};

\node at (3.5,-2.3){\tiny $+\infty$};
\node at (2.6,-1.3){\tiny $-\infty$};

\end{tikzpicture}
\caption{The Lagrangian ($\mathcal{L}$) and Hamiltonian ($\mathcal{H}$) flows stay related by a valid Legendre Transform(LT) across all values of the deformation parameter $\alpha$. As $\alpha$ reaches extreme values, these flows bifurcate into electric and magnetic Carroll theories, both in configuration space and in phase space, with very specific maps between them.}
\label{fig:diagram.intro}
\end{figure}
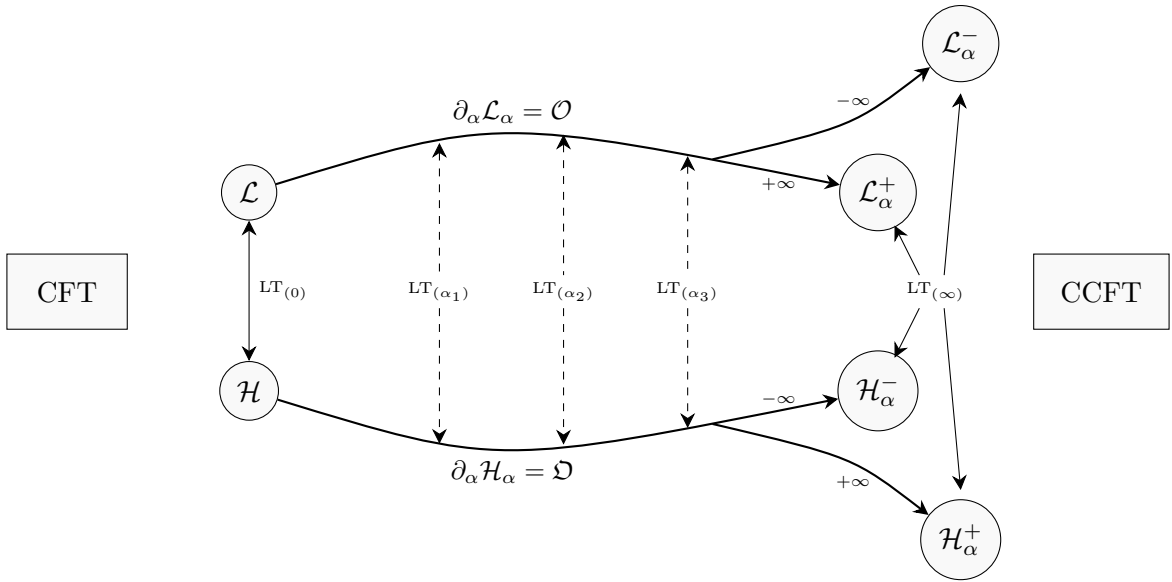

In the current work, we zoom into the flow of CFT to CCFT via $\sqrt{T\overline T}$ operators in a much morerigorous setting. The classical flow equation of the Lagrangian and Hamiltonian for a $\sqrt{T\overline T}$ flow has a usual canonical setting, but the intricacies associated to the operator gives further subtleties to the flowed quantities. Especially at finite flow parameters, connecting the phase space and configuration space objects require special attention. One can still show that the deformed Lagrangians and Hamiltonians are related via a nontrivial Legendre Transformation, as long as the deforming operators in both regimes are related in the appropriate way. Obviously, this relationship entails connecting the velocities and conjugate momenta coming from two descriptions simultaneously at finite deformation.
\medskip

We will exactly try to understand this special relationship between the two descriptions while dialing our deformation parameters towards those special values where the symmetry algebra changes to CCA$_2$. These values reside at the very end of parameter space, i.e. at $\pm\infty$. To make sense of the parameter space at the very end, we will first try to understand this notion of Legendre Transformation at finite values of the deformation as the theories are flowing, focusing on the $N-$scalar case. Since the theory involves a square root, it helps to effectively linearise the theory both in phase space and configuration space, while keeping the equations of motion (EOM) same using some well-defined dynamical variables. These quantities in turn become related across the two sectors when the Legendre Transformation is implied. Once that is understood, we will try to see what happens to these transformations as we gradually (but carefully) approach the infinite values, and how the Legendre Transformations themselves gradually flow to these extreme points. A geometric understanding of the deformed phase space and configuration space mapping will be discussed to make this more intuitive. One can see Fig.\eqref{fig:diagram.intro} for an outline of the main concepts.
\medskip

Starting from the full nonlinearly defined theories, arriving at the extreme ($\pm$ve) values of the deformation parameter in a controlled way will lead to two very different Carroll invariant scalar theories. While one would be the well known \textit{``Electric''} scalar theory discussed in many Carroll literature (see for example \cite{Henneaux:2021yzg,deBoer:2021jej}) involving only temporal derivative of fields, the other one, which we dub as a novel kind of \textit{``Magnetic''} Carroll scalar, involves nonlinear terms in field derivatives. The implications of Carroll boost invariance in this manifestly nonlinear theory will be discussed in detail, and a geometric understanding will also be presented. 
\medskip

Finally, to round up the discussion of such deformations, we focus on worldsheet string theories with such a marginal operator added to the action, which in the phase space setting modifies the Hamiltonian constraint.  In extreme deformation values, this theory realises the CCA$_2$ as the residual gauge symmetry algebra on the worldsheet provided we put in a conformal gauge in the associated action. This case, as a gauge symmetry, is somewhat different from the spacetime symmetry case for scalars, but provides additional insight into the algebraic structure of the Legendre Transformation.   

\medskip

The rest of the paper is structured as follows: we review the intricacies of $\sqrt{T\overline{T}}$ deformations of the free scalar theory, both in Lagrangian and Hamiltonian sense in section \eqref{sect2}. In section \eqref{sect3}, we discuss the linearisation of the dynamics and consequently build up to the Legendre Transformation relation between two perspectives. The extreme parameter space and Carroll Conformal symmetries from different flows are discussed in section \eqref{sect4}. Further, in section \eqref{sect5} we dive in-depth into the structure of these theories near the Carroll Conformal regions, providing a limiting analysis of the actions and discussing the geometric nature of the mapping. In section \eqref{sect6}, we elucidate on our example on flows from string worldsheet theory. Finally section \eqref{sect7} includes a summary and future perspectives. Appendices contain details of various calculations and supplementary material.

\section{Review of $\sqrt{T\overline{T}}$ flows}
\label{sect2}

As described in the introduction, we are interested in a close (but not so close) cousin of $T\overline T$ deformations in two dimensions, where we can define a marginal operator
by taking the square-root of the $T\overline{T}$ operator in the following way:  
\begin{align}
O_{\sqrt{T \overline{T}}}=\sqrt{\frac{1}{2} T^{\mu \nu} T_{\mu \nu}-\frac{1}{4} T_\mu{ }^\mu T_\nu{ }^\nu}.    
\end{align}
For a CFT, the second term inside the square root drops out because the trace vanishes, and also the conformal property of the theory is preserved after the deformation induced by this operator because of its marginal nature.   
Classically, this operator is well-defined with straightforward OPE structures \cite{Borsato:2022tmu}. Moreover, as mentioned before, the classical integrability of the theory still remains intact in this case. Finally, just like the other cousin, $\sqrt{T\overline T}$ flow also satisfies a flow equation (albeit only classically) and thus generates one parameter deformed classes of field theories \cite{Ferko:2022cix} in Lagrangian and Hamiltonian forms, which is central to our discussion. But invariably both these deformed formalisms carry forward the square root they inherited from their parent operator, often leading to some form of non-analyticity\footnote{In 0+1 dimensional version of such theories, authors have reported \cite{Garcia:2022wad} a field dependent redefinition that maps undeformed theories to deformed ones. The equivalent for field theories is not clear yet.}.

\subsection{Flow equations}

Renormalisation Group (RG) flows reveal the structure and fixed points of the space of QFTs in its modern formulation. These flows are parameterised by the relevant energy scales $\mu$ of the theory.
For a theory deformed by an operator $\mathcal{O}$, the flow of the theory can often be expressed as a differential equation involving the action:
\begin{align}
\partial_\alpha \mathcal{S}_\alpha = \mathcal{O}_\alpha,
\label{General flow}\end{align}
Where $\mathcal{S}_\alpha$ is some deformed action functional associated to the theory and the scale is set by a regular function of the energy i.e. $\alpha=f(\mu)$. 
\medskip

In this section, we will quickly review the structure of CFTs in 2d deformed by the $\sqrt{T\overline{T}}$ operator, focusing on a seed theory of $N$ free scalar fields. Our main character of interest in this case in the exact solvabilty of the associated flow equations. The deformation is parameterised by $\alpha$, which defines the flow in the field theory: $\mathcal{L}\rightarrow\mathcal{L}_\alpha$, where $\mathcal{L}$ and $\mathcal{L}_\alpha$ are the Lagrangians of the undeformed and deformed theory respectively, with $\mathcal{L}_\alpha$ being iteratively calculated from the seed lagrangian. The stress-energy tensor also flows with this parameter, i.e. $T^{\mu\nu}\rightarrow T^{\mu\nu}(\alpha)$. Therefore, the marginal deformation operator becomes 
\begin{align}
O_{\sqrt{T\overline T}}^\alpha= \sqrt{\frac{1}{d} T^{\mu \nu}(\alpha) T_{\mu \nu}(\alpha)}=\sqrt{\operatorname{det}|T^{\mu \nu}(\alpha)|},
\end{align}

The deformed Lagrangian and Hamiltonian ($\mathcal{H}_\alpha$ in accordance with our notation, we omit `density' in the nomenclature in what follows) as a function of the flow parameter can be found using the flow equation \cite{Ferko:2023iha}:
\begin{align}
\frac{\partial \mathcal{L}_\alpha}{\partial \alpha}=-\frac{\partial \mathcal{H}_\alpha}{\partial \alpha} = O_{\sqrt{T\overline T}}^\alpha\label{flowequation}.
\end{align}
The sign difference above is crucial here. This allows us to define the deformed Lagrangian and Hamiltonian at each point on the flow, and make a comparison of the two. In the rest of this section we will be looking at both of these perspectives, and investigate more on their subtle connections. 

\subsection{Scalar fields: Lagrangian flows}

We start by looking at a theory of $N\geq 2$ free bosons in two dimensions as an example. 
The undeformed Lagrangian in this case is given by: 
\begin{equation}
\mathcal{L}=-\frac{1}{2} \partial^\mu \phi_i \partial_\mu \phi_i. \label{udefL}
\end{equation}
The summation over the field index $i$ is implied, and we are using diag$(-1,1)$ for the spacetime metric.
Then using the flow equation \eqref{flowequation} we can iteratively write the deformed Lagrangian as \cite{Ferko:2022cix}:
\begin{align}
\mathcal{L}_\alpha = \mathcal{L}\cosh(\alpha)  + \frac{1}{2} \sinh(\alpha) \sqrt{2 \partial^\mu \phi_i \partial_\nu \phi_i \partial^\nu \phi_j\partial_\mu \phi_j  - \left(\partial^\mu \phi_i \partial_\mu \phi_i\right)^2}.
\label{FL1}
\end{align}
Note that we can write \eqref{FL1} in a more compact form  in terms of Lorentz invariant quantities $\mathcal{L}$ and $\mathcal{P}$ as
\begin{align}
\mathcal{L}_\alpha=\mathcal{L} \cosh (\alpha)+\sqrt{\mathcal{L}^2-\mathcal{P}^2} \sinh (\alpha),   
\label{5.1}\end{align}
where $\mathcal{L}$ is the undeformed Lagrangian (at $\alpha=0$), and $\mathcal{P}$ is defined as 
\begin{align}\mathcal{P}^2=2\mathcal{L}^2- \frac{1}{2} \partial^\mu \phi_i \partial_\nu \phi_i \partial^\nu \phi_j \partial_\mu \phi_j.  
\label{P_N^2}\end{align}
Furthermore, $\mathcal{L}$ and $\mathcal{P}$ can be represented within the framework of a general matrix \cite{Conti:2022egv}\footnote{Note that $M$ resembles an induced metric on a $d$-dimensional submanifold embedded in a $N$ dimensional target space.}:
\begin{align}
& M_{\mu\nu}=\sum_{i=1}^{N} \partial_\mu \phi_i \partial_\nu \phi_i, \label{GM1} \\
& \Longrightarrow \mathcal{P}:=-\sqrt{\operatorname{det}\left[\mathbf{M}\right]}, \quad \mathcal{L}^{\mathrm{}}=-\frac{1}{2} \operatorname{tr}\left[\mathbf{M}\right]=-\frac{1}{2} \sum_{i=1}^{N} \eta^{\mu\nu} \partial_\mu \phi_i \partial_\nu \phi_i,
\label{GM2} \end{align}
The Lagrangian flow is non-trivial only when we have more than one scalar field, because for the case of a single scalar field, $\mathcal{P}^2$ is identically zero, which consequently results just in an overall scaling of the undeformed Lagrangian 
\eqref{udefL}, i.e $\mathcal{L}_{\alpha,~ N=1}=\mathcal{L}\exp(\alpha)$. Further, focusing on scalar fields we can cast our \eqref{5.1} in the form\footnote{See appendix \eqref{appA} for some details of algebraic manipulations. This is done purely for computational convenience.} , 
\begin{align}
 \mathcal{L}_\alpha&=\mathcal{L}\cosh (\alpha)+ \sqrt{\sigma^2-\rho^2}  \ \sinh (\alpha),  \qquad \text{where},~~\sigma=\frac{1}{2}\left(\dot\phi_i^2+\phi_i^{\prime 2}\right),~~\rho=\dot\phi_i\phi_i^{\prime}.\label{Lsigmaandrho}
 \end{align}
Note that dots and primes refer to derivatives w.r.t. the spacetime coordinates $t$ and $x$ respectively. Alternatively, the deformed Lagrangian can be written as
 \begin{align}
 \mathcal{L}_\alpha&=\mathcal{L} \cosh (\alpha)+2 \sqrt{\mathcal{T} \overline{\mathcal{T}}} \sinh (\alpha) \label{lttb}. \end{align}
 The above \eqref{lttb} involves $\mathcal{T}$ and $\bar{\mathcal{T}}$, which in our notation henceforth will denote the holomorphic and antiholomorphic components of the \textit{undeformed} stress-energy tensor within the Lagrangian formalism. In various parts of the current work, we will use these above definitions of the deformed theory interchangeably. 

\subsection{Scalar fields: Hamiltonian flows} 

Now, let us look at the flow from the perspective of the Hamiltonian. Solving the Hamiltonian version of the flow equation  \eqref{flowequation} yields  \cite{Ferko:2023iha}:
\begin{align}
\mathcal{H}_\alpha&=\frac{1}{2}\left(\pi_i \pi_i+\phi_i^{\prime } \phi_i^{\prime}\right) \cosh (\alpha)-\sinh (\alpha) \sqrt{\frac{1}{4}\left(\pi_i \pi_i+\phi_i^{\prime } \phi_i^{\prime}\right)^2-\left(\pi_i \phi_i^{ \prime}\right)^2}. \label{Hamiltonion}\end{align}
Here, \(\pi_i\) is the conjugate momentum corresponding to \(\phi_i\), and \((\phi_i, \pi_i)\) together constitute the entire phase space. Note that the momenta as iteratively calculated from the canonical Lagrangian \eqref{Lsigmaandrho} could have implicit $\alpha$ dependence, hence this deformed Hamiltonian above is pretty non-trivial.
 Furthermore, similar to equation \eqref{5.1} we can write the deformed Hamiltonian in a compact form:
\begin{align}
\mathcal{H}_\alpha&=\cosh (\alpha) \mathcal{H}-\sinh (\alpha) \sqrt{\mathcal{H}^2-\mathcal{J}^2},
\label{Hamiltonioncompact} 
\end{align}
where at $\alpha=0$, the $\mathcal{H}_{(0)}=\frac{1}{2}\left(\pi_i \pi_i+\phi_i^{\prime } \phi_i^{\prime}\right)$ is the purely undeformed Hamiltonian of the free theory, and the momentum density $\mathcal{J}$ is defined as $\mathcal{J}=\pi_i \phi_i^{ \prime}$. Similar to the Lagrangian formalism, we can rewrite \eqref{Hamiltonion} as:
\begin{align}
\mathcal{H}_\alpha=\mathcal{H} \cosh (\alpha)-2 \sqrt{T \bar{T}} \sinh (\alpha)\label{HTTD},   
\end{align}
where $T$ and $\bar T$ denote the holomorphic and antiholomorphic components of the stress-energy tensor within the Hamiltonian formalism.
\medskip

This completes our introduction to the deformed theory as viewed from Lagrangian and Hamiltonian perspectives, both of which have their salient features.  But of course, these two formalisms should converge for generic $\alpha$, and that requires the flow equations \eqref{flowequation} to be defined in such a way so as to make sure the flow operators in Lagrangian and Hamiltonian formalism act with opposite signs for all values of $\alpha$. This necessitates the usage of two different scripts for the stress-energy tensors in the two cases, as done above, since the representation of the operators away from $\alpha=0$ will immediately start to differ. Phrased more precisely, the phase space and configuration space operators, at any point on the flow, relate as follows:
\begin{equation}\label{OO}
    \mathcal{O}_{\sqrt{T \bar{T}}}(\phi^i,\pi^i(\dot\phi^i,\phi^{i\prime};\alpha);\alpha) = -\mathcal{O}_{\sqrt{\mathcal{T} \overline{\mathcal{T}}}}(\phi^i,\dot\phi^i,\phi^{i\prime};\alpha).
\end{equation}
Note again that the input momenta we use in the Hamiltonian formalism itself changes with flowing $\alpha$. This operator equivalence will be a remarkably powerful statement, relating evolution in these two sectors, and will be crucial going forward for us. 

\section{Lagrangian vs Hamiltonian deformations}\label{sect3}
Let us now come to one of the main questions we want to ask in this work, that is, how are the Lagrangian and Hamiltonian perspectives for the deformed theories connected to each other? In \cite{Ferko:2023iha, Garcia:2022wad} this question has been discussed for dynamical systems flowing via a general operator and the relation between flow equations were elucidated. 
It is still a bit subtle for a classical field theory as the deformation keeps modifying the relationship between the `velocities' (derivative of fields) and conjugate momenta continuously throughout the flow (see appendix \eqref{appB} for a concrete classical mechanics example).
\medskip

In what follows, we will try to build up to the Legendre Transformation (LT hereafter) relating the Lagrangian and Hamiltonian perspectives for our scalar field theory example, and see what $\alpha$ dependent constraint between `velocities' and conjugate momenta will make this possible. The guiding principle for us, perhaps obviously, would be that both flows are self consistent in their own right, at any $\alpha$. 
\subsection{Configuration space dynamics}
We start with the Lagrangian perspective first. 
We can find the derivatives of the given deformed Lagrangian \eqref{Lsigmaandrho} with respect to fields in the configuration space as:
\begin{align}
\pi_i=\frac{\partial \mathcal{L}_\alpha}{\partial \dot\phi_i}&=\left(\cosh \alpha+\frac{\sigma \sinh \alpha}{\sqrt{\sigma^2-\rho^2}}\right) \dot{\phi}_i  -  \frac{\rho \sinh \alpha}{\sqrt{\sigma^2-\rho^2}}{\phi}_i^\prime,
\label{canonicalmomentum}\\\frac{\partial \mathcal{L}_\alpha}{\partial \phi_i^{\prime}}&=-\phi_i^{\prime} \cosh \alpha+\frac{\sinh \alpha}{\sqrt{\sigma^2-\rho^2}}\left(\sigma \phi_i^{\prime}-\rho \dot{\phi}_i\right),\label{canonicalmomentum2}
\end{align}
where \eqref{canonicalmomentum} is the canonical momentum at some finite value of the deformation $\alpha$, and $(\sigma,\rho)$ has been defined in \eqref{Lsigmaandrho}. 
Given this, we can define two configuration space quantities, 
\begin{align}
C_1 = \cosh \alpha+\frac{\sigma \sinh \alpha}{\sqrt{\sigma^2-\rho^2}}, \quad C_2 =-\frac{\rho \sinh \alpha}{\sqrt{\sigma^2-\rho^2}}.   \label{C_1C_2} 
\end{align}
In the undeformed limit, $C_1\to 1, C_2 \to 0$. Note that for finite $\alpha$, the square root structures persist. To counter that, we can express the deformed Lagrangian using these $C_1$ and $C_2$ in a `linearised' manner:
\begin{align}
\mathcal{L}_\alpha&=C_1 \sigma+C_2 \rho-\phi_i^{\prime 2} \cosh (\alpha) \label{LagranC1C2},
\end{align}
which promptly leads to the derived quantities:
\begin{align}
\pi_i=\frac{\partial \mathcal{L}_\alpha}{\partial \dot{\phi}_i}&=C_1 \dot{\phi}_i+C_2 \phi_i^{\prime},\label{111.5} \\
 \frac{\partial \mathcal{L}_\alpha}{\partial \phi_i^{\prime}}&=C_1 \phi_i^{\prime}+C_2 \dot{\phi}_i-2 \phi_i^{\prime} \cosh (\alpha).\label{111.6}\end{align}
 These can be checked to be exactly equivalent to \eqref{canonicalmomentum} and \eqref{canonicalmomentum2}.
The attentive reader will immediately spot something interesting here. It seems as if taking the variation of \eqref{LagranC1C2} with respect to $\dot{\phi}_i$ and $ \phi_i^{\prime}$, by treating $C_1$ and $C_2$ having no variation with respect to fields, also gives the same dynamics as in \eqref{111.5}-\eqref{111.6}. However, this does not imply that $C_1$ and $C_2$ are constants of motion in a field theory because we cannot eliminate the spatial and temporal (and $\alpha$) dependence of these quantities. This will become clear immediately as we derive the equation of motion in configuration space. Using \eqref{111.5} and \eqref{111.6}
we can find the Lagrangian EOM 
\begin{comment}
\begin{align}
\frac{\partial}{\partial x}\left(\frac{\partial \mathcal{L}_\alpha}{\partial \phi_i^{\prime}}\right)&=C_1 \phi_i^{\prime \prime}+C_1^\prime\phi_i^{\prime}+
C_2 \dot{\phi}_i^{\prime}+C_2^\prime\dot{\phi}_i-2 \phi_i^{\prime \prime} \cosh (\alpha)\nonumber\\\frac{\partial}{\partial t}\left(\frac{\partial \mathcal{L}_\alpha}{\partial \dot\phi_i}\right)&=C_1 \ddot{\phi}_i+\dot C_1 \dot{\phi}_i+C_2 \dot{\phi}_i^{\prime}+\dot C_2 {\phi}_i^{\prime}
\end{align}
This gives the EOM as
\end{comment}
\begin{align}
C_1 \phi_i^{\prime \prime}+C_1^{\prime} \phi_i^{\prime}+C_2 \dot{\phi}_i^{\prime}+C_2^{\prime} \dot{\phi}_i-2 \phi_i^{\prime \prime} \cosh (\alpha)+ C_1 \ddot{\phi}_i+\dot{C}_1 \dot{\phi}_i+C_2 \dot{\phi}_i^{\prime}+\dot{C}_2 \phi_i^{\prime}&=0. \nonumber\end{align}
Collecting the coefficients we get:
\begin{align}
\left(C_1-2 \cosh (\alpha)\right) \phi_i^{\prime \prime}+\left(C_1^{\prime}+\dot{C}_2\right) \phi_i^{\prime}+2 C_2 \dot{\phi}_i^{\prime}+\left(C_2^{\prime}+\dot{C}_1\right) \dot{\phi}_i+C_1 \ddot{\phi}_i&=0 .\label{LEOMNEWWWW}\end{align}
As one can see, the full EOM does contain spatial/temporal derivatives of $C_i$. Now, what we will do is to remember this EOM defines the full dynamics of the fields $\phi_i$ and see if we can make sense of it from the Hamiltonian perspective.

\subsection{Phase space dynamics}\label{sec3.2}
Let's now derive the same dynamics from the phase space perspective. Taking the variation of \eqref{Hamiltonioncompact} with respect to phase space fields $(\pi(\dot\phi,\phi';\alpha), \phi')$ gives the following equations:
\begin{align}
\frac{\partial \mathcal{H}_\alpha}{\partial \pi_i}&=\cosh (\alpha) ~\pi_i-\sinh (\alpha) \frac{\mathcal{H} \pi_i-\mathcal{J} \phi_i^{\prime}}{\sqrt{\mathcal{H}^2-\mathcal{J}^2}},  \label{HE1}\\
\frac{\partial \mathcal{H}_\alpha}{\partial \phi_i^{\prime}}&=\cosh (\alpha) ~\phi_i^{\prime}-\sinh (\alpha) \frac{\mathcal{H} \phi_i^{\prime}-\mathcal{J} \pi_i}{\sqrt{\mathcal{H}^2-\mathcal{J}^2}}.\label{HE2}
\end{align}
Here $\mathcal{H},\mathcal{J}$ are defined around \eqref{Hamiltonioncompact}. Let us define two phase space quantities 
\begin{align}\label{ABdef}
A=\cosh (\alpha)-\frac{\sinh (\alpha) \mathcal{H}}{\sqrt{\mathcal{H}^2-\mathcal{J}^2}}, \quad B= \frac{\sinh (\alpha) \mathcal{J}}{\sqrt{\mathcal{H}^2-\mathcal{J}^2}},    
\end{align}
Substituting these quantities using \eqref{Hamiltonioncompact}, and the EOMs \eqref{HE1} and \eqref{HE2}, we can get a `linearised' version of the Hamiltonian, for which the associated EOM can be equivalently written down:
\begin{align}
\mathcal{H}_\alpha&=A \mathcal{H}+B \mathcal{J},\label{HamilAandB}\\
\frac{\partial \mathcal{H}_\alpha}{\partial \pi_i}&=A \pi_i+B \phi_i^{\prime},~~\frac{\partial \mathcal{H}_\alpha}{\partial \phi_i^{\prime}} =A \phi_i^{\prime}+B \pi_i \label{111.12}\end{align}
We can observe, just like the case of configuration space dynamics, from \eqref{HamilAandB} even without considering the field (and derivatives thereof) dependence of $A$ and $B$, we will still get the same EOM as \eqref{111.12}. But this still does not promise $A$ and $B$ as phase space constants, because both $A$ and $B$ can still evolve in space and time (and in $\alpha$), and we will explicitly see this through by deriving the Hamiltonian equations of motion.
\medskip

From \eqref{111.12} we can find the Hamiltonian EOM, focusing on the linearised version in \eqref{HamilAandB}:
 \begin{align}
    \dot{\phi}_i(x, t)&=\frac{\partial \mathcal{H}_\alpha}{\partial \pi_i}=A(\alpha) \pi_i+B(\alpha) \phi_i^{\prime},    
   \label{H111111111} \\
\dot{\pi}_i(x, t)&=\partial_x\left(\frac{\partial \mathcal{H}_\alpha}{\partial \phi_i^{\prime}}\right)  %=\partial_x\left( A\phi_i^{\prime}+B \pi_i\right) \\
=A\phi_i^{\prime \prime}+A^{\prime}
\phi_i^{\prime}
+B\pi_i^{\prime}+B^{\prime}\pi_i.
  \label{H222}   
    \end{align}
Which again contains the spatial/temporal derivatives of $A,B$, which in reality should not appear in the EOM. Now that we have fixed the dynamics from both phase space and configuration space point of view, we need to fix them further based on consistency conditions so that they are equivalent throughout the flow.

\subsection{Legendre Transformation}\label{LTdef}
With both phase space and configuration space dynamics at hand, let us take two different perspectives to understand the LT in this case. 

\subsubsection*{(a) Via EOM mapping}

The deformed Lagrangian and Hamiltonian perspectives must be mutually connected through the LT if the momenta in the phase space and the configuration space are equal, and remains equal throughout the flow. This is crucial, because if this definition of momenta is only equivalent for the undeformed theory, the Lagrangian and Hamiltonian flows would agree at most upto leading order in $\alpha$ \cite{Ferko:2023iha}.  Therefore, for full consistency, we can find a definition of the LT by equating \eqref{H111111111} and \eqref{111.5} 
\begin{align}
C_1 \dot{\phi}_i+C_2 \phi_i^{\prime}=    \frac{1}{A} \dot{\phi}_i-\frac{B}{A} \phi_i^{\prime}.
\end{align}
Comparing both sides above, this gives the definition of the LT as a map between $A,B$ and $C_{1,2}$:
\begin{align}
\frac{1}{A}=C_1, \quad-\frac{B}{A}=C_2 , \label{L.T}  
\end{align}
which relates phase space and configuration space quantities as they flow in $\alpha$.
We can validate the definition of the LT by observing that it establishes an equivalence between the dynamics in configuration space (governed by \eqref{LEOMNEWWWW}) and the dynamics in phase space (given by \eqref{H111111111},\eqref{H222})
\medskip

\begin{figure}[t]
\centering
\scalebox{0.9}{
\begin{tikzpicture}[node distance=1.8cm]

% Define styles for box and arrow
\tikzstyle{box} = [draw, rounded corners, align=center, fill=gray!5, text width=9cm, minimum height=1cm]
\tikzstyle{arrow} = [thick, ->, >=stealth]

% First box
\node (Lalpha) [box]
{$\mathcal{L}_\alpha = \mathcal{L} \cosh \alpha + \sqrt{\mathcal{L}^2 - \mathcal{P}^2} \sinh \alpha$};

% Arrow to 2nd box
\draw [arrow] (Lalpha) -- ++(0, -2.4)
    node[midway, right=2pt] {$\pi_i = \frac{\partial \mathcal{L}_\alpha}{\partial \dot{\phi}_i}$}
    node[below=2.0cm of Lalpha] (piexpr) [box]
    {$\pi_i \equiv \pi_i(\phi_i, \dot{\phi}_i, \alpha)
    =\left(\cosh \alpha+\frac{\sigma \sinh \alpha}{\sqrt{\sigma^2-\rho^2}}\right) \dot{\phi}_i
    -\frac{\rho \sinh \alpha}{\sqrt{\sigma^2-\rho^2}} \phi_i^{\prime}$};

% Arrow to 3rd box
\draw [arrow] (piexpr) -- ++(0, -2.4)
    node[below=2.0cm of piexpr] (Hblock) [box]
    {
    $\mathcal{H}_\alpha = \cosh \alpha \, \mathcal{H}  - \sinh \alpha \sqrt{\mathcal{H}^2 - \mathcal{J}^2}$ \\[1em]
    $\mathcal{H} = \frac{1}{2} \left( \pi_i \pi_i + \phi^{\prime}_i \phi^{\prime}_i \right)$
    };

% Side connection from first to third box (with arrowheads on both ends)
\draw[thick, Latex-Latex] 
    (Lalpha.east) -- ++(2.5,0) |- (Hblock.east);

% Labels on side connection
\node[right=2.4cm of piexpr.east, align=center]
    {\footnotesize $\dfrac{1}{A} = C_1$ \\[0.9em]\footnotesize $-\dfrac{B}{A} = C_2$};

\end{tikzpicture}}

\caption{The canonical momentum $\pi_i$ derived from the deformed Lagrangian, valid for all values of $\alpha$, serves as conjugate momentum field $\pi_i$ in the deformed Hamiltonian theory. The LT makes sure of the sanctity of this momentum definition throughout the flow.}
\label{fig:diagram1}
\end{figure}
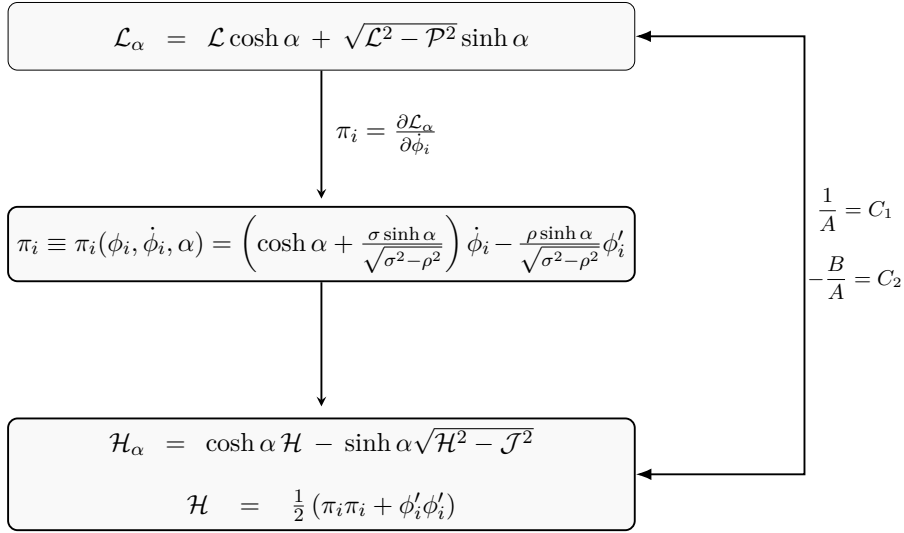

To see this more clearly, let us take the canonical momenta from \eqref{H111111111} and put it back in the Hamiltonian EOM \eqref{H222}, which gives:
\begin{align}
\frac{\partial}{\partial t}  \left(\frac{1}{A} \dot{\phi}_i-\frac{B}{A} \phi_i^{\prime}\right) = A \phi_i^{\prime \prime}+A^{\prime} \phi_i^{\prime}+B\frac{\partial}{\partial x}\left(\frac{1}{A} \dot{\phi}_i-\frac{B}{A} \phi_i^{\prime}\right)+B^{\prime}\left(\frac{1}{A} \dot{\phi}_i-\frac{B}{A} \phi_i^{\prime}\right).\label{HEOMN}
\end{align}
Substituting the LT relations \eqref{L.T} into \eqref{HEOMN} gives
\begin{align}
 %&\frac{\partial}{\partial t}\left(C_1 \dot{\phi}_i+C_2 \phi_i^{\prime}\right)
 %&=\frac{1}{C_1} \phi_i^{\prime \prime}+A^{\prime} \phi_i^{\prime}+B \frac{\partial}{\partial x}\left(C_1 \dot{\phi}_i+C_2 \phi_i^{\prime}\right)+B^{\prime}\left(C_1 \dot{\phi}_i+C_2 \phi_i^{\prime}\right)  \nonumber\\
% =\frac{\phi_i^{\prime \prime}}{C_1}-\frac{C_1^{\prime} \phi_i^{\prime}}{C_1^2}-\frac{C_2}{C_1} \frac{\partial}{\partial x}\left(C_1 \dot{\phi}_i+C_2 \phi_i^{\prime}\right)-\frac{C_1 C_2^{\prime}-C_2 C_1^{\prime}}{C_1^2}\left(C_1 \dot{\phi}_i+C_2 \phi_i^{\prime}\right) \nonumber\\
\dot{C}_1 \dot{\phi}_i+C_1 \ddot{\phi}_i+\dot{C}_2 \phi_i^{\prime}+C_2 \dot{\phi}_i^{\prime}=\frac{\phi_i^{\prime \prime}}{C_1}-\frac{C_1^{\prime} \phi_i^{\prime}}{C_1^2}-&\frac{C_2}{C_1}\left(C_1^{\prime} \dot{\phi}_i+C_1 \dot{\phi}_i^{\prime}+C_2^{\prime} \phi_i^{\prime}+C_2 \phi_i^{\prime \prime}\right)\nonumber\\&\qquad-\frac{C_1 C_2^{\prime}-C_2 C_1^{\prime}}{C_1^2}\left(C_1 \dot{\phi}_i+C_2 \phi_i^{\prime}\right).
\nonumber\end{align}
This can be rearranged and written as
\begin{align}
C_1 \ddot{\phi}_i+2 C_2 \dot{\phi}_i^{\prime}+\left(\dot{C}_1+C_2^{\prime}\right) \dot{\phi}_i+\left(\frac{\dot{C}_2 C_1^2+\left(1-C_2^2\right) C_1^{\prime}+2 C_1 C_2 C_2^{\prime}}{C_1^2}\right) \phi_i^{\prime}+\left(\frac{C_2^2-1}{C_1}\right)\phi_i^{\prime \prime}=0. \label{HEOMNNNNNEW}   \end{align}
From the definitions in \eqref{C_1C_2}, we can easily prove the relation: 
\begin{align}
C_2^2-C_1^2+2 \cosh (\alpha) C_1=1    \implies
C_1^\prime= \frac{\left(2 C_2 C_2^{\prime}\right) C_1-\left(C_2^2-1\right) C_1^{\prime}}{C_1^2},   
\label{3.19}
\end{align}
and substituting these in \eqref{HEOMNNNNNEW} gives
\begin{align}
C_1 \ddot{\phi}_i+2 C_2 \dot{\phi}_i^{\prime}+\left(\dot{C}_1+C_2^{\prime}\right) \dot{\phi}_i+\left(\dot{C}_2+C_1^{\prime}\right) \phi_i^{\prime}+\left(C_1-2 \cosh (\alpha)\right) \phi_i^{\prime \prime}=0. \label{HNNNNNNNNNN}   
\end{align}
We observe immediately that \eqref{HNNNNNNNNNN} and \eqref{LEOMNEWWWW} are identical. Thus, the LT relations, when enforced, ensure the on-shell equivalence between configuration-space dynamics and phase-space dynamics. Therefore, we can confidently conclude that the deformed Hamiltonian \eqref{Hamiltonioncompact} is indeed the LT of the deformed Lagrangian \eqref{Lsigmaandrho}, and this continues to hold throughout the flow as illustrated in Fig.\eqref{fig:diagram1}. 
\medskip

\subsubsection*{(b) Via flow equations}
We can also see the emergence of these LT \eqref{L.T} automatically from the flow equations themselves. This makes sure that the Lagrangian and Hamiltonian solutions of the flow equations are necessarily connected.
We can verify this by putting back \eqref{LagranC1C2} and \eqref{HamilAandB} into the flow equation~\eqref{flowequation} which  gives 
\begin{align}
\frac{1}{2} \frac{\partial A}{\partial \alpha}\left(\pi_i \pi_i\right. & \left.+\phi^{i\, \prime}\phi_i^{\prime}\right)+\frac{\partial B}{\partial \alpha} \pi_i \phi_i^{\prime} \ =-\frac{\partial C_1}{\partial \alpha} \sigma-\frac{\partial C_2}{\partial \alpha} \rho+\phi^{\prime \, 2} \sinh (\alpha). \label{A5}
\end{align}
The quantities on the left-hand side are phase space quantities, so we treat $\left(\pi_i, \phi_i^{\prime}\right)$ as independent of $\alpha$, while those on the right-hand side are configuration space quantities, so we treat $\left(\dot{\phi}_i, \phi_i^{\prime}\right)$ as $\alpha$-independent (see Appendix \eqref{appB}). This works because for the Lagrangian flow, the field derivatives do not really change with $\alpha$, but for each $\alpha$ the Lagrangian itself determines momenta when we take derivatives w.r.t field derivatives. Similarly, focusing on the Hamiltonian flow, the momenta are inherent quantities that do not change with $\alpha$, but the field derivatives are fixed by Hamilton's EOM. In that sense the above equation can also be cast in the form of the operator equivalence of \eqref{OO}\footnote{In fact the form of the LT can be directly inferred from this operator equivalence as well, this is shown in Appendix \eqref{opeqv}.}.
\medskip

Using expression \eqref{H111111111}, we can compute the product of the conjugate momenta in phase space:
\begin{align}
\pi_i\pi_i  =\frac{1}{A^2} \dot{\phi}^2+\frac{B^2}{A^2} \phi^{\prime 2}-\frac{2 B}{A^2} \rho.\label{PIPI}
\end{align}
Replacing it in \eqref{A5} gives
\begin{align}
\frac{1}{2} \frac{\partial A}{\partial \alpha}\left(\frac{1}{A^2} \dot{\phi}^2+\frac{B^2}{A^2} \phi^{\prime \,2}-\frac{2 B}{A^2} \rho+\phi^{\prime\,2}\right)+\frac{\partial B}{\partial \alpha}\left(\frac{1}{A} \rho-\frac{B}{A} \phi^{\prime\,2}\right)\nonumber \\
=-\frac{\partial C_1}{\partial \alpha} \sigma-\frac{\partial C_2}{\partial \alpha} \rho+\phi^{\prime 2} \sinh (\alpha).
\end{align}
After rearranging the terms we arrive at 
\begin{align}
\frac{\partial A}{\partial \alpha}\frac{1}{A^2} \sigma 
- \rho \frac{\partial}{\partial \alpha} \left(\frac{-B}{A}\right)&+ \phi'\,^2 \left( \frac{(B^2 - 1)}{2 A^2} \frac{\partial A}{\partial \alpha} 
+ \frac{1}{2} \frac{\partial A}{\partial \alpha} - \frac{B}{A} \frac{\partial B}{\partial \alpha} \right)\nonumber \\
& = -\frac{\partial C_1}{\partial \alpha} \sigma 
- \frac{\partial C_2}{\partial \alpha} \rho + \phi'^2 \sinh (\alpha),
\end{align}
then matching the coefficients of $\sigma,\rho$, and $\phi'^2$ gives us the identifications:
\begin{align}
\frac{1}{A} &= C_1, \qquad
-\frac{B}{A}= C_2,\label{L2}\\
\sinh (\alpha)&=\frac{\partial A}{\partial \alpha} \left( \frac{(B^2-1)}{2A^2} + \frac{1}{2} \right) 
- \frac{B}{A} \frac{\partial B}{\partial \alpha}.
\label{ID1}\end{align}
Here, \eqref{L2} are the consistency equations for the Legendre transform that we already have \eqref{L.T}, and it can be shown that \eqref{ID1} is just an identity (see appendix \eqref{APPB}) which is automatically satisfied given the form of $A,B$. Thus, we have shown that phase space flows and configuration space flows, under the $\sqrt{T\overline T}$ type operators, are interrelated through a valid LT directly from the flow equation \eqref{flowequation}. This result, i.e. \eqref{L.T} is a central point of this work, and this will be used throughout this manuscript to stress how deep the structure of the classical flow is.

\section{Emergence of Carroll Conformal symmetries}\label{sect4}
The main focus of this work, as alluded in the introduction, is to see how the marginal operator induced flows, starting from a relativistic seed theory, actually lead to Carroll Conformal symmetries at particular points in the $\alpha$ parameter space. With $\sqrt{T\overline T}$ type flows, this rather surprising fact has been discussed before in \cite{Rodriguez:2021tcz,Tempo:2022ndz, Bagchi:2022nvj}. In what follows, let us build on that based on our earlier discussion.

\subsection{Carroll as ultra-boosted structure}
In \cite{Bagchi:2022nvj} a novel perspective of moving from CFT to CCFT was first put forward, based on the idea that CCFTs work as a null limit of their conformal cousin. The main formalism dictates a ``boost'' that mixes the holomorphic and anti-holomorphic coordinates ($z,\bar{z}$) for a 2d CFT, or equivalently mixes the Kac-Moody currents $J,\bar{J}$. This in turn basically rescales the 2d effective speed of light. As this ``boost'' parameter is dialed to infinity, the infinite rescaling between spatial and temporal coordinates takes the symmetry structure to a Carrollian (Galilean) one. In \cite{Bagchi:2024unl} the flow of the quantum spectrum for the same theory was also discussed. 
\medskip

More precisely, this large boost is implied using the hyperbolic rotation of the holomorphic coordinates:

\begin{align}
\left(\begin{array}{c}
z_{\mathrm{L}} \\
z_{\mathrm{R}}
\end{array}\right)=\left(\begin{array}{cc}
\cosh \phi & -\sinh \phi \\
-\sinh \phi & \cosh \phi
\end{array}\right)\left(\begin{array}{l}
z \\
\bar{z}
\end{array}\right),\end{align}
where $\phi=\text{tanh}^{-1}(\beta)$ and $\cosh \phi=\frac{1}{\sqrt{1-\beta^2}},~~ \sinh \phi=\frac{\beta}{\sqrt{1-\beta^2}}$, with $\beta$ being the rapidity parameter. Here and what follows, we will be putting the CFT on the cylinder so that we can choose $z = \tau+\theta$, and $\bar{z}= \tau-\theta$ with $\theta \in [0,2\pi]$. Following through with this deformation at the level of currents, this effectively adds a current-current deformation term to the Hamiltonian density:
\begin{equation}
    \mathcal{H}\longrightarrow \mathcal{H}\pm2\alpha J\bar{J},
\end{equation}
note that for a free scalar CFT, the Hamiltonian density can be written using bilinears of currents via a Sugawara construction, where:
\begin{equation}
    T = \frac{1}{k}\sum_a J^aJ^a,~~~~ \bar{T} = \frac{1}{k}\sum_a \bar{J}^a\bar{J}^a.
\end{equation}
For general values of $\alpha$ this deformed Hamiltonian can again be diagonalised via some rotated currents $J_{L,R}$, but this fails at the particular points $\alpha=\pm 1$, or the points with infinite boosts. These are, intriguingly, the points where the Hamiltonian densities at two spatial points (Poisson) commute, giving rise to Carrollian (or Galilean) degrees of freedom.
\medskip

Now, the reader may wonder, why are we discussing this in our context of $\sqrt{T\bar{T}}$ deformations? And the answer is straightforward: i.e. for a situation where all currents carry the same $U(1)$ indices, we can write:
\begin{equation}
    \sqrt{T\bar{T}} \simeq J\bar{J}.
\end{equation}
So similarly, one should expect the generic $\sqrt{T\bar{T}} $ operator should also instill the same flows that take a Lorentzian CFT to a non-Lorentzian one. In doing so, the natural ``boosted'' CFT concept may get a bit obscured, but in what follows we will try to argue the parallel flow with  $\sqrt{T\bar{T}} $ operator may be viewed somewhat similarly, since, as we will see, the end point of the flow lands one at the same Carrollian theory. In what follows, we will elucidate on this from both Hamilonian and Lagrangian viewpoint.

\subsection{Carroll from Hamiltonian flows}
We have already discussed the structure of the deformed Hamiltonian in preceeding sections.
The main idea for interpolating from relativistic to Carrollian CFT is pretty straightforward, we can write a scaled version of \eqref{HTTD} as
\begin{align}
\hat{\mathcal{H}}_\alpha=\frac{1}{\cosh{\alpha}}\mathcal{H}_\alpha=\mathcal{H} -2 \sqrt{T \bar{T}} \tanh (\alpha)\label{scaledHTTD},    
\end{align}
We can write the undeformed Hamiltonian and momentum density in terms of holomorphic and antiholomorphic stress-energy tensors as\footnote{Recall that in the usual Inönü-Wigner contractions, we would define generators $\mathcal{\tilde H}=\epsilon(T + \bar{T})$ and $\mathcal{\tilde J} = \mathcal{J}=T - \bar{T}$, with $\epsilon \to 0$, which will give rise to the canonical Conformal Carroll algebra. The fourier modes version of these would be $L_n=\mathcal{L}_n-\bar{\mathcal{L}}_{-n}, M_n = \epsilon(\mathcal{L}_n+\bar{\mathcal{L}}_{-n})$, with $\mathcal{L}$'s being the Virasoro generators. Although the effect is similar, the mechanism of the flow via $\sqrt{T\bar{T}}$ is distinguished from that of a direct contraction. }
\begin{align}
\mathcal{H}=T + \bar{T},~~~~~
\mathcal{J}=T - \bar{T} \label{J=T-T}.
\end{align}
To understand the deformed algebra, we recall the algebra of stress tensors, i.e. undeformed conformal algebra
in 2d, written on a cylinder with coordinates $(\tau,\theta)$:
\begin{align}\label{Talgebra}
 \{T(\theta), T(\theta^{\prime})\}_{\mathrm{PB}}&=\left(2 T(\theta)\, \partial_\theta + \partial_\theta T(\theta)\right) \delta(\theta - \theta^{\prime}), \nonumber \\
 \{\bar{T}(\theta), \bar{T}(\theta^{\prime})\}_{\mathrm{PB}}&=-\left(2 \bar{T}(\theta)\, \partial_\theta + \partial_\theta \bar{T}(\theta)\right) \delta(\theta - \theta^{\prime}).
\end{align}
For now we focus only the Poisson brackets, however, an extension to commutators is straightforward.
Now using \eqref{scaledHTTD} and \eqref{J=T-T} together we can form a slightly deformed Conformal algebra involving the scaled Hamiltonian 
\begin{align}
\begin{aligned}
 \left\{\mathcal{J}(\theta), \mathcal{J}\left(\theta^{\prime}\right)\right\}_{\mathrm{PB}}&=\left(2 \mathcal{J}(\theta)\, \partial_\theta + \partial_\theta \mathcal{J}(\theta)\right) \delta\left(\theta-\theta^{\prime}\right), \\
 \left\{\mathcal{J}(\theta), \hat{\mathcal{H}}_\alpha\left(\theta^{\prime}\right)\right\}_{\mathrm{PB}}&=\left(2 \hat{\mathcal{H}}_\alpha(\theta)\, \partial_\theta + \partial_\theta \hat{\mathcal{H}}_\alpha(\theta)\right) \delta\left(\theta-\theta^{\prime}\right), \\
\left\{{\mathcal{\hat H}}_\alpha(\theta), \hat{\mathcal{H}}_\alpha\left(\theta^{\prime}\right)\right\}_{\mathrm{PB}}&=\left(1-\tanh^2 \alpha\right)\left(2 \mathcal{J}(\theta)\, \partial_\theta + \partial_\theta \mathcal{J}(\theta)\right) \delta\left(\theta-\theta^{\prime}\right).
\end{aligned}
\end{align}
This algebra, for finite $\alpha$, can again be made to look like the usual conformal algebra if we undo the rescaling on the generators $\hat{\mathcal{H}}_\alpha$. But of course this can't be done when the scaling becomes singular. At these special  points the Hamiltonians commute with each other at two separate spatial points, the usual Dirac-Schwinger conditions for energy densities in relativistic field theories \cite{Henneaux:2021yzg} vanish, and the algebra eventually transforms to the 2d Conformal Carroll, i.e. $\text{BMS}_3$ algebra\footnote{Written in term of fourier modes, the corresponding Poisson brackets look like: \begin{equation}
    \{L_m,L_n\}=-i(m-n)L_{m+n},~~ \{L_m,M_n\}=-i(m-n)M_{m+n}, ~~\{M_m,M_n\}=0.
\end{equation}}, which we expect at infinite boosts ($\alpha=\infty$ i.e. $\tanh^2\alpha =1$)
\begin{align}\label{interpalgebra}
\begin{aligned}
& \left\{\mathcal{J}(\theta), \mathcal{J}\left(\theta^{\prime}\right)\right\}_{\mathrm{PB}}=\left(2 \mathcal{J}(\theta)\, \partial_\theta + \partial_\theta \mathcal{J}(\theta)\right) \delta\left(\theta-\theta^{\prime}\right), \\
& \left\{\mathcal{J}(\theta), \hat{\mathcal{H}_{\infty}}\left(\theta^{\prime}\right)\right\}_{\mathrm{PB}}=\left(2 {\mathcal{\hat H}_{\infty}}(\theta)\, \partial_\theta + \partial_\theta {\mathcal{\hat H}_{\infty}}(\theta)\right) \delta\left(\theta-\theta^{\prime}\right), \\
& \left\{{\mathcal{\hat H}_{\infty}}(\theta), {\mathcal{\hat H}_{\infty}}\left(\theta^{\prime}\right)\right\}_{\mathrm{PB}}= 0.
\end{aligned}
\end{align}
Where note that the scaled Hamiltonian density at these special points:
\begin{align}
\hat{\mathcal{H}}_\infty=\lim _{\alpha \rightarrow \infty}\frac{\mathcal{H}_\alpha}{\cosh \alpha} =\mathcal{H} -2 \sqrt{T \bar{T}} 
\label{H2TbarT}\end{align} 
is a Carrollian one as it exactly commutes with itself at different spatial points. Note that the limit $\alpha \to -\infty $ where the (``hatted'') Hamiltonian is $\mathcal{H}+2 \sqrt{T \bar{T}} $ is equally valid as a Carroll Hamiltonian, a fact which we will come back to later. 
\medskip

Note that this is a very different way of achieving the Carroll algebra from conformal algebra, which is usually done using an Inönü-Wigner contraction of the generators \cite{Inonu:1953sp}.
Now we are going to see how this limit can be thought of in a different perspective, i.e. how the deformation parameter generates a continuous $\text{SO}(1,1)$ rotation in a plane spanned by the quantities $(\mathcal{H},2\sqrt{T\overline T})$. Let us recall the form of the deformed Hamiltonian from equation \eqref{HTTD}:
\begin{align}
\mathcal{H}_\alpha=\mathcal{H} \cosh (\alpha)-K \sinh (\alpha)\label{HTTK},   
\end{align}
where we have defined $K=2 \sqrt{T \bar{T}}$. Now, going back to the idea of ``boosting'' Lorentz quantities, let us suppose we define a deformed version of the operator, i.e. $K_\alpha$ in a similar way:  
\begin{align}
K_\alpha=-\mathcal{H} \sinh (\alpha)+K \cosh (\alpha)\label{HTK},   
\end{align}
then it is evident that $K_\alpha$ and $\mathcal{H}_\alpha$ together change under some analogue of 2d Lorentz boosts, having some continuous $\text{SO}(1,1)$ (hyperbolic) rotations between these two. One could ask why would we compare this to a ``boost''?
That is since we can define the momentum density as an invariant quantity under these hyperbolic rotations generated by \eqref{HTTK} and \eqref{HTK}:
\begin{align}
\mathcal{H}^2_\alpha- K^2_\alpha =  \mathcal{H}^2-K^2= \mathcal{J}^2\label{invariant j}.
\end{align}
Comparing with Lorentzian boost, we can think of $\mathcal{J}$ as a proper distance in some ambient 2d spacetime. 
The deformed Hamiltonian \eqref{HTTD} can be written as sum of deformed stress-energy tensors $\mathcal{H}_\alpha=T_\alpha+\bar T_\alpha$, 
and with  $K_\alpha=2 \sqrt{T_\alpha\bar T_\alpha}$, we can write the boost operations in a more suggestive way:
\begin{align}
\mathcal{H}_\alpha=&T_\alpha+\bar{T}_\alpha=\cosh (\alpha)(T+\bar{T})-2 \sinh (\alpha) \sqrt{T \bar{T}},  \nonumber\\
K_\alpha=&2\sqrt{T_\alpha \bar T_\alpha}=-\sinh(\alpha)(T+\bar T) + 2\sqrt{T\overline T}\cosh(\alpha).
\label{K}\end{align}
\footnote{ One can also note, as we mentioned before, we can define equivalently $K_\alpha=-2\sqrt{T_\alpha T_\alpha}$, which leads to slightly different rotated objects:
\begin{align}
\mathcal{H}_\alpha=&T_\alpha+\bar{T}_\alpha=\cosh (\alpha)(T+\bar{T})-2 \sinh (\alpha) \sqrt{T \bar{T}},  \nonumber\\
K_\alpha=&-2\sqrt{T_\alpha T_\alpha}=\sinh(\alpha)(T+\bar T) - 2\sqrt{T\overline T}\cosh(\alpha).
\label{Kn}\end{align}}
Then solving the above equations it can be shown that the components of the stress energy tensor transform as:
\begin{align}
& T_\alpha=T \cosh ^2\left(\frac{\alpha}{2}\right)+\bar{T} \sinh ^2\left(\frac{\alpha}{2}\right)-\sqrt{T \bar{T}} \sinh (\alpha), \\
& \bar{T}_\alpha=\bar{T} \cosh ^2\left(\frac{\alpha}{2}\right)+T \sinh ^2\left(\frac{\alpha}{2}\right)-\sqrt{T \bar{T}} \sinh (\alpha).
\end{align}
It is crucial to note that for finite $\alpha$, the $T_\alpha$ together with $\bar T_\alpha$ still constitute the usual conformal algebra in \eqref{Talgebra}, something one can easily verify. Thus, this is nothing but a nonlinear automorphism of the Virasoro algebra, when written in terms of the fourier modes. This was discussed heuristically in \cite{Tempo:2022ndz}, but our motivation, as stated before, is from this hyperbolic rotation between $K_\alpha$ and $\mathcal{H}_\alpha$, which directly brings this about.
\medskip

Now that we understand this ``boosted'' structure of the generators, it is natural to explore the infinite boost regime, given by $\alpha\to\pm\infty$.
Let us consider the case  $\alpha\to+\infty$, where it can be shown that 
\begin{align}
 \lim _{\alpha\to \infty}\mathcal{H}_\alpha  = -\lim _{\alpha\to\infty}K_\alpha. \label{H=K}
\end{align}
So we see $\mathcal{H}_\alpha$ and $K_\alpha$ converge to the same point under infinite boosts, and similar calculation for $\alpha \to -\infty$ can lead to $\mathcal{H}_\alpha \to K_\alpha$. One can then think of these limits as reaching something akin to the ``light-cones'' of the $(\mathcal{H},K)$ plane where we have implemented the infinite boost. To bypass running into infinities, one can also work with the scaled quantities ${\hat{\mathcal{H}}_\alpha}, {\hat{{K}}_\alpha}$, where all hatted quantities are scaled by $\cosh\alpha$. Substituting  \eqref{H=K} in \eqref{invariant j} gives
\begin{align}
\lim _{\alpha\to \infty}\hat{\mathcal{H}}_\alpha^2-{\hat{{K}}_\alpha^2}=\mathcal{\hat J}^2=0.    
\label{J20}\end{align}
This is the peculiar property of infinite boosts. In the usual Lorentzian case, under infinite boosts, the line element becomes zero, and it is the property of a light-like surface that we get through these infinite boosts. In this case the above constraint i.e. $\mathcal{\hat J}^2=0$ acts like the speed of light going to zero limit in that  case\footnote{Equivalently one can think of $c=\frac{1}{\cosh\alpha}$ as the effective ambient speed of light, which goes to zero, and the scaling works accordingly. }. Equivalently, one can also see the emergence of Carroll symmetry in the Hamiltonian case from the vanishing of the $T^{10}$ component of the stress tensor at the extreme points (see Appendix \eqref{StressTvanish}) on the flow. 
\medskip

Note that, for finite $\alpha$ the condition between ${\hat{\mathcal{H}}_\alpha}, {\hat{{K}}_\alpha}$ does not run into any issues. Moreover, in this regime the RHS of \eqref{J20} remains finite, and by demanding it to vanish this condition leads to:
\begin{align}
\mathcal{J}=T-\bar T =0\Longrightarrow T=\bar T .
\label{Chirality}\end{align}
This is a very intriguing case, but could be understood as a degenerate case where the scalar field becomes a chiral one, i.e., $ J_a(\phi) = \bar{J}_a(\phi)$. We will, however, always focus on the infinite $\alpha$ case going forward.

\subsection{Carroll from Lagrangian flows}

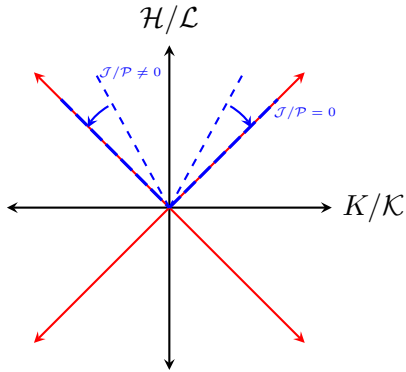
\begin{figure}
\centering
\begin{tikzpicture}[scale=1.8, >=stealth]

%%% -------- LEFT PANEL:  J ≠ 0 -------- %%%
\begin{scope}[shift={(-1.7,0)}]

% Axes (double-headed)
\draw[thick,<->] (-1.2,0) -- (1.2,0) node[right] {$K/\mathcal{K}$};
\draw[thick,<->] (0,-1.2) -- (0,1.2) node[above] {$\mathcal{H}/\mathcal{L}$};

% Red light-cone (double-headed)
\draw[red,thick,<->] (-1,-1) -- (1,1);
\draw[red,thick,<->] (-1,1) -- (1,-1);

% Blue dashed lines (J ≠ 0, inside cone)
\draw[blue,dashed,thick] (0,0) -- (0.55,1.0);
\draw[blue,dashed,thick] (0,0) -- (-0.55,1.0);

% Blue rotation arrows
\draw[blue,->,thick] (0.45,0.75) arc[start angle=60, end angle=28, radius=0.4];
\draw[blue,->,thick] (-0.45,0.75) arc[start angle=120, end angle=152, radius=0.4];

% Label
\node[blue,scale=0.8] at (-0.29,1) {\tiny $\mathcal{J}/\mathcal{P} \neq 0$};
% Blue thick dashed lines overlapping red cone (J = 0)
\draw[blue,thick,dash pattern=on 6pt off 2pt,line width=1.2pt] (0,0) -- (0.8,0.8);
\draw[blue,thick,dash pattern=on 6pt off 2pt,line width=1.2pt] (-0.8,0.8) -- (0,0);

% Label
\node[blue,scale=0.8] at (1,0.70) {\tiny $\mathcal{J}/\mathcal{P} = 0$};
\end{scope}
\end{tikzpicture}
\caption{An illustration of the `boosted' nature of the transformations in $\mathcal{H}-K$ space and $\mathcal{L}-\mathcal{K}$ space. Reaching the corresponding `lightcones' signify Carroll symmetries setting in.  }
\end{figure}

Once the Hamiltonian perspective is well understood, we can look at the Lagrangian point of view for appearence of Carrollian symmetries.
Like in the previous case, we can construct a Lorentz-like 2d plane with the Lagrangian and the configuration space deformation operator $\mathcal{K}=2 \sqrt{\mathcal{T} \overline{\mathcal{T}}}$ as the two axes.
We can then envision a hyperbolic rotation in this plane using these quantities
\begin{align}
  \mathcal{L}_\alpha&=\mathcal{L}\cosh (\alpha)+ \mathcal{K}\sinh (\alpha),
 \label{deforLagra(L-K)}\\
  \mathcal{K}_\alpha&=\mathcal{K} \cosh (\alpha)+\mathcal{L} \sinh (\alpha).
\label{deformedKofLagrangian}\end{align}

The first one is nothing but the deformed Lagrangian here, just like the Hamiltonian picture.
 The $\mathcal{K}_\alpha=2 \sqrt{\mathcal{T}_\alpha \bar{\mathcal{T}}_\alpha}$ is the deformation operator at non-zero value of $\alpha$.
We can define an invariant under these hyperbolic rotations as
\begin{align}
\mathcal{L}^2_\alpha-\mathcal{K}^2_\alpha=   \mathcal{L}^2-\mathcal{K}^2=\mathcal{P}^2.\label{INVARIANT}
\end{align}
This $\mathcal{P}^2$ is defined in \eqref{P_N^2}.
However, now $\mathcal{T}_\alpha$ and $\bar{\mathcal{T}}_\alpha$ are the components of the stress-energy tensor corresponding to the deformed Lagrangian \eqref{FL1}
found by using
\begin{align}
\mathcal{T}^{\mu \nu}_\alpha=\frac{\partial \mathcal{L}_\alpha}{\partial\left(\partial_\mu \phi_i\right)} \partial^\nu \phi_i-g^{\mu \nu} \mathcal{L}_\alpha.\label{86}  \end{align}
The first term on the right hand side can be computed to be
\begin{align}
 \frac{\partial \mathcal{L}_\alpha}{\partial\left(\partial_\mu \phi_i\right)} \partial^\nu \phi_i=\left(\cosh \alpha+\frac{\mathcal{L} \sinh \alpha}{\sqrt{\mathcal{L}^2-\mathcal{P}^2}}\right) \partial^\mu \phi_i \partial^\nu \phi_i-\frac{\mathcal{P}^2 \sinh \alpha}{\sqrt{\mathcal{L}^2-\mathcal{P}^2}} g^{\mu \nu},  
\end{align}
where we have used the following:
\begin{align}
 \frac{\partial \mathcal{P}}{\partial (\partial_\mu\phi_k)} \partial^\nu\phi_k&=\frac{1}{\mathcal{P}}\bigg[\partial_\mu \phi_i \partial^\mu \phi_i \partial^\mu \phi_k \partial^\nu \phi_k-\frac{1}{2} \partial_\alpha \phi_k \partial^\alpha\phi_k\partial^\mu \phi^j  \partial^\nu \phi_k\bigg]=\mathcal{P}g^{\mu\nu}.
\label{83}\end{align}
Note that \eqref{83} can be obtained by utilizing \eqref{GM1}.
Therefore the deformed stress tensor is written as a particularly scaled version of the undeformed stress tensor:
\begin{align}
\mathcal{T}^{\mu \nu}_\alpha &= 
\cosh \alpha\left(\partial^\mu \phi_i \partial^\nu \phi_i-g^{\mu \nu}\mathcal{L}\right)+\frac{\mathcal{L} \sinh \alpha}{\sqrt{\mathcal{L}^2-\mathcal{P}^2}}\bigg(\partial^\mu \phi_i \partial^\nu \phi_i-\frac{1}{\mathcal{L}}\mathcal{P}^2g^{\mu \nu}-\mathcal{L}g^{\mu \nu}+\frac{1}{\mathcal{L}}\mathcal{P}^2g^{\mu \nu}\bigg)
 \nonumber\\
 &=\left(\cosh \alpha+\frac{\mathcal{L} \sinh \alpha}{\sqrt{\mathcal{L}^2-\mathcal{P}^2}}\right)\mathcal{T}^{\mu\nu}.
\label{deformedstressenergytensor}\end{align}
This fits right into the notion of the hyperbolic rotations, and indeed \eqref{deformedKofLagrangian} can be easily verified by directly using \eqref{deformedstressenergytensor}.
\medskip

We now focus again on the $\alpha \to \pm \infty$ case, or what we have been calling ``infinite boosts'', in which case $\mathcal{L}_\alpha$ and $\mathcal{K}_\alpha$ become asymptotically equal and eventually intersect with the light cone of the $(\mathcal{L},\mathcal{K})$ plane:
\begin{align}
    \lim _{\alpha \rightarrow \infty} \mathcal{L}_\alpha=\lim _{\alpha \rightarrow \infty} \mathcal{K}_\alpha.
\label{L=K}\end{align}
As done in the Hamiltonian case before, we would again define certain scaled quantities (w.r.t $\cosh\alpha$), with which one can substitute the above in \eqref{INVARIANT}. Finally, we obtain 
\begin{equation}
\lim _{\alpha \rightarrow \infty}\mathcal{\tilde L}^2_\alpha-\mathcal{\tilde K}^2_\alpha=\mathcal{\tilde P}^2 = 0,
\label{NullP}\end{equation}
which gives the null hypersurface in this 2d plane. 
\medskip

This limit in Lagrangian formalism can also be thought of in an intuitive way. Note that $\mathcal{\tilde P}^2 = 0$, looking at \eqref{GM2}, also means a scaled $\operatorname{det}\left[\mathbf{M}\right]=0$, i.e. the field space pullback metric becomes degenerate in this case. This would be important when we discuss a similar degeneration of string worldsheet metric in the later parts of this work, which brings about Carrollian structures as well.

\subsection{Carroll via Legendre Transformation}
It should be clear to the reader now that both the phase space and configuration space dynamics leads to Carrollian structures at the edges of the parameter space. This also leads to the question of whether the Legendre transformation between these two pictures, as discussed in \eqref{L.T}, stays valid as one takes $\alpha \to \pm \infty$. To see this, lets go back to the Lagrangian in \eqref{Lsigmaandrho}
\begin{align}
\mathcal{L}_\alpha=\mathcal{L} \cosh (\alpha)+\sqrt{\sigma^2-\rho^2} \sinh (\alpha).    \end{align}
Let us scale this Lagrangian in a way such that
$\mathcal{L}_\alpha\to\frac{1}{\cosh\alpha}\mathcal{L}_\alpha$, and also $\dot\phi_i\to \cosh \alpha~ \dot{\phi}_i$, which finally gives
\begin{align}
\hat{\mathcal{L}}_\alpha=\hat{\sigma} +\sqrt{\hat{\sigma}^2-\hat{\rho}^2} \tanh (\alpha)-\phi_i^{\prime\,2},  \label{ElectricLagrangiannew}\end{align} 
where the hatted quantities are derived via scaling the \textit{time direction} implicitly:
\begin{align}
 \hat{\sigma}=\frac{1}{2}\left[(\cosh\alpha~\dot\phi_i)^2+\phi_i^{\prime\,2}\right],  \qquad \hat\rho = \cosh \alpha ~\dot{\phi}_i\phi_i^\prime.
\end{align}
These are not to be confused with the naively scaled Lagrangian quantities in \eqref{NullP}.
Taking the limit $\alpha\to\pm\infty$ should now smoothly take the theory to some sector of the $\text{CCFT}_2$. For a single scalar field, it's easy to see that \eqref{ElectricLagrangiannew} reduces to
 \begin{align}
\hat{\mathcal{L}}_\alpha=\frac{\dot\phi^2}{2(1-\tanh \alpha)}-\phi^{\prime 2} \frac{1+\tanh \alpha}{2},     
 \label{Electric.lagrangian.single.scalar}\end{align}
 and one can easily verify that \eqref{Electric.lagrangian.single.scalar} is found by usual LT of $\sqrt{T\overline T}$ deformed scaled Hamiltonian for a single scalar field, given by:
 \begin{align}\label{singH}
\hat{\mathcal{H}}_\alpha=\left(\frac{\tilde{\pi}^2+\phi^{\prime\,2}}{2}\right)-\left(\frac{\tilde{\pi}^2-\phi^{\prime 2}}{2}\right)\tanh (\alpha),  
 \end{align}
where $\tilde{\pi}$ is defined by the field derivatives of redefined Lagrangian \eqref{ElectricLagrangiannew}:
\begin{align}
\tilde{\pi}_i=\frac{\partial \hat{\mathcal{L}}_\alpha}{\partial \dot{\phi}_i}&=\left(1+\frac{\hat{\sigma} \tanh \alpha}{\sqrt{\hat{\sigma}^2-\hat{\rho}^2}}\right) \cosh ^2 \alpha \dot{\phi}_i-\frac{\hat{\rho} \tanh \alpha}{\sqrt{\hat{\sigma}^2-\hat{\rho}^2}} \phi_i^{\prime} \cosh \alpha,\label{Electric.Ca.Mom.1}\\
\frac{\partial \hat{\mathcal{L}}_\alpha}{\partial \phi_i^{\prime}}&=-\phi_i^{\prime} +\frac{\tanh \alpha}{\sqrt{\hat{\sigma}^2-\hat{\rho}^2}}\left(\hat{\sigma} \phi_i^{\prime}-\hat{\rho} \dot{\phi}_i\cosh \alpha\right).
\label{Electric.Ca.Mom.2}\end{align}
From \eqref{singH} it can be checked that the Hamiltonian interpolates between $\mathcal{H}_{-\infty}=\tilde{\pi}^2$ to $\mathcal{H}_{+\infty}=\phi'^2$ as $\alpha \to \pm\infty$ \footnote{Note that if we had not done the explicit time scaling beforehand, the action of single deformed scalar field would just have been $${\mathcal{L}}_\alpha=\frac{1}{2}\left[{\dot\phi^2}{(1-\tanh \alpha)}-\phi^{\prime 2} ({1+\tanh \alpha)}\right]$$
 but our motivation has been to connect the action to the Hamiltonian.}, somewhat akin to a direct electric or magnetic limit of, for example, \cite{Henneaux:2021yzg}.
Let us define the quantities closely related to \eqref{C_1C_2}
\begin{align}
\hat{C}_1=1+\frac{\hat{\sigma} \tanh (\alpha)}{\sqrt{\hat{\sigma}^2-\hat{\rho}^2}}, \quad \hat{C}_2=-\frac{\hat{\rho} \tanh (\alpha)}{\sqrt{\hat{\sigma}^2-\hat{\rho}^2} \label{6.10}
}.  \end{align}
Using this we can rewrite \eqref{ElectricLagrangiannew}, \eqref{Electric.Ca.Mom.1}, \eqref{Electric.Ca.Mom.2} in their linearised avatars
\begin{align}
\begin{aligned}
\hat{\mathcal{L}}_\alpha & =\hat{C}_1 \hat{\sigma}+\hat{C}_2 \hat{\rho}-\phi_i^{\prime 2},  \\
\tilde{\pi}_i=\frac{\partial \hat{\mathcal{L}}_\alpha}{\partial \dot{\phi}_i} & =\hat{C}_1 \dot{\phi}_i\cosh^2\alpha+\hat{C}_2 \phi_i^{\prime} \cosh\alpha \label{Ele.Canon.Momentum}\\
\frac{\partial \hat{\mathcal{L}}_\alpha}{\partial \phi_i^{\prime}} & =\hat{C}_1 \phi_i^{\prime}+\hat{C}_2 \dot{\phi}_i\cosh\alpha-2 \phi_i^{\prime}.
\end{aligned}
\end{align}
One can also verify that the $\hat{C}_1$ and $\hat{C}_2$, like their unhatted counterparts, do not participate in the process of taking derivatives w.r.t. fields for the scaled Lagrangian.
\medskip

Now recall, the Hamiltonian after scaling by $\frac{1}{\cosh\alpha}$ is 
\begin{align}
\hat{\mathcal{H}}_\alpha=\mathcal{H}-\tanh (\alpha) \sqrt{\mathcal{H}^2-\mathcal{J}^2}.    \label{El.Hamiltonion}
\end{align}
This quantity, as discussed before, is important for us to see the interpolating symmetry algebra. In fact, the single scalar field version of this is indeed \eqref{singH}. For the general multi-field case, we can define two phase space quantities $\hat{A}$ and $\hat{B}$, closely related to \eqref{ABdef}:
\begin{align}
\hat{A}=1-\frac{\tanh (\alpha) \mathcal{H}}{\sqrt{\mathcal{H}^2-\mathcal{J}^2}}, \quad 
\hat{B}=\frac{\tanh (\alpha) \mathcal{J}}{\sqrt{\mathcal{H}^2-\mathcal{J}^2}}.   
\label{4.41}
\end{align}

Here, what we find is the the EOM are just a scaled version of those in section~\eqref{sec3.2}, where $A$ gets replaced by $\hat{A}$ and all $B$ gets replaced by $\hat{B}$. This, in turn, should be equivalent to the EOM coming from the scaled Lagrangian above, if the LT still works. For example:
\begin{align}
\dot{\phi}_i(x, t)=\frac{\partial \hat{\mathcal{H}}_\alpha}{\partial \tilde{\pi}_i}=\hat{A}(\alpha) \tilde{\pi}_i+\hat{B}(\alpha) \phi_i^{\prime}.  \label{4..42}  
\end{align}
This can be rearranged to fix the momenta:
\begin{align}
\tilde{\pi}_i=\frac{1}{\hat{A}} \dot{\phi}_i-\frac{\hat{B}}{\hat{A}} \phi_i^{\prime}. \label{Electricmom}   
\end{align}
Equating \eqref{Electricmom} and \eqref{Ele.Canon.Momentum} gives the final condition that relates the Lagrangian $\hat{\mathcal{L}}$ in \eqref{ElectricLagrangiannew} and Hamiltonian $\hat{\mathcal{H}}$ in \eqref{El.Hamiltonion} given finite values of $\alpha$:
\begin{align}
\hat{C}_1 \cosh ^2 \alpha =\frac{1}{\hat{A}}, \quad \hat{C}_2\cosh\alpha =-\frac{\hat{B}}{\hat{A}}    
\label{Electric.LT}.
\end{align}
One can immediately compare these two with our original LT relations in \eqref{L.T}.
We would like to validate these LT relations, between the ``hatted'' Lagrangian and the ``hatted'' Hamiltonian, through a stronger analysis at the level of EOM in phase space and configuration space. This can be done following the steps in section \eqref{LTdef} for the ``unhatted'' dynamical quantities to confirm that the above are the right definition of the LT. For brevity we omit the details of that computation here. 
 
%\subsection{Equivalence of flows at Carroll point}

\section{Electric and Magnetic limits}\label{sect5}

Now that we understand how the LT between the two pictures work with the ``hatted'' variables, the one best suited to be taken Carroll limits on, we need to discuss how these LT actually hold up at the very edge of the parameter space. Since the LT themselves are explictly and implictly dependent on $\alpha$, we may expect that putting $\alpha \to \pm\infty$ may lead to two mutually exclusive theories. This is a highly subtle idea, and in what follows, we will delve more into this matter. 

\subsection{Flowing via Legendre transform}
\subsubsection*{Actions on the edge}
In this section, our focus is on explicitly taking the  $|\alpha|\to\infty$ on our actions. Let's start from the ``hatted'' form of the Lagrangian \eqref{ElectricLagrangiannew}:
\begin{equation}
\hat{\mathcal{L}_\alpha}=\hat{\sigma}+\tanh(\alpha)\sqrt{\hat{\sigma}^2-\hat{\rho}^2}-\phi^{\prime 2}_i. \label{7.1}
\end{equation}
In terms of $\hat{C}_1$ and $\hat{C}_2$, this can be compactly written in the linearised manner as before seen in \eqref{LagranC1C2}:
\begin{equation}
\hat{\mathcal{L}_\alpha}=\hat{C_1}\hat{\sigma}+\hat{C_2}\hat{\rho}-\phi^{\prime 2}_i. \label{7.2}
\end{equation}
To simplify the expansion process, we consider
\begin{equation}
    P=\frac{\hat{\sigma}}{\sqrt{\hat{\sigma}^2-\hat{\rho}^2}};\qquad Q=\frac{\hat{\rho}}{\sqrt{\hat{\sigma}^2-\hat{\rho}^2}}. \label{7.3}
\end{equation}
In this language $\hat{C}_1$ and $\hat{C}_2$ are defined as
\begin{equation}
\hat{C}_1=1+\tanh(\alpha)P;\qquad \hat{C}_2=-\tanh(\alpha)Q. \label{7.4a}
\end{equation}
We will do another redefinition to make better sense of the  $|\alpha|\to\infty$ regime. Consider $\epsilon^{-1}=\cosh(\alpha)$, then $ \tanh(\alpha)=\pm\sqrt{1-\epsilon^2}$ \footnote{Note that considering $\frac{1}{\cosh\alpha}$ acts as an ambient speed of light, we can think of $\epsilon$ as akin to a contraction parameter.}. For $\alpha$ being very large (i.e. $\epsilon$ very small), we could expand:
\begin{equation}
\tanh(\alpha)\approx\pm(1-\frac{1}{2}\epsilon^2).
\end{equation}
Here now comes the difference between the two large values of $\alpha$.  This means for $\alpha\to+\infty$, we choose the positive branch and for $\alpha\to-\infty$, the negative branch. Just following through with this logic gives us:
\begin{equation}
\hat{\sigma}=\frac{1}{2}\Big(\frac{\dot{\phi}^2_i}{\epsilon^2}+\phi^{\prime 2}_i\Big);\qquad \hat{\rho}=\frac{\dot{\phi}_i \phi^{\prime}_i}{\epsilon}. \label{7.5}
\end{equation}
Then $P$ and $Q$ are written up to $\epsilon^2$ order as
\begin{equation}
P=1+2\epsilon^2\Big(\frac{\dot{\phi}_i \phi^{\prime}_i}{\dot{\phi}^2_j}\Big)^2;\qquad 
Q=2\epsilon\Big(\frac{\dot{\phi}_i \phi^{\prime}_i}{\dot{\phi}^2_j}\Big), \label{7.6}
\end{equation}
Note that the sum is separately implied over denominator and numerator and all indices are contracted. Let us now define the quantity $\mathcal{X}$ as 
\begin{equation}\label{Chi}
\mathcal{X}=\frac{\dot{\phi}_i \phi^{\prime}_i}{\dot{\phi}^2_j},
\end{equation}
then, $P=1+2\epsilon^2\mathcal{X}^2$ and $Q=2\epsilon\mathcal{X}$. The equations \eqref{7.4a} can be shown to become
\begin{equation}
\hat{C}_1=1\pm\Big[1+2\epsilon^2\Big(\mathcal{X}^2-\frac{1}{4}\Big)\Big];\qquad \hat{C}_2=\mp2\epsilon \mathcal{X}. \label{7.7}
\end{equation}
We have kept upto $\mathcal{O}(\epsilon^2)$ terms in the above expressions for instructive purposes.
It can be seen easily, given a curve on which the $\hat C$s evolve, that at $\alpha\to+\infty$, $(\hat{C}_1,\hat{C}_2)$ approaches the point $(2,0)$ and at $\alpha\to-\infty$, $(\hat{C}_1,\hat{C}_2)$ approaches the point $(0,0)$. 
Plugging these expansions into the Lagrangian, we get 
\begin{align}
\hat{\mathcal{L}}_\alpha^{\pm}&=\hat{\sigma}\pm\hat{\sigma}\Big[1+\epsilon^2\Big(2\mathcal{X}^2-\frac{1}{2}\Big)\Big]\mp2\hat{\rho}\epsilon \mathcal{X}-\phi^{\prime 2}_i \nonumber \\
&=\frac{1}{2}\Big[\frac{\dot{\phi}^2_i}{\epsilon^2}\pm\frac{\dot{\phi}^2_i}{\epsilon^2}-\phi^{\prime 2}_i\mp\Big(2\dot{\phi}^2_i \mathcal{X}^2+\frac{1}{2}\dot{\phi}^2_i-\phi^{\prime 2}_i\Big)+\mathcal{O}(\epsilon^2)\Big]. %\nonumber 
\label{7.8}
\end{align}
Following our logic, we see when $\alpha\to+\infty$, the Lagrangian up to $\epsilon^0$ order is 
\begin{align}
\hat{\mathcal{L}}^{+}_\alpha
%&=\frac{\dot{\phi}^2_i}{\epsilon^2}-\frac{(\dot{\phi}_i \phi^{\prime}_i)^2}{\dot{\phi}^2_i}-\frac{1}{4}\dot{\phi}^2_i+\mathcal{O}(\epsilon^2)\nonumber \\
&=\frac{\dot{\phi}^2_i}{\epsilon^2}-\mathcal{O}(\epsilon^0)+...
\label{7.9}
\end{align}
while $\alpha\to-\infty$ gives in the leading order:
\begin{align}
\hat{\mathcal{L}}^{-}_\alpha&=-\phi^{\prime 2}_i+\frac{(\dot{\phi}_i \phi^{\prime}_i)^2}{\dot{\phi}^2_j}+\frac{1}{4}\dot{\phi}^2_i+\mathcal{O}(\epsilon^2).
\label{7.10}
\end{align}

Now this is an interesting result, we have two explicit Lagrangians at the edge of the $\alpha$ parameter space. 
Note the leading term in $\hat{\mathcal{L}}^{+}_\alpha$ is $\frac{\dot{\phi}^2_i}{\epsilon^2}$, so this action can be compared to the ``Electric'' Carroll scalar action, discussed for example in \cite{Henneaux:2021yzg,deBoer:2021jej,Bagchi:2022eav}, where the time derivative dominates over all terms.\footnote{We need to remember that we had scaled the $\dot{\phi}_i$s with $\cosh(\alpha)$. If we dial back the scaling, i.e. $\dot{\phi}_i=\epsilon\dot{\varphi}_i$, 
\begin{align}
\hat{\mathcal{L}}^{+}_\alpha=\dot{\phi}^2_i-\frac{(\dot{\phi}_i \phi^{\prime}_i)^2}{\dot{\phi}^2_j}+\mathcal{O}(\epsilon^2),~~~~
\hat{\mathcal{L}}^{-}_\alpha=-\phi^{\prime 2}_i+\frac{(\dot{\phi}_i \phi^{\prime}_i)^2}{\dot{\phi}^2_j}+\mathcal{O}(\epsilon^2).
\label{7.11}
\end{align}
This makes them look like being on the same footing. Clearly scaling before taking limits creates the difference.}
Clearly this Lagrangian is Carroll boost invariant.
We now need to understand what $\hat{\mathcal{L}}^{-}_\alpha$, containing both kinds of derivatives, means in this case. For a single scalar field it is trivial since the non-linear term cancels the spatial derivative, and we are left with $\hat{\mathcal{L}}^{-}_\alpha\propto \dot\phi^2$, which is again Carroll boost invariant. But the question persists for more than one field.
\medskip

To this effect, let us first note that the full Lagrangian \eqref{7.10}, sans the next leading order terms, can be written as:
\begin{equation}\label{Xaction}
 \hat{\mathcal{L}}^{-}_\alpha =   -\phi_i^{\prime 2} + \dot{\phi}_i^{2}\!\left(\mathcal{X}^2+\dfrac{1}{4}\right), 
\end{equation}
where $\mathcal{X}$, defined in \eqref{Chi} is very much a dynamical quantity. Observe that under Carroll boosts given by:
\begin{equation}
    \delta\dot\phi_i =0 ,~~\delta\phi'_i  = b\dot\phi_i, 
\end{equation}
where $b$ is the boost paramter, the above action is manifestly invariant for all $\mathcal{X}$. Under such boosts this quantity acquires a shift:
\begin{equation}\label{Xshift}
    \mathcal{X} \to \mathcal{X}+b.
\end{equation}
So $\mathcal{X}$ needs to have this freedom to make sure the full action is Carroll boost invariant. Carroll boosts will map any theory defined on a fixed $\mathcal{X}$ to another defined at another fixed $\mathcal{X}$ on the real line orbit. To appreciate the dynamical nature of $\mathcal{X}$, one can naively fix the value $\mathcal{X}^2 = \frac{3}{4}$, at which the action \eqref{Xaction} regains Lorentz boost invariance. 
\medskip

So, all in all, we have recovered two different Carroll invariant actions from our parent $\sqrt{T\overline T}$ deformed one, the usual electric Carroll scalar, and an absolutely novel ``magnetic-like'' non-linear scalar action. One should obviously spot that this one is very different from the magnetic scalar actions in \cite{Henneaux:2021yzg,deBoer:2021jej}. In fact, in \eqref{Xaction} the field dependent factor $\left(\mathcal{X}^2+\dfrac{1}{4}\right)^{-1}$ acts as some kind of squared speed of light necessarily giving ultralocal structure!\footnote{One could have just chosen $\mathcal{X}^{-1}$ as the velocity field here as evidently the pure $\dot\phi_i^2$ part in \eqref{Xaction} is directly Carroll boost invariant, so the extra $1/4$ is just redundant. Using just a plane wave solution like $\phi = f(x-vt)$ one could simply show that $\mathcal{X} = \frac{1}{v}$. As per \eqref{Xshift}, Carroll boost can just change this effective ``velocity'', contrary to our Lorentz invariant ideas. In fact this shift in the velocity implies:
$$ \frac{1}{v'} = \frac{1}{v}+b \implies v'= \frac{v}{1+bv},$$
which is nothing but Carroll boost addition law \cite{deBoer:2021jej}.} This is a remarkable find from the expanded deformed theory. 
\medskip 

To round up the discussion here, we must talk more about the physical nature of the non-linear player $\mathcal{X}$, which has not appeared in any known Carroll literature as per our current knowledge. We have already discussed that it plays the role of a purported (inverse) `velocity', 
however this velocity is clearly a pure gauge object which can be naively put to zero by appropriate Carroll boost\footnote{A related interpretation of $T\overline{T}$ deformations as arising from a change of gauge, particularly in connection with uniform light-cone gauge, has been discussed in \cite{Baggio:2018gct}.}, although the condition $\mathcal{X} = 0$ is not Carroll invariant.
Also note that at special values $\mathcal{X} = \pm 1$ we can have purely right-moving or purely left-moving chiral solutions for the field, and the shift under Carroll boost 
\eqref{Xshift} simply mixes between chiralities. The naive special point of $\mathcal{X} =\pm \frac{\sqrt{3}}{2}$, where relativistic conformal symmetries are restored, lies beyond this regime bounded by pure chiral theories.

\subsubsection*{Fate of the LT}
Now we come back to another important question: does the prowess of the LT survive the Carroll limit?
Let us then check the asymptotic limits of $\hat{A}$ and $\hat{B}$ from the `scaled' LT  as in \eqref{Electric.LT}:
\begin{align}
\hat{A}=\frac{1}{\hat{C_1}\cosh^2\alpha};\qquad \hat{B}=-\frac{\hat{C}_2}{\hat{C_1}\cosh\alpha}.
\label{7.12}
\end{align}
Recall that these equations provided a one-to-one map between configuration space and phase space quantities at a given value of $\alpha$. Then our question can be rephrased: does this map still persist at $\alpha \to \pm \infty$?
Using the expression of $\hat{C_1}$ from \eqref{7.7} we can expand $\hat{A}$ in leading orders of $\epsilon$:
\begin{align}
\hat{A}&=\frac{1}{\hat{C_1}\cosh^2\alpha}\nonumber \\
&=\epsilon^2\hat{C}^{-1}_1
%=\epsilon^2\Big[1\pm\Big(1+\epsilon^2(2\mathcal{X}^2-\frac{1}{2})\Big)\Big]^{-1} \nonumber\\&
=\frac{1}{2}\epsilon^2\Big[\frac{1}{2}\pm\Big(\frac{1}{2}+\epsilon^2(\mathcal{X}^2-\frac{1}{4})\Big)\Big]^{-1}.
%&=\frac{1}{2}\epsilon^2\Big[1+\epsilon^2\Big(\mathcal{X}^2-\frac{1}{4}\Big)\Big]^{-1}\text{ (for +ve branch)}\nonumber\\
%&=\frac{1}{2}\epsilon^2\Big[1-\epsilon^2\Big(\mathcal{X}^2-\frac{1}{4}\Big)\Big]=\mathcal{O}(\epsilon^2)\text{ (for +ve branch)}
\label{7.13}
\end{align}
The values for the positive and negative branches $(\alpha\to\pm\infty)$ are:
\begin{align}
\hat{A}&=\dfrac{1}{2}\,\epsilon^{2}+\mathcal{O}(\epsilon^4), \ &(\text{$+$ve branch}); \\
&=-\dfrac{1}{2\!\left(\mathcal{X}^{2}-\tfrac{1}{4}\right)}+\epsilon^2f_1+\mathcal{O}(\epsilon^4), \ & (\text{$-$ve branch}).
\label{7.14}
\end{align}
Here, $f_1$ is also a field-dependent dimensionless quantity, which has the form:
\begin{align}
    f_1=\frac{6\mathcal{X}^4-(4r+1)\mathcal{X}^2-\tfrac{1}{8}}{4(\mathcal{X}^2-\frac{1}{4})^2},\quad \Big(r=\frac{\phi^{\prime2}_i}{\dot{\phi}^2_i}\Big).
\end{align}
This term arises if one carefully expands $\hat{C}_1$ upto $\epsilon^4$ order, such that there is a higher order correction, and we can expand $\hat{A}$ upto $\epsilon^2$ order.
\medskip

Note that in leading order, the $\mathcal{X}$ dependence survives for the (-ve) branch.
Now, let us look at the expression for $\hat{B}$:
\begin{align}
\hat{B}&=-\frac{\hat{C}_2}{\hat{C_1}\cosh\alpha}\nonumber \\
&=-\epsilon\hat{C}_2\hat{C}^{-1}_1=-\epsilon\Big(\mp2\epsilon \mathcal{X}\Big)\Big[1\pm\Big(1+2\epsilon^2(\mathcal{X}^2-\frac{1}{4})\Big)\Big]^{-1}, 
%&=\mp\epsilon^2 \mathcal{X}\Big[\frac{1}{2}\pm\Big(\frac{1}{2}+\epsilon^2(\mathcal{X}^2-\frac{1}{4})\Big)\Big]^{-1}\nonumber \\
%&=-\epsilon^2 \mathcal{X}\Big[1-\epsilon^2\Big(\mathcal{X}^2-\frac{1}{4}\Big)\Big]\text{ (for +ve branch)}\nonumber\\
%&=-\mathcal{O}(\epsilon^2) 
\label{7.15}
\end{align}
which give us
\begin{align}
\hat{B}&=\epsilon^{2}\mathcal{X}+\mathcal{O}(\epsilon^4), \ &  (\text{$+$ve branch});\\
&=-\frac{\mathcal{X}}{\mathcal{X}^{2}-\tfrac{1}{4}}\Big[1+\epsilon^2f_2\Big]+\mathcal{O}(\epsilon^4), \  & (\text{$-$ve branch}),
\label{7.14n}
\end{align}
where,
\begin{align}
f_2=2(\mathcal{X}^{2}-\tfrac{1}{4})-r+\frac{6\mathcal{X}^4-(4r+1)\mathcal{X}^2-\tfrac{1}{8}}{2(\mathcal{X}^2-\frac{1}{4})}.
\end{align}
Which has similar dependence like $\hat A$ as well. These two correction terms depending on $f_1$ and $f_2$, however, will not be as important in the rest of the discussion.  Note that although $\hat{A}$ and $\hat{B}$ are phase space quantities, we do not try to write the inherent $\mathcal{X}$ dependence in terms of phase space variables, for reasons to be discussed later.
\medskip

\paragraph{Mini summary:}
We already have had some crucial inputs from the algebraic structures of the limits above. Let us summarise the results so far for the sake of the reader:
When we take the limits $(\alpha\to\pm\infty)$, we get the theory to split into $+$ve and the $-$ve branches. The limiting Lagrangian becomes
\begin{align}
\hat{\mathcal{L}}_{\alpha} &=
\begin{cases}
+\dfrac{\dot{\phi}_i^2}{\epsilon^{2}} - \mathcal{O}(\epsilon^{0}), & (\alpha \to +\infty); \\[8pt]
-\phi_i^{\prime 2} + \dot{\phi}_i^{2}\!\left(\mathcal{X}^2+\dfrac{1}{4}\right) + \mathcal{O}(\epsilon^{2}), & (\alpha \to -\infty),
\end{cases}
\label{7.20-21}
\end{align}
while the expanded values of $\hat{C}_1$, $\hat{C}_2$ take the form
\begin{align}
\hat{C}_1 &=
\begin{cases}
2\!\left[1+\epsilon^{2}\!\left(\mathcal{X}^{2}-\dfrac{1}{4}\right)\right], & (\alpha \to +\infty); \\[8pt]
-2\epsilon^{2}\!\left(\mathcal{X}^{2}-\dfrac{1}{4}\right), & (\alpha \to -\infty),
\end{cases}
\qquad
\hat{C}_2 =
\begin{cases}
-2\epsilon\,\mathcal{X}, & (\alpha \to +\infty); \\[8pt]
+2\epsilon\,\mathcal{X}, & (\alpha \to -\infty),
\end{cases}
\label{7.18-19}
\end{align}
which, in the phase space, gets mapped to 
\begin{align}
\hat{A} &=
\begin{cases}
+\frac{1}{2}\epsilon^{2}, & (\alpha \to +\infty); \\[8pt]
-\frac{1}{2\big(\mathcal{X}^{2}-\frac{1}{4})}+\epsilon^2f_1, & (\alpha \to -\infty),
\end{cases}
\qquad
\hat{B} =
\begin{cases}
+\epsilon^{2}\mathcal{X}, & (\alpha \to +\infty); \\[8pt]
\dfrac{\mathcal{X}}{\mathcal{X}^{2}-\tfrac{1}{4}}[1+\epsilon^2 f_2], & (\alpha \to -\infty).
\end{cases}
\label{7.22-23}
\end{align}
One can easily check these expansions in $\epsilon$ are in accordance with \eqref{Electric.LT} even for small $\epsilon.$
Now let us talk about the strict Carroll limit where we exactly put $\epsilon = 0$.
We can see that for the positive branch ($\alpha\to+\infty$), all the coefficients  ($\hat{C}_1$, $\hat{C}_2$) and ($\hat{A}$, $\hat{B}$) converge to specific values $(2,0)$ and $(0,0)$, meanwhile the Lagrangian in leading order gives a $\dot{\phi}_i^2$ term. For the negative branch ($\alpha\to-\infty$), our ($\hat{C}_1$, $\hat{C}_2$) converges to $(0,0)$, however in the phase space ($\hat{A}$, $\hat{B}$) still depends on $\mathcal{X}$. The Lagrangian in this branch also gets the non-linear term that depends on $\mathcal{X}$, as we had discussed earlier. But both cases lead to Carroll boost invariant theories.  

\medskip

Clearly these two limits are very different from each other, the latter being more involved than the former. In the (+ve) branch, even at the exact limit, we can get a one-to-one mapping between configuration space and phase space quantities, but the uniqueness of the map changes in the (-ve) branch. To make sense of this, we can think of this from a geometric point of view. In the next section, we will start out with the idea that the equations governing the coefficients ($\hat{C}_1$, $\hat{C}_2$) or ($\hat{A}$, $\hat{B}$) represents two separate classes of hyperbolas. We will see how these hyperbolas in the $\hat{C}_1$-$\hat{C}_2$ plane map to those in the $\hat{A}$-$\hat{B}$ plane and how the value of $\mathcal{X}$ dictates the different sectors of the hyperbolic map.

\subsection{Geometric flows}
\label{sec8}
\subsubsection*{Hyperbolic parameter space}
\begin{figure}[h!]
 \centering
\begin{tikzpicture}[scale=0.9] %\label{Figure 8}
    % Draw axes
    \draw[->] (-3,0) -- (8,0) node[right] {$C_1$};
    \draw[->] (0,-3) -- (0,3) node[above] {$C_2$};

    % Colors
    \definecolor{c0}{rgb}{0.8,0,0}
    \definecolor{c1}{rgb}{0,0.5,0.8}
    \definecolor{c2}{rgb}{0,0.6,0}

    % ---- Alpha = 0  (two lines)
    \draw[c0,thick,domain=-1.9:3.9] plot (\x,{\x - 1});
    \draw[c0,thick,domain=-1.9:3.9] plot (\x,{-(\x - 1)});

    % ---- Alpha = 1  (example intermediate)
    \pgfmathsetmacro{\a}{cosh(1)}
    \draw[c2,thick,domain=2.72:4.3,smooth,variable=\x]
    plot ({\x},{sqrt(\x*\x - 2*\x*\a + 1)});
    \draw[c2,thick,domain=2.72:4.3,smooth,variable=\x]
    plot ({\x},{-sqrt(\x*\x - 2*\x*\a + 1)});
    \draw[c2,thick,domain=-1.7:0.366,smooth,variable=\x]
    plot ({\x},{sqrt(\x*\x - 2*\x*\a + 1)});
    \draw[c2,thick,domain=-1.7:0.365,smooth,variable=\x]
    plot ({\x},{-sqrt(\x*\x - 2*\x*\a + 1)});

    % ---- Alpha → ∞  (tanh = 1)
    \pgfmathsetmacro{\a}{cosh(3)}
    %\draw[c1,thick,domain=(1+\t):4,smooth,variable=\x]
       % plot ({\x},{sqrt((\x-1)^2 - \t*\t)});
    %\draw[c1,thick,domain=(1+\t):4,smooth,variable=\x]
       % plot ({\x},{-sqrt((\x-1)^2 - \t*\t)});
    \draw[c1,thick,domain=-0.47:0.049,smooth,variable=\x]
    plot ({\x},{sqrt(\x*\x - 2*\x*\a + 1)});
    \draw[c1,thick,domain=-0.47:0.049,smooth,variable=\x]
    plot ({\x},{-sqrt(\x*\x - 2*\x*\a + 1)});

    % ---- Mark points
    \fill (0,0) circle (2pt) node[below left] {$(0,0)$};
    \fill (1,0) circle (2pt) node[below] {$(1,0)$};
    %\fill (7.9,0) circle (2pt) node[below] {$(\infty,0)$};
    \fill (2.715,0) circle (2pt); %node[below right] {$(a,0)$};
    \fill (0.368,0) circle (2pt); %node[below right] {$(a,0)$};
    
    % Labels
    \node[text=c0] at (5.8, 2.5) {$\alpha=0$ };
    \node[text=c2] at (5.8, 1.8) {$|\alpha|>0$};
    \node[text=c1] at (5.9, 1.1) {$|\alpha|\to\infty$};

    %Arrows
    %\node[text=c2] at (2,0.15) {$\longrightarrow$};
    %\node[text=c1] at (3.6,0.15) {$\longrightarrow\infty$};

     %arrows
     \draw[->,thick,c2] 
     (1.2,0.15) -- (2.6,0.15);

     \draw[<-,thick,c2] 
     (0.4,0.15) -- (0.8,0.15);

     \draw[->,thick,c1] 
     (3,0.15) -- (4,0.15);
     \node[text=c1] at (4.3,0.15) {$\infty$};

     \draw[<-,thick,c1] 
     (0,0.15) -- (0.3,0.15);
  
\end{tikzpicture}
\caption{Hyperbolas in the $C_1$-$C_2$ plane for different $\alpha$. Note that the branch moving right signifies $\alpha \to \infty$ and the one moving left signifies $\alpha \to -\infty$. }
  \label{fig2}
\end{figure}
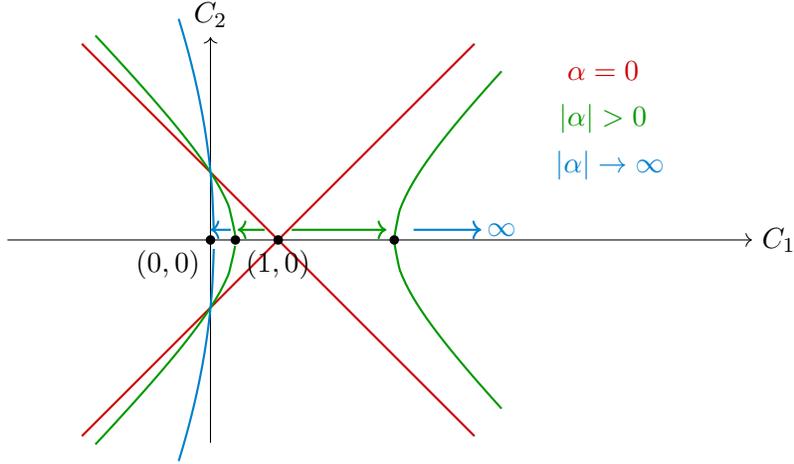

For a moment let us go back to our original ``unhatted'' variables.
Let us recall the relations between $C_1$ and $C_2$ from equation \eqref{3.19}
\begin{equation}
   C^2_1-2C_1\cosh(\alpha)+1-C^2_2=0. \label{8.1}
\end{equation}
Attentive readers can notice, this trajectory gives equations for two hyperbolas in the $C_1$-$C_2$ plane, characterised by the parameter $\alpha$. At $\alpha=0$, the equation is
\begin{equation}
   (C_1-1)^2-C^2_2=0,
\end{equation}
which are two straight lines intersecting at $C_1=1$ and $C_2=0$. For us, this marks the trivial case of no deformation. As we proceed to non-zero $\alpha$, the straight lines immediately become two disjoint hyperbolas. Note that, due to the symmetric nature of \eqref{8.1} for a value of $|\alpha|$ we always get two branches of the hyperbolas. As the value of $|\alpha|$ increases, the two branches gets separated, and as $\alpha\rightarrow\infty$, the node of the left branch moves gradually to $(0,0)$, while that of the right branch moves out to infinity. See Fig.\eqref{fig2} for a description.

From the Hamiltonian analysis \eqref{ABdef}, we also find that $A$ and $B$ follow a similar relationship, giving exactly the same equations for the hyperbolas at finite values of $\alpha$:
\begin{equation}
   A^2-2A\cosh(\alpha)+1-B^2=0. \label{8.3}
\end{equation}
The mapping, given directly by the LT \eqref{L.T}, transforms the hyperbola in the $C_1$-$C_2$ plane to the hyperbola in the $A$-$B$ plane and vice versa, making it a geometric relation. Since we are interested in the region where $\alpha$ is very large, we are going to see the hyperbolas for the scaled system given by ($\hat{C}_1$, $\hat{C}_2$, $\hat{A}$, $\hat{B}$) and how the mapping works there. 

\subsubsection*{Mapping at the Carroll points}

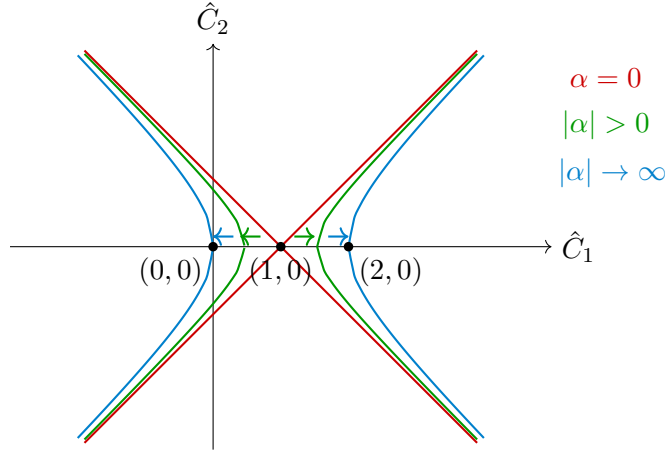
\begin{figure}[h!]
 \centering
\begin{tikzpicture}[scale=0.9] %\label{Figure 8}
    % Draw axes
    \draw[->] (-3,0) -- (5,0) node[right] {$\hat{C}_1$};
    \draw[->] (0,-3) -- (0,3) node[above] {$\hat{C}_2$};

    % Colors
    \definecolor{c0}{rgb}{0.8,0,0}
    \definecolor{c1}{rgb}{0,0.5,0.8}
    \definecolor{c2}{rgb}{0,0.6,0}

    % ---- Alpha = 0  (two lines)
    \draw[c0,thick,domain=-1.9:3.9] plot (\x,{\x - 1});
    \draw[c0,thick,domain=-1.9:3.9] plot (\x,{-(\x - 1)});

    % ---- Alpha = 1  (example intermediate)
    \pgfmathsetmacro{\t}{tanh(0.6)}
    \draw[c2,thick,domain=(1+\t):3.9,smooth,variable=\x]
    plot ({\x},{sqrt((\x-1)^2 - \t*\t)});
    \draw[c2,thick,domain=(1+\t):3.9,smooth,variable=\x]
    plot ({\x},{-sqrt((\x-1)^2 - \t*\t)});
    \draw[c2,thick,domain=-1.9:(1-\t),smooth,variable=\x]
    plot ({\x},{sqrt((\x-1)^2 - \t*\t)});
    \draw[c2,thick,domain=-1.9:(1-\t),smooth,variable=\x]
    plot ({\x},{-sqrt((\x-1)^2 - \t*\t)});

    % ---- Alpha → ∞  (tanh = 1)
    \pgfmathsetmacro{\t}{1}
    \draw[c1,thick,domain=(1+\t):4,smooth,variable=\x]
        plot ({\x},{sqrt((\x-1)^2 - \t*\t)});
    \draw[c1,thick,domain=(1+\t):4,smooth,variable=\x]
        plot ({\x},{-sqrt((\x-1)^2 - \t*\t)});
    \draw[c1,thick,domain=-2:(1-\t),smooth,variable=\x]
        plot ({\x},{sqrt((\x-1)^2 - \t*\t)});
    \draw[c1,thick,domain=-2:(1-\t),smooth,variable=\x]
        plot ({\x},{-sqrt((\x-1)^2 - \t*\t)});
    
    % ---- Mark points
    \fill (0,0) circle (2pt) node[below left] {$(0,0)$};
    \fill (1,0) circle (2pt) node[below] {$(1,0)$};
    \fill (2,0) circle (2pt) node[below right] {$(2,0)$};
    
    % Labels
    \node[text=c0] at (5.8, 2.5) {$\alpha=0$ };
    \node[text=c2] at (5.8, 1.8) {$|\alpha|>0$};
    \node[text=c1] at (5.9, 1.1) {$|\alpha|\to\infty$};

     %arrows
     \draw[->,thick,c2] 
     (1.2,0.15) -- (1.5,0.15);

     \draw[<-,thick,c2] 
     (0.4,0.15) -- (0.7,0.15);

     \draw[->,thick,c1] 
     (1.7,0.15) -- (2.0,0.15);
    % \node[text=c1] at (4.3,0.15); %{$\infty$};

     \draw[<-,thick,c1] 
     (0,0.15) -- (0.3,0.15);

\end{tikzpicture}
\caption{Hyperbolas in the \textit{scaled} $\hat{C}_1$-$\hat{C}_2$ plane for different $\alpha$. }
   \label{fig3}
\end{figure}

As we have learnt before, to understand the dynamics at the Carroll points, it is better to  consider the ``hatted'' quantities $\hat{C}_1$ and $\hat{C}_2$ given by \eqref{6.10}. They satisfy an equivalent hyperbolic trajectory:
%\eqref{6.44}, which can be rewritten as
\begin{align}
(\hat{C}_1-1)^2-\hat{C}_2^2=\tanh^2\alpha. \label{8.4}   
\end{align}
The following plot i.e. Fig.\eqref{fig3} depicts the hyperbolas at the trivial case ($\alpha$=0) and at the Carroll points ($\alpha\to\pm\infty$).
\medskip

%FIGURE 4: HYPERBOLA FOR SCALED SYSTEM

We can see from the plot that as $\alpha$ is dialed from $0$ to $\pm\infty$, the nodes of the hyperbolas shift from $(1,0)$ to $(0,0)$ and $(2,0)$. Let us recall equation \eqref{Electric.LT}, the map between $\hat{C}_1$, $\hat{C}_2$ and $\hat{A}$, $\hat{B}$ : 
\begin{align} \label{8.5}
\hat{C}_1 \cosh ^2 (\alpha) =\frac{1}{\hat{A}}; \quad \hat{C}_2\cosh(\alpha) =-\frac{\hat{B}}{\hat{A}},    
\end{align}
From the above relations we can obtain the same equation \eqref{8.4} for the scaled quantities $\hat{A}$ and $\hat{B}$:
\begin{align}
(\hat{A}-1)^2-\hat{B}^2=\tanh^2\alpha. \label{8.4n}   
\end{align}
 The LT in \eqref{8.5} now maps a region of the $\hat{C}_1$-$\hat{C}_2$ hyperbola to a region of the $\hat{A}$-$\hat{B}$ hyperbola.  
Let us understand the map step by step.
\medskip

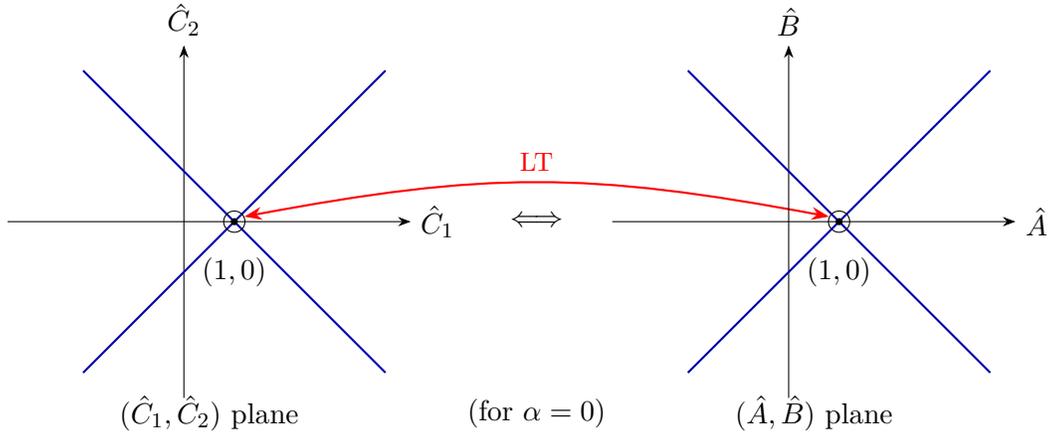
\begin{figure}[H]
  \centering
\begin{tikzpicture}[scale=0.67,>=Stealth]
% ===== LEFT: C1,C2 plane (center at (1,0), hyperbola (x-1)^2 - y^2 = 1) =====
\begin{scope}[shift={(0,0)}]
  % axes
  \draw[->] (-3.5,0) -- (4.5,0) node[right] {$\hat{C}_1$};
  \draw[->] (0,-3.5) -- (0,3.5) node[above] {$\hat{C}_2$};

%straight lines
  \draw[blue!70!black,thick,domain=-2:4,smooth,variable=\x]
    plot ({\x},{(\x-1)});
  \draw[blue!70!black,thick,domain=-2:4,smooth,variable=\x]
    plot ({\x},{-(\x-1)});

 % special points (vertices)
  %\fill (0,0) circle (2pt) node[below left] {$(0,0)$};
  \fill (1,0) circle (2pt) node at (1,-1) {$(1,0)$};
  \draw (1,0) circle (6pt);
  %\fill (2,0) circle (2pt) node[below right] {$(2,0)$};

  % title
  \node at (0.5,-3.8) {$(\hat{C}_1,\hat{C}_2)$ plane};
  \node at (7,0) {$\Longleftrightarrow$ };
  \node at (7,-3.8) {$(\text{for } \alpha=0)$ };

  %arrows
  \draw[<->,thick,red]
  (1.2,0.1) .. controls (6,1) and (8,1) .. (12.8,0.1);
  \node at (7,1.2) {\small\textcolor{red}{LT}};
\end{scope}

% ===== RIGHT: A,B plane (color blue) =====
\begin{scope}[shift={(12,0)}] % shift horizontally
  % axes
  \draw[->] (-3.5,0) -- (4.5,0) node[right] {$\hat{A}$};
  \draw[->] (0,-3.5) -- (0,3.5) node[above] {$\hat{B}$};

   %straight lines
  \draw[blue!70!black,thick,domain=-2:4,smooth,variable=\x]
    plot ({\x},{(\x-1)});
  \draw[blue!70!black,thick,domain=-2:4,smooth,variable=\x]
    plot ({\x},{-(\x-1)});

   % special points (vertices)
  %\fill (0,0) circle (2pt) node[below left] {$(0,0)$};
  \fill (1,0) circle (2pt) node at (1,-1) {$(1,0)$};
  \draw (1,0) circle (6pt);
  %\fill (2,0) circle (2pt) node[below right] {$(2,0)$};
  % title
  \node at (0.5,-3.8) {$(\hat{A},\hat{B})$ plane};
\end{scope}

\end{tikzpicture}
\caption{Plot for $(\hat{C}_1 $-$\hat{C}_2)$ and $(\hat{A}$-$\hat{B})$ planes for $\alpha=0$}
  \label{fig4}
\end{figure}

Initially when $\alpha=0$, we have two straight lines that intersect at $(1,0)$. This is the trivial case of no deformation having $\hat{C}_1=\hat{A}=1$, and $\hat{C}_2=\hat{B}=0$. One can see that (Fig.\eqref{fig4}) the point of intersection in one plane maps to the other via the LT relations.

\begin{figure}[h!]
  \centering
\begin{tikzpicture}[scale=0.7,>=Stealth]
% ===== LEFT: C1,C2 plane (center at (1,0), hyperbola (x-1)^2 - y^2 = 1) =====
\begin{scope}[shift={(0,0)}]
  % axes
  \draw[->] (-3.5,0) -- (4.5,0) node[right] {$\hat{C}_1$};
  \draw[->] (0,-3.5) -- (0,3.5) node[above] {$\hat{C}_2$};

  % parameter (the value inside your original sqrt was (x-1)^2 - 0.6)
    \pgfmathsetmacro{\a}{0.1}

    % left and right intersection x-values: 1 +/- sqrt(a)
    \pgfmathsetmacro{\rootA}{1 - sqrt(\a)} % ≈ 0.2254
    \pgfmathsetmacro{\rootB}{1 + sqrt(\a)} % ≈ 1.7746

    % define function safely: Ctwo(x) = sqrt(max(0,(x-1)^2 - a))
    \pgfmathdeclarefunction{Ctwo}{1}{%
      \pgfmathparse{max(0,(#1-1)^2 - \a)}%
      \pgfmathparse{sqrt(\pgfmathresult)}%
    }

    % left branch: colored blue (x > 0)
  \draw[blue!70!black,thick,domain=-2:0.683,smooth,variable=\x]
    plot ({\x},{sqrt((\x-1)^2 - 0.1)});
  \draw[blue!70!black,thick,domain=-2:0.683,smooth,variable=\x]
    plot ({\x},{-sqrt((\x-1)^2 - 0.1)});

  % right branch: colored blue (x >= 2)
  \draw[blue!70!black,thick,domain=1.317:4,smooth,variable=\x]
    plot ({\x},{sqrt((\x-1)^2 - 0.1)});
  \draw[blue!70!black,thick,domain=1.317:4,smooth,variable=\x]
    plot ({\x},{-sqrt((\x-1)^2 - 0.1)});

  % special points (vertices)
  \fill (0,0) circle (2pt) node[below left] {$(0,0)$};
  \fill (1,0) circle (2pt) node[below ] {$(1,0)$};
  \fill (2,0) circle (2pt) node[below right] {$(2,0)$};

  % title
  \node at (0.5,-3.8) {$(\hat{C}_1,\hat{C}_2)$ plane};
  \node at (7,0) {$\Longleftrightarrow$ };
  \node at (7.1,0.5) {\small{LT}};
  %\node at (7,-3.8) {$(\text{for } 0<|\alpha|<\infty)$ };

   % --- shaded area between the left blue branch and the y-axis (for x>0)
    \begin{scope}
      \clip (0,-3.5) rectangle (4.5,3.5); % restrict to x>0 region
      \fill[blue!70!black,opacity=0.15]
        % upper edge from x=0 to x=rootA
        plot[smooth,domain=0:\rootA,variable=\x,samples=120] 
          (\x,{sqrt(max(0,(\x-1)^2 - \a))})
        % back along the lower edge from x=rootA to x=0
        -- plot[smooth,domain=\rootA:0,variable=\x,samples=120] 
          (\x,{-sqrt(max(0,(\x-1)^2 - \a))})
        -- cycle;
    \end{scope}
\end{scope}

% ===== RIGHT: A,B plane  =====
\begin{scope}[shift={(12,0)}] % shift horizontally
  % axes
  \draw[->] (-3.5,0) -- (4.5,0) node[right] {$\hat{A}$};
  \draw[->] (0,-3.5) -- (0,3.5) node[above] {$\hat{B}$};

  % parameter (the value inside your original sqrt was (x-1)^2 - 0.6)
    \pgfmathsetmacro{\a}{0.1}

    % left and right intersection x-values: 1 +/- sqrt(a)
    \pgfmathsetmacro{\rootA}{1 - sqrt(\a)} % ≈ 0.2254
    \pgfmathsetmacro{\rootB}{1 + sqrt(\a)} % ≈ 1.7746

    % define function safely: Ctwo(x) = sqrt(max(0,(x-1)^2 - a))
    \pgfmathdeclarefunction{Ctwo}{1}{%
      \pgfmathparse{max(0,(#1-1)^2 - \a)}%
      \pgfmathparse{sqrt(\pgfmathresult)}%
    }

     % left branch: colored blue 
  \draw[blue!70!black,thick,domain=-2:0.683,smooth,variable=\x]
    plot ({\x},{sqrt((\x-1)^2 - 0.1)});
  \draw[blue!70!black,thick,domain=-2:0.683,smooth,variable=\x]
    plot ({\x},{-sqrt((\x-1)^2 - 0.1)});

  % right branch: colored blue
  \draw[blue!70!black,thick,domain=1.317:4,smooth,variable=\x]
    plot ({\x},{sqrt((\x-1)^2 - 0.1)});
  \draw[blue!70!black,thick,domain=1.317:4,smooth,variable=\x]
    plot ({\x},{-sqrt((\x-1)^2 - 0.1)});

  % special points (vertices)
  \fill (0,0) circle (2pt) node[below left] {$(0,0)$};
  \fill (1,0) circle (2pt) node[below ] {$(1,0)$};
  \fill (2,0) circle (2pt) node[below right] {$(2,0)$};
  % title
  \node at (0.5,-3.8) {$(\hat{A},\hat{B})$ plane};

  % --- shaded area between the right blue branch and the y-axis (for x>0)
    \begin{scope}
      \clip (0,-3.5) rectangle (4.5,3.5); % restrict to x>0 region
      \fill[blue!70!black,opacity=0.15]
        % upper edge from x=0 to x=rootA
        plot[smooth,domain=0:\rootA,variable=\x,samples=120] 
          (\x,{sqrt(max(0,(\x-1)^2 - \a))})
        % back along the lower edge from x=rootA to x=0
        -- plot[smooth,domain=\rootA:0,variable=\x,samples=120] 
          (\x,{-sqrt(max(0,(\x-1)^2 - \a))})
        -- cycle;
    \end{scope}
  \end{scope}
\end{tikzpicture}
\caption{Mapping between $(\hat{C}_1,\hat{C}_2)$ and $(\hat{A},\hat{B})$ planes for $|\alpha|>0$}
  \label{fig5}
\end{figure}
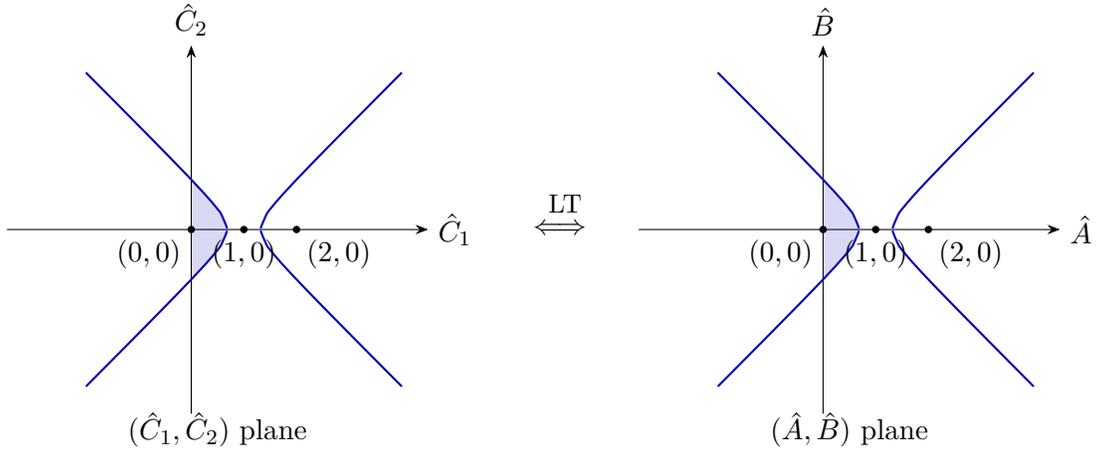

The moment we have a finite $\alpha$ ($0<|\alpha|<\infty$), the resulting hyperbolas (Fig.\eqref{fig5}) separate out into left and right branches, the vertices moving away from the  $(1,0)$ point, slowly towards extreme values i.e. $(0,0$) and $(2,0)$.

%FIGURE 6: INTERMEDIATE HYPERBOLA

On the other hand, when $\alpha$ is very large ($\epsilon=1/\cosh(\alpha)$ being very small), the vertices shift closer to $(0,0$) and $(2,0)$. Note that the area enclosed between the left branch and the vertical axis shrinks gradually as it approaches the $(0,0)$ point. Let us analyse the regions of the hyperbola, and consequently the mapping, very carefully for this case. We will track leading corrections in $\epsilon$ to be sure. The Fig.\eqref{fig7}  depicts the plots at very large values of $\alpha$, albeit just shy of the exact limit. Each plot in this figure (in the $\hat{C}_1$-$\hat{C}_2$ and $\hat{A}$-$\hat{B}$ planes) have 6 distinct regions in the limiting case:

%%%%%%% FIGURE 7: COULORED HYPERBOLA %%%%%%%

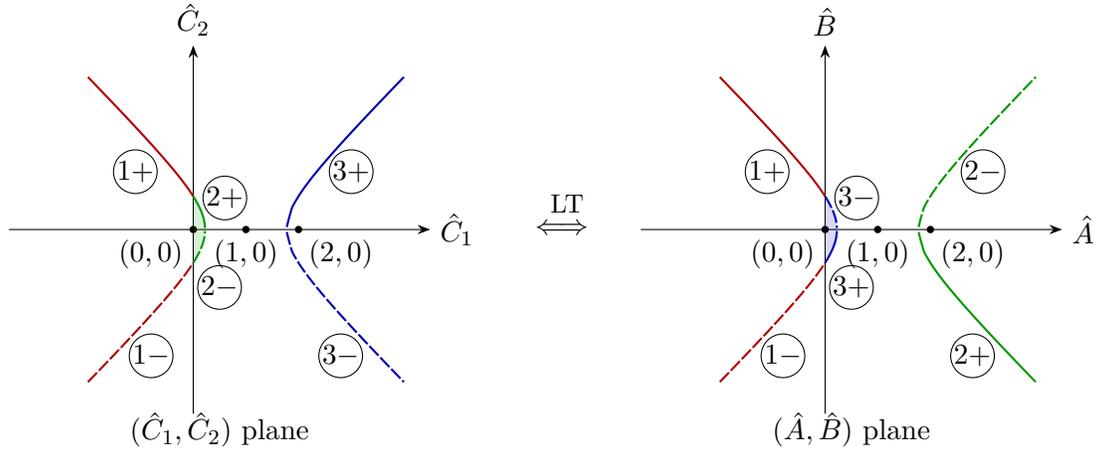
\begin{figure}[h!]
  \centering
\begin{tikzpicture}[scale=0.7,>=Stealth]
% ===== LEFT: C1,C2 plane (center at (1,0), hyperbola (x-1)^2 - y^2 = 1) =====
\begin{scope}[shift={(0,0)}]
  % axes
  \draw[->] (-3.5,0) -- (4.5,0) node[right] {$\hat{C}_1$};
  \draw[->] (0,-3.5) -- (0,3.5) node[above] {$\hat{C}_2$};

  % parameter (the value inside your original sqrt was (x-1)^2 - 0.6)
    \pgfmathsetmacro{\a}{0.6}

    % left and right intersection x-values: 1 +/- sqrt(a)
    \pgfmathsetmacro{\rootA}{1 - sqrt(\a)} % ≈ 0.2254
    \pgfmathsetmacro{\rootB}{1 + sqrt(\a)} % ≈ 1.7746

    % define function safely: Ctwo(x) = sqrt(max(0,(x-1)^2 - a))
    \pgfmathdeclarefunction{Ctwo}{1}{%
      \pgfmathparse{max(0,(#1-1)^2 - \a)}%
      \pgfmathparse{sqrt(\pgfmathresult)}%
    }

    % left branch: colored red (x < 0)
  \draw[red!70!black,thick,domain=-2:0,smooth,variable=\x]
  plot ({\x},{sqrt((\x-1)^2 - 0.6)});    \draw[red!70!black,thick,dashed,dash pattern=on 5pt off 1pt,domain=-2:0,smooth,variable=\x]
  plot ({\x},{-sqrt((\x-1)^2 - 0.6)});

    % left branch: colored green (x > 0)
  \draw[green!60!black,thick,domain=0:0.225,smooth,variable=\x]
    plot ({\x},{sqrt((\x-1)^2 - 0.6)});
  \draw[green!60!black,thick,dashed,dash pattern=on 5pt off 1pt,domain=0:0.225,smooth,variable=\x]
    plot ({\x},{-sqrt((\x-1)^2 - 0.6)});

  % right branch: colored blue (x >= 2)
  \draw[blue!70!black,thick,domain=1.776:4,smooth,variable=\x]
    plot ({\x},{sqrt((\x-1)^2 - 0.6)});
  \draw[blue!70!black,thick,dashed,dash pattern=on 5pt off 1pt,domain=1.776:4,smooth,variable=\x]
    plot ({\x},{-sqrt((\x-1)^2 - 0.6)});

  % special points (vertices)
  \fill (0,0) circle (2pt) node[below left] {$(0,0)$};
  \fill (1,0) circle (2pt) node[below ] {$(1,0)$};
  \fill (2,0) circle (2pt) node[below right] {$(2,0)$};

  % title
  \node at (0.5,-3.8) {$(\hat{C}_1,\hat{C}_2)$ plane};
  \node at (7,0) {$\Longleftrightarrow$ };
  \node at (7.1,0.5) {\small{LT}};
  %\node at (7,-3.8) {$(\text{for } 0<|\alpha|<\infty)$ };

   % --- shaded area between the left blue branch and the y-axis (for x>0)
    \begin{scope}
      \clip (0,-3.5) rectangle (4.5,3.5); % restrict to x>0 region
      \fill[green!60!black,opacity=0.15]
        % upper edge from x=0 to x=rootA
        plot[smooth,domain=0:\rootA,variable=\x,samples=120] 
          (\x,{sqrt(max(0,(\x-1)^2 - \a))})
        % back along the lower edge from x=rootA to x=0
        -- plot[smooth,domain=\rootA:0,variable=\x,samples=120] 
          (\x,{-sqrt(max(0,(\x-1)^2 - \a))})
        -- cycle;
    \end{scope}

    % --- Label numbers 1-6 inside small circles --- C1C2
  \node[circle,draw,inner sep=0.2pt] at (-1.1,1.1) {$1+$};
  \node[circle,draw,inner sep=0.2pt] at (-0.8,-2.4) {$1-$};
  \node[circle,draw,inner sep=0.2pt] at (0.6,0.6) {$2+$};
  \node[circle,draw,inner sep=0.2pt] at (0.5,-1.1) {$2-$};
  \node[circle,draw,inner sep=0.2pt] at (3.0,1.1) {$3+$};
  \node[circle,draw,inner sep=0.2pt] at (2.8,-2.4) {$3-$};
 
 \end{scope}

% ===== RIGHT: A,B plane  =====
\begin{scope}[shift={(12,0)}] % shift horizontally
  % axes
  \draw[->] (-3.5,0) -- (4.5,0) node[right] {$\hat{A}$};
  \draw[->] (0,-3.5) -- (0,3.5) node[above] {$\hat{B}$};

  % parameter (the value inside your original sqrt was (x-1)^2 - 0.6)
    \pgfmathsetmacro{\a}{0.6}

    % left and right intersection x-values: 1 +/- sqrt(a)
    \pgfmathsetmacro{\rootA}{1 - sqrt(\a)} % ≈ 0.2254
    \pgfmathsetmacro{\rootB}{1 + sqrt(\a)} % ≈ 1.7746

    % define function safely: Ctwo(x) = sqrt(max(0,(x-1)^2 - a))
    \pgfmathdeclarefunction{Ctwo}{1}{%
      \pgfmathparse{max(0,(#1-1)^2 - \a)}%
      \pgfmathparse{sqrt(\pgfmathresult)}%
    }

     % left branch: colored red (x < 0)
  \draw[red!70!black,thick,domain=-2:0,smooth,variable=\x]
    plot ({\x},{sqrt((\x-1)^2 - 0.6)});
  \draw[red!70!black,thick,dashed,dash pattern=on 5pt off 1pt,domain=-2:0,smooth,variable=\x]
    plot ({\x},{-sqrt((\x-1)^2 - 0.6)});

     % left branch: colored blue (x > 0)
  \draw[blue!70!black,dashed,dash pattern=on 5pt off 1pt,thick,domain=0:0.225,smooth,variable=\x]
    plot ({\x},{sqrt((\x-1)^2 - 0.6)});
  \draw[blue!70!black,thick,domain=0:0.225,smooth,variable=\x]
    plot ({\x},{-sqrt((\x-1)^2 - 0.6)});

  % right branch: colored green (x >= 2)
  \draw[green!60!black,thick,dashed,dash pattern=on 5pt off 1pt,domain=1.776:4,smooth,variable=\x]
    plot ({\x},{sqrt((\x-1)^2 - 0.6)});
  \draw[green!60!black,thick,domain=1.776:4,smooth,variable=\x]
    plot ({\x},{-sqrt((\x-1)^2 - 0.6)});

  % special points (vertices)
  \fill (0,0) circle (2pt) node[below left] {$(0,0)$};
  \fill (1,0) circle (2pt) node[below ] {$(1,0)$};
  \fill (2,0) circle (2pt) node[below right] {$(2,0)$};
  % title
  \node at (0.5,-3.8) {$(\hat{A},\hat{B})$ plane};

  % --- shaded area between the right green branch and the y-axis (for x>0)
    \begin{scope}
      \clip (0,-3.5) rectangle (4.5,3.5); % restrict to x>0 region
      \fill[blue!70!black,opacity=0.15]
        % upper edge from x=0 to x=rootA
        plot[smooth,domain=0:\rootA,variable=\x,samples=120] 
          (\x,{sqrt(max(0,(\x-1)^2 - \a))})
        % back along the lower edge from x=rootA to x=0
        -- plot[smooth,domain=\rootA:0,variable=\x,samples=120] 
          (\x,{-sqrt(max(0,(\x-1)^2 - \a))})
        -- cycle;
    \end{scope}

    % --- Label numbers 1-6 inside small circles --- AB
  \node[circle,draw,inner sep=0.2pt] at (-1.1,1.1) {$1+$};
  \node[circle,draw,inner sep=0.2pt] at (-0.8,-2.4) {$1-$};
  \node[circle,draw,inner sep=0.2pt] at (0.6,0.6) {$3-$};
  \node[circle,draw,inner sep=0.2pt] at (0.5,-1.1) {$3+$};
  \node[circle,draw,inner sep=0.2pt] at (3.0,1.1) {$2-$};
  \node[circle,draw,inner sep=0.2pt] at (2.8,-2.4) {$2+$};
 
  \end{scope}
  %\label{fig7}
\end{tikzpicture}
\caption{Mapping between $(\hat{C}_1,\hat{C}_2)$ and $(\hat{A},\hat{B})$ planes for $|\alpha|>>0$}
  \label{fig7}
\end{figure}

\begin{itemize}
    \item[$\bigstar$] The regions \circled{$1\pm$} are obtained when the $X$ axis ($\hat{C}_1$ or $\hat{A}$) is negative. This is only applicable to the left branches. The $\pm$ denotes positive and negative values of the $Y$ axis ($\hat{C}_2$ or $\hat{B}$) in this region. They get mapped to each other through the LT.

\item[$\bigstar$] The small portion of the left branch of the hyperbola, trapped infinitesimal neighbourhood between the node and the $Y$ axis is denoted by Region \circled{$2\pm$}. This is for positive values of $\hat{C}_1$  for the left branch. This region of the hyperbola gradually shrinks to the point $(0,0)$ as one approaches the limit. This also gets mapped to the right branch hyperbola in the $\hat{A}$-$\hat{B}$ plane, but the $\hat{B}$ values take the opposite sign to that of $\hat{C}_2$, so the relative orientation flips.

\item[$\bigstar$] The right branch in the $\hat{C}_1$-$\hat{C}_2$ plane is denoted by Region \circled{$3\pm$}, which gets mapped to the small (trapped) region in the $\hat{A}$-$\hat{B}$ plane, with positive $\hat{A}$ values in the left branch, consequently trapped by the $Y$ axis.
\end{itemize}

\begin{table}[htb]
\centering
\renewcommand{\arraystretch}{1.3}
\begin{tabular}{|c|c|c|c|c|}
\hline
 & $\hat{C}_1$ & $\hat{C}_2$ & $\hat{A}$ & $\hat{B}$ \\ \hline
$\alpha\to+\infty$ &
$2\!\big[1+\epsilon^{2}\!\big(\mathcal{X}^{2}-\frac{1}{4}\big)\big]$ &
$-2\epsilon\,\mathcal{X}$ &
$\frac{\epsilon^{2}}{2}$ &
$\epsilon^{2}\,\mathcal{X}$ \\ \hline
$\alpha\to-\infty$ &
$-2\epsilon^{2}\!\big(\mathcal{X}^{2}-\frac{1}{4}\big)$ &
$+2\epsilon\,\mathcal{X}$ &
$-\frac{1}{2\big(\mathcal{X}^{2}-\frac{1}{4})}+\epsilon^2f_1$ &
$\frac{\mathcal{X}}{\mathcal{X}^{2}-\frac{1}{4}}[1+\epsilon^2f_2]$ \\ \hline
\end{tabular}
\caption{Summary of limiting values of parameters for \(\alpha\to\pm\infty\).}
\label{tab:C1C2AB_transposed}
\end{table}

For all transformations on $\hat C_1$-$\hat C_2$ plane pertaining to $\alpha \to \infty$ we need to focus on how the right hyperbola \circled{$3\pm$} changes, and for all pertaining to $\alpha \to -\infty$, how the left  hyperbola \big(\circled{$1\pm$}$~\bigcup$~\circled{$2\pm$}\big) changes.
To understand how the regions are mapped by the LT, let us recall the values of $\hat{C}_1$, $\hat{C}_2$, $\hat{A}$ and $\hat{B}$ as a function of the quantity $\mathcal{X}$ from equations \eqref{7.18-19} and \eqref{7.22-23}, as tabulated in Table\eqref{tab:C1C2AB_transposed}. The different regions of the hyperbolas in the limit are clearly dependent on  different configurations of $\mathcal{X}$, which determines the fate of the map at the two extreme values.
Looking at the specific values, we outline the algebraic structure the following cases:

\paragraph{Case 1: \(\mathcal{X}^2<\tfrac{1}{4}\):}
\begin{table}[h!]
\centering
\renewcommand{\arraystretch}{1.3}
\setlength{\tabcolsep}{8pt}
\begin{tabular}{c c c c c c}
\toprule
 Branch & \(\hat{C}_1\) %& Region(\(\hat{C}_1\)-\(\hat{C}_2\))
 & \ & \(\hat{A}\) & Region mapping at extreme limit \\%(\(\hat{C}_1\)-\(\hat{C}_2\) or \(\hat{A}\)-\(\hat{B}\)) \\
\midrule
\(\alpha \to +\infty\) &
\(2\!\left[1-\epsilon^2\Delta\right]\) %&(\(3\!\pm\)) 
&\(\Leftarrow\text{LT}\Rightarrow\) &
\(+\dfrac{\epsilon^2}{2}\) &
(\(3\!\pm\))$_{(2,0)\to(0,0)}$ \\[4pt]
\(\alpha \to -\infty\) &
\(+2\epsilon^2\Delta\) %&(\(2\!\pm\)) 
&\(\Leftarrow\text{LT}\Rightarrow\) &
\(+\dfrac{1}{2\Delta}\) &
(\(2\!\pm\))$_{(0,0)\to(\hat A(\mathcal{X}),\hat B(\mathcal{X}))}$ \\
\bottomrule
\end{tabular}
\end{table}
\begin{itemize}
\item[$\blacksquare$] In the above $\Delta=|\mathcal{X}^2-\tfrac{1}{4}|$ has been used to simplify the notation.
Here we only have to worry about +ve values of $\hat C_1$ (but only $\leq 2$), which correlates to +ve values of $\hat A$, so the region \circled{$1\pm$} does not appear here. The sign of $C_2$ dictates if the hyperbola is in region $(n+)$ or $(n-)$ (where $n=2,3$). Subsequently the sign of $\hat{B}$ will be opposite to that of $C_2$. 

\item[$\blacksquare$] For the positive branch, as $\epsilon\to0$, we approach $(\hat{C}_1,\hat{C}_2)=(2,0)$ and $(\hat{A},\hat{B})=(0,0)$, as both the small regions \circled{$2\pm$} and \circled{$3\pm$} coalesce to points,  providing an one-to-one map. However, for the negative branch, $(\hat{C}_1,\hat{C}_2)$ approaches $(0,0)$ but $(\hat{A},\hat{B})$ keeps depending on $\mathcal{X}$, giving an one-to-many map. Note the closer $\mathcal{X}^2$ goes to $0$, the closer $(\hat{A}, \hat{B})$ is to $(2, 0)$, whereas the the points diverges (to $\hat{A}\to+\infty$) if $\mathcal{X}^2$ approaches $\tfrac{1}{4}$. 

\item[$\blacksquare$] We also see that there is a criss-cross in the map, i.e. the left branch of the \(\hat{C}_1\)-\(\hat{C}_2\) hyperbola (i.e. \circled{$2\pm$}) gets mapped to the right branch of the \(\hat{A}\)-\(\hat{B}\) hyperbola and vice versa, for both $\alpha\to\pm\infty$.
\end{itemize}
%\anote{See that anything beyond $\hat{C}_1 = 2$ is not available in this branch. So in Fig 8 the \circled{$3\pm$}) on the left is of the same size roughly as of \circled{$2\pm$}), just a small semicircle encompassing the node. So on this case the map to phase space is almost area preserving?}

\paragraph{Case 2: \(\mathcal{X}^2>\tfrac{1}{4}\):}
\begin{table}[h!]
\centering
\renewcommand{\arraystretch}{1.3}
\setlength{\tabcolsep}{8pt}
\begin{tabular}{c c c c c c}
\toprule
 Branch & \(\hat{C}_1\) %& Region(\(\hat{C}_1\)-\(\hat{C}_2\)) 
 & \ & \(\hat{A}\) & Region mapping at extreme limit \\%(\(\hat{A}\)-\(\hat{B}\)) \\
\midrule
\(\alpha \to +\infty\) &
\(2\!\left[1+\epsilon^2\Delta\right]\) &
%(\(3\!\pm\)) &
\(\Leftarrow\text{LT}\Rightarrow\) &
\(+\dfrac{\epsilon^2}{2}\) &
(\(3\!\pm\))$_{(2,0)\to(0,0)}$ \\[4pt]
\(\alpha \to -\infty\) &
\(-2\epsilon^2\Delta\) &
%(\(1\!\pm\)) &
\(\Leftarrow\text{LT}\Rightarrow\) &
\(-\dfrac{1}{2\Delta}\) &
(\(1\!\pm\))$_{(0,0)\to(\hat A(\mathcal{X}),\hat B(\mathcal{X}))}$ \\
\bottomrule
\end{tabular}
\end{table}
\begin{itemize}
\item[$\blacksquare$] Note here $\hat{C_1}$ is either $\geq 2$ or $\leq0$, so the region \circled{$2\pm$} is not of importance in both the planes.

\item[$\blacksquare$] In this case, the positive branch gives the same results as Case 1, as $\epsilon\to0$ we get a point-to-point map. For the negative branch, on  \(\hat{C}_1\)-\(\hat{C}_2\) plane we focus on an infinitesimal neighbourhood in the $0^-$ regime of \circled{$1\pm$}, which coalesces to the node $(0,0)$ in the absolute limit, but the corresponding points on the \(\hat{A}\)-\(\hat{B}\) plane, depends on $\mathcal{X}$. These points approach the extreme ends ($\hat{A}\to-\infty$) as $\mathcal{X}$ gets closer to $\frac{1}{4}$.

\item[$\blacksquare$]In this negative branch mapping, there is no crossover in the map, and the sign of $\hat{B}$ is the same as $\hat{C_2}$.
\end{itemize}
Note that in both of the above cases, we are not fixing the value of  $\mathcal{X}$, so that the limiting Lagrangian for $\alpha\to -\infty$ as in \eqref{Xaction} stays manifestly Carroll boost invariant when higher orders in $\epsilon$ are neglected. But in any case, we should look at couple of intriguing fixed values, viz:

\paragraph{Case 3: Special point: \(\mathcal{X}^2=\frac{1}{4}\) i.e. $\Delta=0$:}
\begin{table}[h!]
\centering
\renewcommand{\arraystretch}{1.3}
\setlength{\tabcolsep}{8pt}
\begin{tabular}{c c c c c c}
\toprule
 Branch & \(\hat{C}_1\) & \(\hat{C}_2\) & \ & \(\hat{A}\) & \(\hat{B}\) \\
\midrule
\(\alpha \to +\infty\) &
2 &
\(\pm\epsilon\) &
\(\Leftarrow\text{LT}\Rightarrow\) &
\(+\dfrac{\epsilon^2}{2}\) &
\(\mp\dfrac{\epsilon^2}{2}\) \\[4pt]
\(\alpha \to -\infty\) &
0 &
\(\mp\epsilon\) &
\(\Leftarrow\text{LT}\Rightarrow\) &
\(\pm\infty\) &
\(\pm\infty\) \\
\bottomrule 
\end{tabular}
\end{table}
\noindent
The positive branch once again gives a consistent map from $(2,0)$ to $(0,0)$ points as $\epsilon\to0$, while the negative branch gives $(0,\mp\epsilon)$ on the \(\hat{C}_1\)-\(\hat{C}_2\) plane, that maps to the extreme end-points of the both branches of the \(\hat{A}\)-\(\hat{B}\) hyperbolas. 

%%%%%% FIGURE 8 : CASE 3 PLOT %%%%%

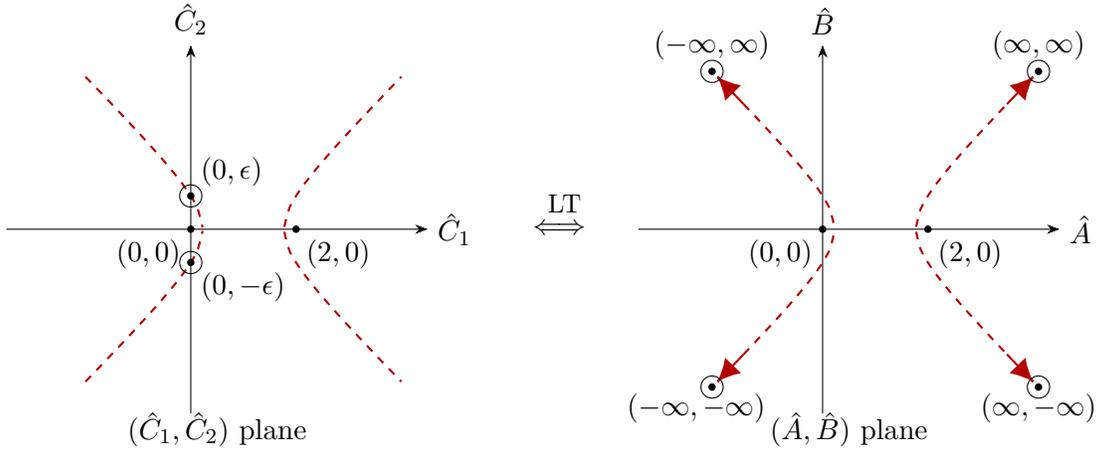
\begin{figure}[h!]
  \centering
\begin{tikzpicture}[scale=0.7,>=Stealth]
% ===== LEFT: C1,C2 plane (center at (1,0), hyperbola (x-1)^2 - y^2 = 1) =====
\begin{scope}[shift={(0,0)}]
  % axes
  \draw[->] (-3.5,0) -- (4.5,0) node[right] {$\hat{C}_1$};
  \draw[->] (0,-3.5) -- (0,3.5) node[above] {$\hat{C}_2$};

  % parameter (the value inside your original sqrt was (x-1)^2 - 0.6)
    \pgfmathsetmacro{\a}{0.6}

    % left and right intersection x-values: 1 +/- sqrt(a)
    \pgfmathsetmacro{\rootA}{1 - sqrt(\a)} % ≈ 0.2254
    \pgfmathsetmacro{\rootB}{1 + sqrt(\a)} % ≈ 1.7746

    % define function safely: Ctwo(x) = sqrt(max(0,(x-1)^2 - a))
    \pgfmathdeclarefunction{Ctwo}{1}{%
      \pgfmathparse{max(0,(#1-1)^2 - \a)}%
      \pgfmathparse{sqrt(\pgfmathresult)}%
    }

    % left branch: colored red (x < 0)
  \draw[red!70!black,thick,dashed,dash pattern=on 3pt off 3pt,domain=-2:0.225,smooth,variable=\x]
  plot ({\x},{sqrt((\x-1)^2 - 0.6)});    \draw[red!70!black,thick,dashed,dash pattern=on 3pt off 3pt,domain=-2:0.225,smooth,variable=\x]
  plot ({\x},{-sqrt((\x-1)^2 - 0.6)});

     % right branch: colored blue (x >= 2)
  \draw[red!70!black,thick,dashed,dash pattern=on 3pt off 3pt,domain=1.776:4,smooth,variable=\x]
    plot ({\x},{sqrt((\x-1)^2 - 0.6)});
  \draw[red!70!black,thick,dashed,dash pattern=on 3pt off 3pt,domain=1.776:4,smooth,variable=\x]
    plot ({\x},{-sqrt((\x-1)^2 - 0.6)});

  % special points (vertices)
  \fill (0,0) circle (2pt) node[below left] {$(0,0)$};
  %\fill (1,0) circle (2pt) node[below ] {$(1,0)$};
  \fill (2,0) circle (2pt) node[below right] {$(2,0)$};
  \fill (0,0.633) circle (2pt) node[above right] {$(0,\epsilon)$};
  \fill (0,-0.633) circle (2pt) node[below right] {$(0,-\epsilon)$};
  \draw (0,0.633) circle (6pt);
  \draw (0,-0.633) circle (6pt);

  % title
  \node at (0.5,-3.8) {$(\hat{C}_1,\hat{C}_2)$ plane};
  \node at (7,0) {$\Longleftrightarrow$ };
  \node at (7.1,0.5) {\small{LT}};
  %\node at (7,-3.8) {$(\text{for } 0<|\alpha|<\infty)$ };

       % --- Label numbers 1-6 inside small circles --- C1C2
  %\node[circle,draw,inner sep=0.2pt] at (-1.1,1.1) {$1+$};
  %\node[circle,draw,inner sep=0.2pt] at (-0.8,-2.4) {$1-$};
  %\node at (0.6,0.6) {$(+\epsilon,0)$};
  %\node at (0.5,-1.1) {$(-\epsilon,0)$};
  %\node[circle,draw,inner sep=0.2pt] at (3.0,1.1) {$3+$};
  %\node[circle,draw,inner sep=0.2pt] at (2.8,-2.4) {$3-$};
 
 \end{scope}

% ===== RIGHT: A,B plane  =====
\begin{scope}[shift={(12,0)}] % shift horizontally
  % axes
  \draw[->] (-3.5,0) -- (4.5,0) node[right] {$\hat{A}$};
  \draw[->] (0,-3.5) -- (0,3.5) node[above] {$\hat{B}$};

  % parameter (the value inside your original sqrt was (x-1)^2 - 0.6)
    \pgfmathsetmacro{\a}{0.6}

    % left and right intersection x-values: 1 +/- sqrt(a)
    \pgfmathsetmacro{\rootA}{1 - sqrt(\a)} % ≈ 0.2254
    \pgfmathsetmacro{\rootB}{1 + sqrt(\a)} % ≈ 1.7746

    % define function safely: Ctwo(x) = sqrt(max(0,(x-1)^2 - a))
    \pgfmathdeclarefunction{Ctwo}{1}{%
      \pgfmathparse{max(0,(#1-1)^2 - \a)}%
      \pgfmathparse{sqrt(\pgfmathresult)}%
    }

     % left branch: colored red (x < 0)
   \draw[red!70!black,thick,-{Latex[length=3mm,width=3mm]},domain=-1.5:-2,smooth,samples=100,variable=\x]
  plot ({\x},{sqrt((\x-1)^2 - 0.6)});
  \draw[red!70!black,thick,dashed,dash pattern=on 3pt off 3pt,domain=-1.5:0.225,smooth,variable=\x]
    plot ({\x},{sqrt((\x-1)^2 - 0.6)});
    \draw[red!70!black,thick,-{Latex[length=3mm,width=3mm]},domain=-1.5:-2,smooth,samples=100,variable=\x]
  plot ({\x},{-sqrt((\x-1)^2 - 0.6)});
  \draw[red!70!black,thick,dashed,dash pattern=on 3pt off 3pt,domain=-1.5:0.225,smooth,variable=\x]
    plot ({\x},{-sqrt((\x-1)^2 - 0.6)});

     % left branch: colored blue (x > 0)
  
  % right branch: colored green (x >= 2)
  \draw[red!70!black,thick,dashed,dash pattern=on 3pt off 3pt,domain=1.776:3.5,smooth,variable=\x]
    plot ({\x},{sqrt((\x-1)^2 - 0.6)});
  \draw[red!70!black,thick,dashed,dash pattern=on 3pt off 3pt,domain=1.776:3.5,smooth,variable=\x]
    plot ({\x},{-sqrt((\x-1)^2 - 0.6)});
     \draw[red!70!black,thick,-{Latex[length=3mm,width=3mm]},domain=3.5:4,smooth,samples=100,variable=\x]
  plot ({\x},{sqrt((\x-1)^2 - 0.6)});
  \draw[red!70!black,thick,-{Latex[length=3mm,width=3mm]},domain=3.5:4,smooth,samples=100,variable=\x]
  plot ({\x},{-sqrt((\x-1)^2 - 0.6)});

      % special points (vertices)
  \fill (0,0) circle (2pt) node[below left] {$(0,0)$};
  %\fill (1,0) circle (2pt) node[below ] {$(1,0)$};
  \fill (2,0) circle (2pt) node[below right] {$(2,0)$};
  \node at (-2.1,3.5) {$(-\infty,\infty)$};
  \fill (-2.1,3) circle (2pt);
  \draw (-2.1,3) circle (6pt);
  \node at (-2.4,-3.4) {$(-\infty,-\infty)$};
  \fill (-2.1,-3) circle (2pt);
  \draw (-2.1,-3) circle (6pt);
  \node at (4.1,3.5) {$(\infty,\infty)$};
  \fill (4.1,3) circle (2pt);
  \draw (4.1,3) circle (6pt);
  \node at (4.1,-3.4) {$(\infty,-\infty)$};
  \fill (4.1,-3) circle (2pt);
  \draw (4.1,-3) circle (6pt);
  % title
  \node at (0.5,-3.8) {$(\hat{A},\hat{B})$ plane};

      % --- Label numbers 1-6 inside small circles --- AB
  %\node[circle,draw,inner sep=0.2pt] at (-1.1,1.1) {$1+$};
  %\node[circle,draw,inner sep=0.2pt] at (-0.8,-2.4) {$1-$};
  %\node[circle,draw,inner sep=0.2pt] at (0.6,0.6) {$3-$};
  %\node[circle,draw,inner sep=0.2pt] at (0.5,-1.1) {$3+$};
  %\node[circle,draw,inner sep=0.2pt] at (3.0,1.1) {$2-$};
  %\node[circle,draw,inner sep=0.2pt] at (2.8,-2.4) {$2+$};
 
  \end{scope}
\end{tikzpicture}
\caption{Mapping between $(\hat{C}_1,\hat{C}_2)$ and $(\hat{A},\hat{B})$ planes for the negative branch, at $\mathcal{X}^2=\frac{1}{4}$}
  \label{fig8}
\end{figure}

\paragraph{Case 4: Special point: \(\mathcal{X}=0\):}
\begin{table}[h!]
\centering
\renewcommand{\arraystretch}{1.3}
\setlength{\tabcolsep}{8pt}
\begin{tabular}{c c c c c c}
\toprule
 Branch & \(\hat{C}_1\) & \(\hat{C}_2\) & \ & \(\hat{A}\) & \(\hat{B}\) \\
\midrule
\(\alpha \to +\infty\) &
\(2-\frac{1}{2}\epsilon^2\) &
0 &
\(\Leftarrow\text{LT}\Rightarrow\) &
\(+\dfrac{\epsilon^2}{2}\) &
0 \\[4pt]
\(\alpha \to -\infty\) &
\(+\dfrac{\epsilon^2}{2}\) &
0 &
\(\Leftarrow\text{LT}\Rightarrow\) &
\(2-\frac{1}{2}\epsilon^2\)&
0 \\
\bottomrule
\end{tabular}
\end{table}
\(\mathcal{X}=0\) denotes the points, where the $X$ axis intersects the hyperbolas, i.e., the nodes $(\tfrac{\epsilon^2}{2},0)$ and $(2-\tfrac{\epsilon^2}{2},0)$. In both the branches we have a clear map between the nodes for either planes. 
The small regions \circled{$2\pm$} and \circled{$3\pm$} (as in Fig.\ref{fig7}) in the $\hat{C}_1$-$\hat{C}_2$ and $\hat{A}$-$\hat{B}$ planes, respectively are given by the points $(0,\pm\epsilon)$ and the node $(\tfrac{\epsilon^2}{2},0)$. We observe that the region converges toward the node $(0,0)$ in the limit. %This will be important while considering the electric and magenttic limits.
\medskip

So what does all of these discussions mean for actual Carroll theories? In our quest of extreme limits, we have unearthed the two inequivalent situations for $|\alpha|\to \infty$ where in the first case we always get an one-to-map between configuration space and phase space, even at bang-on the limit! The second, $\mathcal{X}$ dependent case, would not have such a map, but unfixed  $\mathcal{X}$ will always provide Carroll boost invariance. A single limiting point on the configuration space will map to a finite region in the phase space. One can check the canonical momenta coming from \eqref{Xaction}:
\begin{equation}
    \pi_i = \frac{\partial \hat{\mathcal{L}}^{-}_\alpha}{\partial \dot{\phi}_i}
= 2\mathcal{X} \phi_i' + \dot{\phi}_i\left(\frac12 - 2\mathcal{X}^2\right),
\end{equation}
which remains unfixed for the limiting case of the negative branch. Note this momenta clearly distinguishes between the two cases $\mathcal{X}^2<1/4$ and $\mathcal{X}^2>1/4$, with the latter pointing towards an instability.
Most importantly, the positive branch leads to a leading action with only temporal derivatives, while the one from negative branch contains spatial derivatives as well.

\subsubsection*{Geometrising Electric/Magnetic limits}
We are now led to the culmination of all the intuitive understanding we've gathered so far.
Exactly at the limiting value ($\epsilon=0$), the positive branch in the $\hat{C}_1$-$\hat{C}_2$ plot, given by region \circled{$3\pm$} collapses to the point $(2,0)$. The Lagrangian is given by equation \eqref{7.9}, and we identify this limit as the ``Electric'' Carrollian limit, having same form as the similar actions in \cite{deBoer:2021jej,Henneaux:2021yzg} for example.
\medskip

The negative branch is subsequently identified as the ``Magnetic'' Carrollian limit. One intuitive way to argue for this comes from the definition of the momentum in equation \eqref{Ele.Canon.Momentum}, since as $\alpha \to -\infty$, we have the region \circled{$2\pm$} in the $\hat{C}_1$-$\hat{C}_2$ plane collapsing to the point $(0,0)$ which also coincides with the node of the left hyperbola. At this exact limit, the canonical momenta in \eqref{Ele.Canon.Momentum} becomes just proportional to $\phi'$. 
\medskip

The above discussion was from the Lagrangian point of view.
In the Hamiltonian picture (\(\hat{A}\)-\(\hat{B}\) plane), the ``Magnetic'' case is given by the point $(0,0)$ while the ``Electric'' is the opposite, due to how the associated regions are mapped into each other. This can be verified from equation \eqref{Electricmom}. 
The corresponding maps are reversed. The electric limit in $\hat{C}_1$-$\hat{C}_2$ gives the magnetic limit in $\hat{A}$-$\hat{B}$ and vice versa. 
\medskip

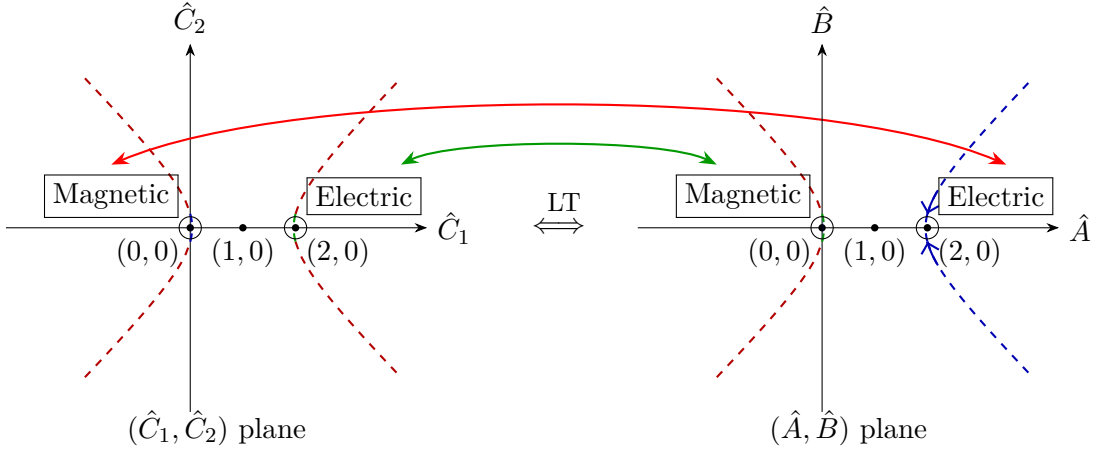
\begin{figure}[h!]
  \centering
\begin{tikzpicture}[scale=0.7,>=Stealth]
% ===== LEFT: C1,C2 plane (center at (1,0), hyperbola (x-1)^2 - y^2 = 1) =====
\begin{scope}[shift={(0,0)}]
  % axes
  \draw[->] (-3.5,0) -- (4.5,0) node[right] {$\hat{C}_1$};
  \draw[->] (0,-3.5) -- (0,3.5) node[above] {$\hat{C}_2$};

  % parameter (the value inside your original sqrt was (x-1)^2 - 0.6)
    \pgfmathsetmacro{\a}{0.93}

    % left and right intersection x-values: 1 +/- sqrt(a)
    \pgfmathsetmacro{\rootA}{1 - sqrt(\a)} % ≈ 0.2254
    \pgfmathsetmacro{\rootB}{1 + sqrt(\a)} % ≈ 1.7746

    % define function safely: Ctwo(x) = sqrt(max(0,(x-1)^2 - a))
    \pgfmathdeclarefunction{Ctwo}{1}{%
      \pgfmathparse{max(0,(#1-1)^2 - \a)}%
      \pgfmathparse{sqrt(\pgfmathresult)}%
    }

   % left branch: colored red (x < 0)
  \draw[red!70!black,thick,dashed,dash pattern=on 3pt off 3pt,domain=-2:0,smooth,variable=\x]
    plot ({\x},{sqrt((\x-1)^2 - 0.93)});
  \draw[red!70!black,thick,dashed,dash pattern=on 3pt off 3pt,domain=-2:0,smooth,variable=\x]
    plot ({\x},{-sqrt((\x-1)^2 - 0.93)});

    % left branch: colored blue (x > 0)
  \draw[blue!70!black,thick,domain=0:0.03,smooth,variable=\x]
    plot ({\x},{sqrt((\x-1)^2 - 0.93)});
  \draw[blue!70!black,thick,domain=0:0.03,smooth,variable=\x]
    plot ({\x},{-sqrt((\x-1)^2 - 0.93)});

  % right branch: colored green (x < 2)
  \draw[green!60!black,thick,domain=1.97:2,smooth,variable=\x]
    plot ({\x},{sqrt((\x-1)^2 - 0.93)});
  \draw[green!60!black,thick,domain=1.97:2,smooth,variable=\x]
    plot ({\x},{-sqrt((\x-1)^2 - 0.93)});

    % right branch: colored red (x > 2)
  \draw[red!70!black,thick,dashed,dash pattern=on 3pt off 3pt,domain=2:4,smooth,variable=\x]
    plot ({\x},{sqrt((\x-1)^2 - 0.93)});
  \draw[red!70!black,thick,dashed,dash pattern=on 3pt off 3pt,domain=2:4,smooth,variable=\x]
    plot ({\x},{-sqrt((\x-1)^2 - 0.93)});

  % special points (vertices)
  \fill (0,0) circle (2pt) node[below left] {$(0,0)$};
  \fill (1,0) circle (2pt) node[below ] {$(1,0)$};
  \fill (2,0) circle (2pt) node[below right] {$(2,0)$};
  \draw (0,0) circle (6pt);
  \draw (2,0) circle (6pt);

  % title
  \node at (0.5,-3.8) {$(\hat{C}_1,\hat{C}_2)$ plane};
  \node at (7,0) {$\Longleftrightarrow$ };
  \node at (7.1,0.5) {\small{LT}};
  \node at (3.3,0.6) {\boxed{\text{Electric}}};
  \node at (-1.5,0.6) {\boxed{\text{Magnetic}}};
  %\node at (7,-3.8) {$(\text{for } 0<|\alpha|<\infty)$ };

   % --- shaded area between the left blue branch and the y-axis (for x>0)
    \begin{scope}
      \clip (0,-3.5) rectangle (4.5,3.5); % restrict to x>0 region
      \fill[blue!70!black,opacity=0.15]
        % upper edge from x=0 to x=rootA
        plot[smooth,domain=0:\rootA,variable=\x,samples=120] 
          (\x,{sqrt(max(0,(\x-1)^2 - \a))})
        % back along the lower edge from x=rootA to x=0
        -- plot[smooth,domain=\rootA:0,variable=\x,samples=120] 
          (\x,{-sqrt(max(0,(\x-1)^2 - \a))})
        -- cycle;
    \end{scope}

    \draw[<->,thick,green!60!black]
  (4,1.2) .. controls (5,1.7) and (9,1.7) .. (10,1.2);

  \draw[<->,thick,red]
  (-1.5,1.2) .. controls (2,2.7) and (12,2.7) .. (15.5,1.2);

\end{scope}

% ===== RIGHT: A,B plane  =====
\begin{scope}[shift={(12,0)}] % shift horizontally
  % axes
  \draw[->] (-3.5,0) -- (4.5,0) node[right] {$\hat{A}$};
  \draw[->] (0,-3.5) -- (0,3.5) node[above] {$\hat{B}$};

  % parameter (the value inside your original sqrt was (x-1)^2 - 0.6)
    \pgfmathsetmacro{\a}{0.93}

    % left and right intersection x-values: 1 +/- sqrt(a)
    \pgfmathsetmacro{\rootA}{1 - sqrt(\a)} % ≈ 0.2254
    \pgfmathsetmacro{\rootB}{1 + sqrt(\a)} % ≈ 1.7746

    % define function safely: Ctwo(x) = sqrt(max(0,(x-1)^2 - a))
    \pgfmathdeclarefunction{Ctwo}{1}{%
      \pgfmathparse{max(0,(#1-1)^2 - \a)}%
      \pgfmathparse{sqrt(\pgfmathresult)}%
    }

    % left branch: colored red (x < 0)
  \draw[red!70!black,thick,dashed,dash pattern=on 3pt off 3pt,domain=-2:0,smooth,variable=\x]
    plot ({\x},{sqrt((\x-1)^2 - 0.93)});
  \draw[red!70!black,thick,dashed,dash pattern=on 3pt off 3pt,domain=-2:0,smooth,variable=\x]
    plot ({\x},{-sqrt((\x-1)^2 - 0.93)});

    % left branch: colored blue (x > 0)
  \draw[green!60!black,thick,domain=0:0.03,smooth,variable=\x]
    plot ({\x},{sqrt((\x-1)^2 - 0.93)});
  \draw[green!60!black,thick,domain=0:0.03,smooth,variable=\x]
    plot ({\x},{-sqrt((\x-1)^2 - 0.93)});

  % right branch: colored green (x < 2)
  \draw[blue!70!black,thick,domain=1.97:2,smooth,variable=\x]
    plot ({\x},{sqrt((\x-1)^2 - 0.93)});
  \draw[blue!70!black,thick,domain=1.97:2,smooth,variable=\x]
    plot ({\x},{-sqrt((\x-1)^2 - 0.93)});

    % right branch: colored blue (x > 2)
  \draw[blue!70!black,thick,dashed,dash pattern=on 3pt off 3pt,domain=2:4,smooth,variable=\x]
    plot ({\x},{sqrt((\x-1)^2 - 0.93)});
  \draw[blue!70!black,thick,dashed,dash pattern=on 3pt off 3pt,domain=2:4,smooth,variable=\x]
    plot ({\x},{-sqrt((\x-1)^2 - 0.93)});
    \draw[blue!70!black,thick,-{Straight Barb[length=1.2mm]},domain=2.2:2,smooth,samples=100,variable=\x]
  plot ({\x},{sqrt((\x-1)^2 - 0.93)});
    \draw[blue!70!black,thick,-{Straight Barb[length=1.2mm]},domain=2.2:2,smooth,samples=100,variable=\x]
  plot ({\x},{-sqrt((\x-1)^2 - 0.93)});

  % special points (vertices)
  \fill (0,0) circle (2pt) node[below left] {$(0,0)$};
  \fill (1,0) circle (2pt) node[below ] {$(1,0)$};
  \fill (2,0) circle (2pt) node[below right] {$(2,0)$};
  \draw (0,0) circle (6pt);
  \draw (2,0) circle (6pt);

  \node at (3.3,0.6) {\boxed{\text{Electric}}};
  \node at (-1.5,0.6) {\boxed{\text{Magnetic}}};
  
  \node at (0.5,-3.8) {$(\hat{A},\hat{B})$ plane};

  % --- shaded area between the right blue branch and the y-axis (for x>0)
    \begin{scope}
      \clip (0,-3.5) rectangle (4.5,3.5); % restrict to x>0 region
      \fill[blue!70!black,opacity=0.15]
        % upper edge from x=0 to x=rootA
        plot[smooth,domain=0:\rootA,variable=\x,samples=120] 
          (\x,{sqrt(max(0,(\x-1)^2 - \a))})
        % back along the lower edge from x=rootA to x=0
        -- plot[smooth,domain=\rootA:0,variable=\x,samples=120] 
          (\x,{-sqrt(max(0,(\x-1)^2 - \a))})
        -- cycle;
    \end{scope}
  \end{scope}
\end{tikzpicture}
\caption{Mapping between $(\hat{C}_1,\hat{C}_2)$ and $(\hat{A},\hat{B})$ planes at the Carrollian limit ($|\alpha|\to\infty$). The swap between Electric and Magnetic pictures have been shown above.}
  \label{fig9}
\end{figure}

From Table \eqref{tab:C1C2AB_transposed} we see that the map of the region \circled{$2\pm$} shrinks to a point in $\hat{C}_1$-$\hat{C}_2$, while in $\hat{A}$-$\hat{B}$ gives the entire right branch, depending on $\mathcal{X}$. However the nodes of the hyperbolas always map to each other. Since at the limit the ``Magnetic'' region collapses to the node, therefore, exactly at the limit, this must correspond to the other node $(2,0)$ in the $\hat{A}$-$\hat{B}$: which is the ``Electric'' limit. In summary, \textit{configuration space electric limit corresponds to phase space magnetic one, and vice versa}. This is represented by the map in Fig.\eqref{fig9}. 

%%%%%%%%%%%%%%%%% FIGURE 9: CASE 4 PLOT %%%%%%%%%%%%%%%%%%%%%

%\color{magenta}*** Place the figure correctly\color{black}

\section{A String Theory example}\label{sect6}

To appreciate even better what we have done in this work so far, let us consider the example of a bosonic string worldsheet propagating in flat spacetime. In conformal gauge, the theory is that of pullback scalars on the worldsheet, and the residual gauge symmetry on the worldsheet is given by the 2d conformal algebra, i.e. two copies of the Virasoro algebra. Therefore, we can go down the same road as we did for 2d CFTs in the previous sections. 
\medskip

We deform the worldsheet gauge fixed theory by a $\sqrt{T\overline{T}}$ operator, with the deformation parameter $\alpha$. For finite $\alpha$ we will show that it is equivalent to choosing a different gauge, however equivalent to the conformal one, as the worldsheet metric becomes dependent on the parameter. As $|\alpha|\to\infty$ we get a string theory that has BMS$_3$ as the worldsheet residual gauge symmetry. Depending on the sign of the parameter, we obtain the ``electric'' or the ``magentic'' Carrollian limits of the tensile string. But the worldsheet Carroll limit also affects the tension, that goes to zero or infinity respectively.  To study this deformation explicitly, we begin with the Polyakov action for the bosonic string in a flat target space, which is given by

\begin{equation}\label{Polk}
S_{P}=-\frac{{1}}{4\pi\alpha'}\int d^2\sigma\ \sqrt{-g}g^{ab}\partial_a X^\mu \partial_b X^\nu \eta_{\mu\nu},
\end{equation}
where $\alpha'\sim l_s^2$ denotes the string length (not to be confused with our coupling parameter $\alpha$), and $\frac{1}{2\pi\alpha'}=\mathbb{T}$ is the string tension. 
The action \eqref{Polk} can be written in the Hamiltonian form:
\begin{equation}
S=\int d^2\sigma\ \Big[\Pi\cdot\dot{X}-N\mathcal{H}-N_\phi \mathcal{J}].
\label{6.2}
\end{equation}
with $\Pi_\mu=\frac{\delta \mathcal{L}}{\delta \dot{X}^\mu}$ being the canonical momentum. Note that $\mathcal{H}=\Pi^2+\mathbb{T}^2X^{\prime2}$ and $\mathcal{J}=\Pi\cdot X^\prime$ are the Hamiltonian and momentum constraints whose classical algebra generates the conformal Posisson brackets. $N$ and $N_\phi$ are the Lagrange multipliers (lapse and shift, respectively). The Polyakov action \eqref{Polk} can be recovered from the Hamiltonian form by redefining the intrinsic metric
\begin{equation}
g^{ab}=\begin{pmatrix}
-1 & N_\phi \\ N_\phi & \ \ \ -N^2_\phi+4 N^2 \mathbb{T}^2  
\end{pmatrix},\label{metric}
\end{equation}
and the action in the configuration space being 
\begin{equation}
S=\frac{1}{2}\int d^2\sigma \frac{1}{2N}\Big[\Pi\cdot \dot{X}-2N_\phi\Pi\cdot X^\prime+(N_\phi^2-4N^2\mathbb{T}^2)X^{\prime2}\Big]. \label{7.4} 
\end{equation}
It would be useful to work with the conformal gauge, which is given by $(g^{ab}=\eta^{ab})$. This corresponds to putting $N=\frac{1}
{2\mathbb{T}}$ and $N_\phi=0$, and the action simplifies to the multiple scalar field action:
\begin{equation}\label{confaction}
S=\frac{\mathbb{T}}{2}\int d^2\sigma\big[  \dot{X}^2 - X^{\prime 2}\big],
\end{equation}
where a field index has been suppressed.
Therefore, we can use our formulation for N scalar fields for the case of bosonic strings to define the deformations. In this gauge, the contraints are given by
\begin{align} \label{9.6}
\dot{X}^2+X^{\prime2}=0; \quad 
\dot{X}\cdot X^{\prime}=0,
\end{align}
while the EOM are given by 
\begin{equation}
    \ddot{X}^\mu-X^{''\mu}=0.
    \label{6.7}
\end{equation}
The constraints \eqref{9.6} would prove to be very useful when we apply the formalism used in the previous sections, for the $\sqrt{T\bar{T}}$ deformation of the bosonic string.

\subsection{Deformation at finite $\alpha$}

\subsubsection*{Deformation of the Hamiltonian}

Let us directly apply the deformation to the gauge fixed theory we have. 
The $\sqrt{T\overline{T}}$ deformation can be applied by considering the deformed Hamiltonian \eqref{Hamiltonioncompact}
\begin{align}
\mathcal{H}_\alpha&=\cosh (\alpha) \mathcal{H}-\sinh (\alpha) \sqrt{\mathcal{H}^2-\mathcal{J}^2}, \label{9.8}
\end{align}
For the string case, this can be expanded in terms of undeformed constraints
\begin{align}
\mathcal{H}_\alpha&=\cosh (\alpha)\left(\Pi^2 + \mathbb{T}^2X^{\prime2}\right) -\sinh (\alpha) \sqrt{\left(\Pi^2 + \mathbb{T}^2X^{\prime2}\right)^2-\left(\Pi \cdot X^{\prime}\right)^2}, \label{9.9}\end{align}
The action in phase space for the deformed bosonic string can then equivalently be written using deformed constraints and associated Lagrange multipliers:
\begin{align}
S&=\int d^2\sigma\big[\Pi\cdot \dot{X}-{N}_\alpha\mathcal{H}_\alpha-N_\phi \mathcal{J}\big],
%&=\int d^2\sigma\big[\Pi\cdot \dot{X}-\hat{N}_\alpha\big(\Pi^2 + \mathbb{T}^2X^{\prime2}\big)-N_\phi \Pi\cdot X^\prime \nonumber \\
%&\qquad \qquad \qquad \qquad +\hat{N}_\alpha\tanh (\alpha) \sqrt{\left(\Pi^2 + \mathbb{T}^2X^{\prime2}\right)^2-\left(\Pi \cdot X^{\prime}\right)^2} \big]
\label{6.10n}
\end{align}
where $N_\alpha$ is the ``deformed'' lapse function. It is trivial to check that the momentum constraint remains invariant:
\begin{equation}
\mathcal{J}_\alpha=\mathcal{J}=\Pi\cdot X^\prime.\label{6.11}\end{equation}
We can therefore define the full action in phase space for the deformed bosonic string:
\begin{align}
S&=\int d^2\sigma\Big[\Pi\cdot \dot{X}-N_\alpha\cosh(\alpha)\big(\Pi^2 + \mathbb{T}^2X^{\prime2}\big)-N_\phi \Pi\cdot X^\prime \nonumber \\
&\qquad \qquad \qquad \qquad \qquad \qquad \qquad +N_\alpha\sinh (\alpha) \sqrt{\left(\Pi^2 + \mathbb{T}^2X^{\prime2}\right)^2-\left(\Pi \cdot X^{\prime}\right)^2} \Big].
\label{6.12}
\end{align}
The equivalent of taking the conformal gauge would be taking $N_\alpha=\lambda$ and $\ N_\phi=0$, where $\lambda$ is a constant. In the conformal gauge, this becomes
\begin{align}
S&=\int d^2\sigma\left[\Pi\cdot \dot{X}-\lambda\cosh(\alpha)\big(\Pi^2 + \mathbb{T}^2X^{\prime2}\big)+\lambda\sinh{(\alpha)}\sqrt{\left(\Pi^2 + \mathbb{T}^2X^{\prime2}\right)^2-\left(\Pi \cdot X^{\prime}\right)^2} \right]. \label{6.13}
\end{align}
The presence of the square root in the deformation term prevents us from integrating out the momenta and writing it in configuration space. However, we will see that using the LT \eqref{L2} it is possible to simplify the expression and write an action in configuration space.
\medskip

One should note here that the changed Hamiltonian constraint coupled to the changed Lagrange multiplier $N_\alpha  \cosh\alpha$ would take the same form as in \eqref{scaledHTTD}. Hence, the constraint algebra, in the same vein as in \eqref{interpalgebra}, would simply flow to the BMS one in the $\alpha \to \pm \infty$ limit.

\subsection*{Deformation of the Lagrangian}

We can apply the deformation to the Lagrangian by using \eqref{5.1}:
\begin{align}
\mathcal{L}_\alpha=\cosh (\alpha)\mathcal{L}+\sinh (\alpha)\sqrt{\mathcal{L}^2-\mathcal{P}^2}.
\label{6.14}
\end{align}
Considering the conformal gauge \eqref{confaction}, the deformed Lagrangian is written as
\begin{align}
\mathcal{L}_\alpha = \frac{\mathbb{T}}{2} \cosh(\alpha) \big[  \dot{X}^2 - X^{\prime 2}\big]+ \frac{\mathbb{T}}{2} \sinh(\alpha) \sqrt{(\dot{X}^2 - X^{\prime 2})^2-\det|\gamma_{ab}|}.
\label{6.15}
\end{align}
where $\mathcal{P}^2=\det|\gamma|$, and $\gamma_{ab}$ is the induced metric defined by
\begin{equation} 
\gamma_{ab}=\begin{pmatrix}
  \dot{X}^2 & \dot{X}\cdot X^\prime \\
  \dot{X}\cdot X^\prime & X^{\prime 2}\\
  \end{pmatrix}.
  \label{6.16}
  \end{equation}
We can rewrite the deformed Lagrangian in linearised form exactly in the same fashion as \eqref{LagranC1C2} :
\begin{equation}
    \mathcal{L}_\alpha=\mathbb{T}\left[\frac{C_1}{2}(\dot{X}^2 + X^{\prime 2}) + C_2 \dot{X}\cdot X^\prime - \cosh(\alpha) X^{\prime 2} \right],
    \label{6.17}
\end{equation}
where $C_1$ and $C_2$ are defined by
\begin{eqnarray}
\begin{split}
C_1&=\cosh(\alpha)+\frac{\sinh(\alpha)(\dot{X}^2 + X^{\prime 2})}{\sqrt{(\dot{X}^2 + X^{\prime 2})^2-(\dot{X}\cdot X^{\prime })^2}}, \\
C_2&=-\frac{\sinh(\alpha)\dot{X}\cdot X^{\prime}}{\sqrt{(\dot{X}^2 + X^{\prime 2})^2-(\dot{X}\cdot X^{\prime })^2}}.
\end{split}
\end{eqnarray}
%{\color{red} Check the factors of $\mathbb{T}$.}
Note that $C_1$ and $C_2$ are directly related to the constraints \eqref{9.6}, hence provided we can still fix the conformal gauge on the worldsheet, they can be treated as ($\alpha$ dependent) constants in subsequent analysis. This is connected to the Hamiltonian \eqref{6.13} which can be rewritten from \eqref{HamilAandB}
\begin{align}
\mathcal{H}_\alpha=A\Big(\Pi^2+\mathbb{T}^2X^{\prime2}\Big)+B\ \Pi\cdot X^\prime, \label{9.19}
\end{align}
where $A$ and $B$ are given by 
\begin{align}
A&=\cosh(\alpha)-\frac{\sinh(\alpha)(\Pi^2 + \mathbb{T}^2X^{\prime 2})}{\sqrt{(\Pi^2 + \mathbb{T}^2X^{\prime 2})^2-(\Pi\cdot X^{\prime })^2}},\label{6.20} \\
B&=\frac{\sinh(\alpha)\Pi\cdot X^{\prime}}{\sqrt{(\Pi^2 + \mathbb{T}^2X^{\prime 2})^2-(\Pi\cdot X^{\prime })^2}}.
\label{6.21}
\end{align}
Similar to $C_1$ and $C_2$, the quantities $A$ and $B$ can also be treated as constants of motion, as they are directly related to the Hamiltonian and momentum constraints. The connection $A=\frac{1}{C_1}$ and $B=-\frac{C_2}{C_1}$ represents the LT between the Lagrangian and the Hamiltonian formalisms. 

\subsubsection*{Deformation of the intrinsic metric}

In section \eqref{sec8} we have seen how both the Lagrangian constants $(C_1, C_2)$ and the Hamiltonian constants $(A,B)$ obey the equations 
\begin{align}
C_1^2-2C_1\cosh(\alpha)+1&=C_2^2; \\
A^2-2A\cosh(\alpha)+1&=B^2 \label{9.23},
\end{align}
which means that the constraints are not completely independent from each other, but always related by the set of hyperbolae, depicted by the above equations.  
%From this point, we will be working with rescaled quantities $\hat{A}=A/\cosh(\alpha)$  and $\hat{B}=B/\cosh(\alpha)$. 
The linearised form allows us to integrate out the momenta from the phase space action:
\begin{align}
S&=\int d^2\sigma\left[\Pi\cdot \dot{X}-{N}_\alpha A\big(\Pi^2 + \mathbb{T}^2X^{\prime2}\big)-\big(N_\phi+N_\alpha B\big)\Pi\cdot X^\prime\right]. \label{6.24}
\end{align}
By writing the momenta as %\eqref{hamiltonionmomentum}
\begin{equation}
\Pi_\mu=\frac{1}{2N_\alpha A}\big[\dot{X}_\mu-(N_\phi+N_\alpha B)X^\prime_\mu \big],
\label{6.25}
\end{equation}
we get the action in the configuration space as
\begin{align}
S&=\frac{1}{2}\int d^2\sigma\frac{1}{2N_\alpha A}\big[\dot{X}^2-2(N_\phi+N_\alpha B)\dot{X}\cdot X^\prime \nonumber \\
&\qquad\qquad\qquad\qquad+(N_\phi^2+2N_\phi N_\alpha B+N_\alpha^2 {B}^2-4N_\alpha^2 A^2 \mathbb{T}^2)X^{\prime2}\big]. \label{6.27}
\end{align}
So now we can see how the deformation modifies the string worldsheet from the metric 
\begin{equation}
g^{ab}_{(\alpha)}=\begin{pmatrix}
-1 & N_\phi+N_\alpha B \\ N_\phi+N_\alpha B & \ \ \ -(N_\phi+N_\alpha B)^2+4 N_\alpha^2 A^2 \mathbb{T}^2  
\end{pmatrix}.
\end{equation}
If we redefine $\varrho=N_\phi+N_\alpha B$ and $\Lambda=N_\alpha A$, 
\begin{equation}
g^{ab}_{(\alpha)}=\begin{pmatrix}
-1 & \varrho \\ \varrho & \ \ \ -\varrho^2+4\Lambda^2 \mathbb{T}^2  
\end{pmatrix},
\label{6.28}
\end{equation}
and the action becomes 
\begin{align}
S&=\frac{1}{2}\int d^2\sigma\frac{1}{2\Lambda}\big[\dot{X}^2-2\varrho\dot{X}\cdot X^\prime+(\varrho^2-4\Lambda^2 \mathbb{T}^2)X^{\prime2}\big]. \label{6.29}
\end{align}
which is the usual expression for the action of the string \eqref{7.4} in configuration space. To see the effect of the deformation, we consider the conformal gauge, where $(N_\alpha=\lambda,N_\phi=0)$. Then the deformed gauge parameters become $\Lambda=\lambda A$ and $\varrho=\lambda B$. Thus we see that the role of the lapse and shift are taken by $A$ and $B$, respectively. The trivial case of $\alpha=0$, leads us to the familiar action in the conformal gauge \eqref{confaction}. Once the deformation is applied it takes the string to a different gauge. 
\medskip

In the limit $\alpha\to\pm\infty$ the metric \eqref{6.28} becomes degenerate ($4\Lambda^2\mathbb{T}^2=0$), corresponding to the Carroll points, as we will discuss in the following sections. In one case, we obtain the tensionless limit of the bosonic string ($\mathbb{T}=0$), while in the other case, we get $\Lambda=0$, which makes the action \eqref{6.29} blow up. This happens because the momentum \eqref{6.25} is also ill-defined in this case. Therefore, the standard route in the Hamiltonian formalism needs to be modified.     
\medskip

The gauge parameters in \eqref{6.29} also become related to each other here. In fact \eqref{9.23} can be modified to
\begin{equation}
\Lambda^2-2\lambda\Lambda\cosh(\alpha)+\lambda^2=\varrho^2.
\end{equation}
The degrees of freedom doesn't change, because now we need to choose a particular $\lambda$ and one of either $\Lambda$ or $\varrho$. Choosing a particular $\lambda$, the above equation gives two hyperbolas parameterised by $\alpha$ in the $\Lambda$-$\varrho$ plane, and $\varrho=0$ represents the major axis.  

\subsubsection*{Effective tension}

Let us see how the tension is affected by the deformation. To see this, at first we look at equations \eqref{6.20} and \eqref{6.21}. These consist of the Hamiltonian and momentum constraints $\mathcal{H}$ and $\mathcal{J}$. For a finite $\alpha$, the Hamiltonian constraint changes to $\mathcal{H}_\alpha$, while the momentum constraint remains the same. Thus, if we use the constraint $\Pi\cdot X^\prime=0$, $A$ and $B$ becomes
\begin{align}
A &=\cosh(\alpha)-\frac{\sinh(\alpha)(\Pi^2 + \mathbb{T}^2X^{\prime 2})}{|(\Pi^2 + \mathbb{T}^2X^{\prime 2})|};\quad B=0.
\label{6.31}
\end{align}
Depending on the sign of $(\Pi^2 + \mathbb{T}^2X^{\prime 2})$, we can have two values: $A = e^{\pm\alpha}$. This in turn translates to fixing $\Lambda=\lambda e^{\pm\alpha}$ and $\varrho=0$. Replacing it into the action we get
\begin{align}
S&=\frac{1}{2}\int d^2\sigma\frac{1}{2\lambda e^{\pm\alpha}}\big[\dot{X}^2-4\lambda^2e^{\pm2\alpha}\  \mathbb{T}^2X^{\prime2}\big]. \label{6.32}
\end{align}
We can recast this into the familiar Polyakov form, by considering $\lambda=(2\mathbb{T})^{-1}f(\alpha)$ (remembering the conformal gauge choice) and considering an arbitrary function of $\alpha$. This gives 
\begin{align}
S&=-\frac{\mathbb{T}e^{\mp\alpha}}{2}\int d^2\sigma\sqrt{-|g_{(\alpha)}|}g_{(\alpha)}^{ab}\partial_a X^\mu\partial_b X^\nu\eta_{\mu\nu}. \label{6.33}
\end{align}
where the intrinsic metric $g^{ab}$ depends on $f(\alpha)$:

\begin{equation}
g^{ab}_{(\alpha)}=\begin{pmatrix}
-1 & 0 \\ 0 & f^2(\alpha)e^{\pm2\alpha}   
\end{pmatrix}.
\label{6.34}
\end{equation}
Therefore, we can see that the deformation rescales the tension, and changes the worldsheet metric. The effective tension is given by $\mathbb{T}e^{\mp\alpha}$. We can see at a glance that the tension goes to zero as $\alpha\to+\infty$, while it goes to infinity for the other limit.      

\subsection{The deformed string at the Carrollian limits}

We mentioned before, as we approach the Carrollian limit ($\alpha\to\pm\infty$) asymptotically, it is useful to consider the quantities that are scaled (``hatted''). We consider the scaled version of the Lagrangian \eqref{6.17} and write down the action:
\begin{equation}
   S=\int d\hat{\tau}d\sigma\ \hat{\mathcal{L}_\alpha}=\frac{\mathbb{T}}{2}\int d\hat{\tau}d\sigma \left[\hat{C}_1(\hat{\dot{X}}^2 + X^{\prime 2}) + 2\hat{C}_2 \hat{\dot{X}}\cdot X^\prime -  2X^{\prime 2} \right],
\end{equation}
where $\hat{C}_1$ and $\hat{C}_2$ are the scaled $C_1$ and $C_2$ \eqref{7.4a}. To keep things on the same footing, the $\hat{\tau}$ and consequently $\hat{\dot{X}}$ also originate from the scaled version of the time-like coordinate on the worldsheet. At the verge of Carrollian limit, these are given by
\begin{equation}
\hat{\tau}=\frac{\tau}{\cosh\alpha}=\epsilon\tau;\qquad \hat{\dot{X}}=\dot{X}\cosh\alpha=\frac{\dot{X}}{\epsilon}.    
\end{equation}
We have again used $\epsilon=\frac{1}{\cosh\alpha}\to 0$ for $\alpha$ being very large. %as we did in section \ref{}, 
We can use the asymptotic form of $\hat{C}_1$ and $\hat{C}_2$ from \eqref{7.7} which gives us the action for the positive and negative branches. Using those, the limiting actions take the now familiar forms:
\begin{align}
S^{(+)} &=
\mathbb{T}\int \frac{d\tau d\sigma}{\epsilon}
\Big[\dot{X}^2 - \mathcal{O}(\epsilon^2)\Big], & (\alpha \to +\infty)
\label{6.37}
\\S^{(-)} &=
\mathbb{T}\int d\tau d\sigma\ \epsilon\ \Big[-X^{\prime 2} + \dot{X}^{2}\!\left(\mathcal{X}^2+\dfrac{1}{4}\right) + \mathcal{O}(\epsilon^{2})\Big],& (\alpha \to -\infty)
\label{6.38}
\end{align}
where $\mathcal{X}=\frac{\dot{X}\cdot X^{\prime}}{\dot{X}^2}$ is the dimensionless nonlinear factor that we had seen before in the case of field theory. To make the action finite, the tension needs to be scaled accordingly for the two separate branches:  
\begin{align}
\mathbb{T}=\frac{1}{2\pi\alpha^\prime}\to\frac{\epsilon}{2\pi c^\prime}, \quad (\alpha \to +\infty); \\
\mathbb{T}=\frac{1}{2\pi\alpha^\prime}\to\frac{\epsilon^{-1}}{2\pi c^\prime}, \quad (\alpha \to -\infty), 
    \end{align}
 where $c^\prime$ is finite and has been kept to match the dimensions of length. As per our discussion in the scalar field case, both of these actions are clearly invariant under Carroll boosts on the worldsheet. But noticeably, the tensions for these two kind of strings are different. 
%%%%
The electric-like deformation is the standard tensionless string. This can be given by the ILST action \cite{Isberg:1993av}:
\begin{equation}
    S=\int d^2\sigma\ V^a V^b\partial_a X^\mu\partial_b X^\nu\eta_{\mu\nu}, 
\end{equation}
where $V^a$ is the vector density that replaces the intrinsic metric $g^{ab}$ in the tensionless limit. The action \eqref{6.37} is recovered by choosing the gauge: $V^a=(\frac{1}{\sqrt{2\pi c^\prime}},0)$. This type of tensionless string, from the perspective of Carrollian symmetries on the worldsheet, has been well studied in the past \cite{Isberg:1993av,Bagchi:2013bga,Bagchi:2015nca,Bagchi:2019cay,Bagchi:2020fpr,Bagchi:2021rfw,Banerjee:2024fbi,Bagchi:2024qsb}. This ``electric'' limit has also been observed in marginal current-current deformations of the bosonic string \cite{Parekh:2023xms} where the marginal deformation takes the conformal theory to a Carrollian one. 

 \subsubsection*{A comment on the magnetic Carrollian string}
 
The magnetic limit given by \eqref{6.38} can be interpreted as a string theory where the tension becomes infinite, with action depending on the nonlinear factor $\mathcal{X}$: 
\begin{equation}
    S=\frac{1}{2\pi c^\prime}\int d^2\sigma\Big[-X^{\prime 2} + \dot{X}^{2}\left(\mathcal{X}^2+\dfrac{1}{4}\right)\Big].
\label{6.42}
\end{equation}
A different kind of magnetic Carrollian string was observed in \cite{Parekh:2023xms}, and now we will briefly touch upon how it is related to our formalism. Recall the action \eqref{6.42} was obtained in the Lagrangian formalism by choosing $\hat{C}_1$ and $\hat{C}_2$ asymptotically for $\alpha\to-\infty$. 
\medskip

If we remember Fig.\eqref{fig7}, this regime in parameter space is given by the region \circled{$2\pm$} of the $\hat{C}_1$-$\hat{C}_2$ hyperbola (the green region). However if we look at the similar region for the $\hat{A}$-$\hat{B}$ hyperbola, region \circled{$3\pm$} (the blue region), this gives a magnetic limit, albeit in the Hamiltonian formalism. %NEW EXPLANATION
To see this, we consider the hatted equivalent of the Hamiltonian \eqref{9.19}
\begin{align}
\hat{\mathcal{H}}_\alpha=\hat{A}\Big(\tilde{\Pi}^2+\mathbb{T}^2X^{\prime2}\Big)+\hat{B}\ \tilde{\Pi}\cdot X^\prime.\label{6.43}
\end{align}Plugging in the values of $\hat{A}$ and $\hat{B}$ for this region, and augmented by the fact $\mathbb{T}\to\infty$, we obtain the Hamiltonian $\hat{\mathcal{H}}_{(-)}\sim X^{\prime2}$, which is precisely the magnetic Carroll string in \cite{Parekh:2023xms}. One can also find %the hatted version of 
the canonical momenta \eqref{6.25} in terms of the hatted quantities (string equivalent of \eqref{Electricmom}), and find that the definition of $\tilde{\Pi}_\mu$ breaks down, as expected in the magnetic Carrollian limit. 
%Plugging in the values of $\hat{A}=\frac{1}{2}\epsilon^2$ and $\hat{B}=\epsilon^2\mathcal{X}$, augmented by the fact $T\to\infty$, we obtain the Hamiltonian $H_{(-)}\sim X^{\prime2}$, which is precisely the magnetic Carroll string in \cite{Parekh:2023xms}. 

\section{Discussions and Conclusions}\label{sect7}
\subsection*{Summary}
In this work, we investigated an intriguing regime for $\sqrt{T\overline{T}}$ operators for a two dimensional massless scalar field theory. Being a marginal deformation, $\sqrt{T\overline{T}}$ operators are already important in trying to understand new classes of CFTs. But our focus has been on a degenerate limit of the coupling space where the seed CFT smoothly changes into a Carroll Conformal Field Theory(CCFT) in the same dimension, as the Dirac-Schwinger conditions dictate. This, we believe, offers a more physical understanding of this change in symmetry, than what one can grasp following the usual contraction process. We set out to understand this phenomena riding on the flow equations for $\sqrt{T\overline{T}}$ operators both in Lagrangian and Hamiltonian formalisms, and via their interconnection. This connection via LT relations became much more illuminating when we replaced the square root both in Lagrangian and Hamiltonian by establishing a linearised structure, which led to an algebraic map.
\medskip

We further judiciously followed this LT map throughout the purported flow, even to the singular values in the moduli space. Our demands have been simple: these LT must be true everywhere in parameter space. This led to some striking results for the scalar field action as we pushed it to the edge of the conformal parameter space. Investigating the positive and negative branches of infinite valued couplings, and consistently expanding the actions, we arrived at a well-known electric type Carroll field theory, and a new magnetic type one, which curiously involved nonlinear terms in field derivatives. The global symmetry algebra for both these branches are that of CCA, which reflects in the structure of these actions. Furthermore, we delved into qualitative understanding of the LT, via a geometric formalism to see how configuration space maps into phase space as we move towards the edge of parameter space. This gave us clear vision on these maps and how they are different for given extreme values of the coupling. We ended on a well-chosen example of worldsheet string theory in conformal gauge that has been deformed with  $\sqrt{T\overline{T}}$, and how it flows the residual symmetry algebra to that of CCA.
\medskip

All in all, we asked a simple but unique question, given the inevitable flow from CFT to CCFT via the full nonlinear $\sqrt{T\overline{T}}$ deformations, how could we view the change in dynamics from both phase space and configuration space? We also asked, whether viewing this flow from different `pathways' creates some subtlety in terms of their canonical structure, given any value of the coupling, and given the endpoint always leads to a Carrollian realm. We hope the reader will agree with us that this work provides a comprehensive answer for the same.  

\subsection*{The road ahead}

Given the highly interesting directions this work opens up, one could think of a myriad of extensions going forward in future. Let us discuss a select few here:

\begin{itemize}
\item \textit{A `Quantum' version:} It has been technically challenging to think about proper quantum analogs of $\sqrt{T\overline{T}}$ deformations due to the square root. Hence understanding the flow of the spectrum from CFT to Carrollian theories are still beyond reach in this paradigm. However, given our understanding of the linearisation of the deformed theory in Lagrangian and Hamiltonian formalism, one may be able to have a better idea along these directions. Quantisation of a quantum mechanical system with $\sqrt{T\overline{T}}$ deformation was attempted in \cite{Ferko:2023iha} via a path integral formalism. One could hope that the same is possible for the scalar field case. 

\item \textit{Fermionic theories:} The fermionic version of $\sqrt{T\overline{T}}$ deformations were already discussed in \cite{Ferko:2022cix}. However, there are still various facets of the deformed spinors that needs to be explored. Most importantly for us, such a marginal deformation of relativistic spinors to Carroll spinors (as discussed in \cite{Bagchi:2022eui, Banerjee:2022ocj, Bergshoeff:2023vfd, Grumiller:2025rtm} for example) in suitable limits has not been spelled out in the literature. It will be instructive to understand this regime of the deformation flows, especially how the representations of the Clifford algebra changes with the deformation.

    \item \textit{Deformation of observables:} Since we now understand the flows from CFT to CCFT using the full operator from various directions, we can focus on computing observables in the CFT that flows to their CCFT cousins. For example, $n$-point functions in the deformed theory would be an interesting object to explore, as the primary operators flow from the conformal scaling to their Carroll avatars.  

    \item \textit{`New' Magnetic theories:} A very important takeaway from this work was the novel magnetic kind of scalar field theory that we deduced from the expansion of the full field theory. Despite having a manifestly non-linear term in fields, a clear progeny of the square root in the full action, this theory is Carroll boost invariant. Although we call it a magnetic theory due to the presence of spatial derivatives, it does not fit the bill for known magnetic Carroll theories. The EOM for such a theory is highly non-linear and coupled set of PDEs. One needs to understand the implication of such a theory, and the physics of the field dependent `velocity' $\mathcal{X}$ in a better manner. 
\end{itemize}

Surely, there are many a paths to take from where we stand in this work, and we hope to come back to some of the above mentioned problems in future communications.

\section*{Acknowledgments}
The authors would like to thank Rodrigo Aros, Arjun Bagchi, Eric Bergeshoeff, Joaquim Gomis, David Tempo and Ricardo Troncoso for helpful discussions and comments on the manuscript.

\medskip
ABan is supported in part by an OPERA grant and a seed grant NFSG/PIL/2023/P3816 from BITS-Pilani, and further an early career research grant ANRF/ECRG/2024/002604/PMS
from ANRF India. He also acknowledges financial support from the Asia Pacific Center for Theoretical Physics (APCTP) via an Associate Fellowship. 

PP acknowledges support from ANID Fondecyt grants N° 1211226, 1221624, 1250487 and 3210558, and thank Simons Center for Geometry and Physics for hospitality during the Summer Workshop 2025.

RRaj acknowledges financial support from the Universidad Andrés Bello doctoral fellowship and from Chilean National Science Foundation (Fondecyt) Regular Grant No. 1220335 (STAGE 2024 and STAGE 2025).
\appendix

\section{Representations of $\sqrt{T\overline T}$ operator}\label{appA}

The holomorphic and anti-holomorphic coordinates can be write in terms of space-time co-ordinates as $z=t+ x,~~\bar z=t- x$. 
%\Longrightarrow t=\frac{1}{2}\left(z+\bar z\right)\qquad x=\frac{1}{2}\left(z-\bar z\right)\\\Longrightarrow
%\frac{\partial t}{\partial z}=\frac{1}{2}\quad \frac{\partial x}{\partial z}=\frac{1}{2} \quad \frac{\partial t}{\partial \bar z}=\frac{1}{2}\quad \frac{\partial x}{\partial \bar z}=-\frac{1}{2}\label{27} 
We have covariant tensor transformation laws as
\begin{align}
T_{zz}= \frac{\partial x^\mu}{\partial z }\frac{\partial x^\nu}{\partial z} T_{\mu\nu},~~~\quad
T_{\bar z\bar z}=
 \frac{\partial x^\mu}{\partial \bar z }\frac{\partial x^\nu}{\partial \bar z} T_{\mu\nu}. 
\end{align}
Using the above, we can write
\begin{align}
T_{zz}=\frac{1}{4}T_{tt}+\frac{1}{4}T_{xx}+\frac{1}{2}T_{tx},~~
T_{\bar z \bar z}=\frac{1}{4}T_{tt}+\frac{1}{4}T_{xx}-\frac{1}{2}T_{tx}.
\end{align}
Hence,
\begin{align}
\sqrt{T \bar T}   = \ \sqrt{T_{zz}T_{\bar z \bar z}}=\sqrt{\frac{1}{16}\left(T_{tt}^2+T_{xx}^2+2T_{tt}T_{xx}\right)-\frac{1}{4}T_{tx}T_{tx}}\label{STT},
\end{align} 
and
\begin{align}
 T+\bar T=\frac{1}{2}\left(T_{tt}+T_{xx}\right).  
\end{align}
In the Hamiltonian formulation, the components of the stress-energy tensor can be identified as $T ^{tt}=T^{xx}=\mathcal{H},~~ T_{tx}=T_{xt}= \mathcal{J}$. 
Thus \eqref{STT} becomes 
\begin{align}
\sqrt{T \bar{T}}= \frac{1}{2}\sqrt{\mathcal{H}^2-\mathcal{J}^2}. \label{AppA-5}
\end{align}
In the Lagrangian formulation, 
\begin{align}
\mathcal{T}^{\mu \nu}=\partial^\mu \phi_i\partial^\nu \phi_i-g^{\mu \nu}\left(\frac{1}{2}\left(\partial_\rho \phi_i \partial^\rho \phi_i\right)\right),\\
\implies
 \begin{array}{r}
\mathcal{T}^{t t}=\mathcal{T}^{x x}=\frac{1}{2}(\dot\phi_i^2+\phi_i^{\prime\,2} ),~~~
\mathcal{T}_{t x}=\mathcal{T}_{x t}=\dot\phi_i\phi_i^\prime.
\end{array}    
 \end{align}
 Thus \eqref{STT} becomes
\begin{align}
2\sqrt{\mathcal{T}\bar{\mathcal{T}}}    
=\sqrt{\frac{1}{4}\left(\dot{\phi}_i^2+\phi_i^{\prime 2}\right)^2-(\dot{\phi}_i \phi_i^{\prime})^2}=\sqrt{\sigma^2-\rho^2}=\sqrt{\mathcal{L}^2-\mathcal{P}^2},\label{Appendix1}\end{align}
where $\sigma,\rho,\mathcal{L}$ and $\mathcal{P}$ are defined as in \eqref{Lsigmaandrho}.
%\begin{align}
%\sigma=\frac{1}{2}\left(\dot{\phi}_i^2+\phi_i^{\prime 2}\right)\quad \rho=\dot{\phi}_i %\phi_i^{\prime};
%\quad\mathcal{L}=
%\frac{1}{2}\left(\dot{\phi}_i^2-\phi_i^{\prime 2}\right)  \qquad \mathcal{P}^2=2 \mathcal{L}^2-%\frac{1}{2} \partial^\mu \phi_i \partial_\nu \phi_i \partial^\nu \phi_j \partial_\mu \phi_j
%\end{align}

\section{Nontrivial LT: A free particle example}\label{appB}

To illustrate the structure of the deformation flows discussed in the main text, let us consider the simple case of a free particle of unit mass, following discussions in \cite{Ferko:2023iha}. Lagrangians and Hamiltonians are given by:
\begin{align}
L_0 = \frac{1}{2} \dot{x}^2,
\qquad
H_0 = \frac{p^2}{2}.\label{freeparticle,Initial.conditions}
\end{align}
The connection between the undeformed velocity and momentum is given by the simple relation 
$p=\dot{x}.$ Now, let us examine the corresponding set of flow equations where the flow operator is simply the momentum:
\begin{align}
\frac{\partial L_\lambda}{\partial \lambda}=\mathcal{O}(x, \dot{x})= \dot{x}, \quad \frac{\partial H_\lambda}{\partial \lambda}=\mathfrak{O}(x, p)=-p,   \end{align}
with solutions giving deformed actions:
\begin{align}
L_\lambda=\frac{1}{2}  \dot{x}^2+\lambda \dot{x}, \quad H_\lambda=\frac{p^2}{2 }-\lambda p.   
\end{align}
One can immediately note, these Lagrangian $L_\lambda$ and Hamiltonian $H_\lambda$ are not connected through Legendre transform at finite $\lambda$, because the operator equivalence
\begin{align}
\mathcal{O}(x, \dot{x}) =-\mathfrak{O}(x, p)\label{operatorequiv}   
\end{align}
only holds for the undeformed theory (i.e. $\dot x\neq p \quad\forall \lambda\neq 0$).
For a finite parameter value, the momentum operator is shifted:
\begin{align}
p_\lambda=\frac{\partial L_\lambda}{\partial \dot{x}}= \dot{x}+\lambda.     
\end{align}
The flow generated by this deformed momentum can be expressed as
\begin{align}
\frac{\partial L_\lambda}{\partial \lambda}
= \mathcal{O}(x, \dot{x})
= \frac{\partial L_\lambda}{\partial \dot{x}},
\qquad
\frac{\partial H_\lambda}{\partial \lambda}
= \mathfrak{O}(x, p)
= -p.
\label{momentumflows}
\end{align}
Solving these flow equations with the initial conditions \eqref{freeparticle,Initial.conditions} gives the following deformed quantities
\begin{align}
L_\lambda = \frac{1}{2} \dot{x}^2 + \lambda \dot{x} + \frac{1}{2} \lambda^2,
\qquad
H_\lambda = \frac{p^2}{2} - \lambda p.
\end{align}
One can directly verify that these deformed quantities are now connected through the LT because now the operator equivalence \eqref{operatorequiv} is valid everywhere, confirming the consistency of the flow. Note even more crucially that the operator equivalence would mean when we do $\frac{\partial L_\lambda}{\partial \lambda}$, we take $\dot x$ to be $\lambda$ independent, and while doing $\frac{\partial H_\lambda}{\partial \lambda}$, we take $p$ to be $\lambda$ independent.  As a more nontrivial example, one may consider a $\sqrt{T\bar{T}}$-type deformation of the free particle. For similar constructions in the context of $T\bar{T}$-deformed systems, we refer the reader to \cite{Gross:2019ach}.

It is often useful to view classical mechanics as a $(0+1)$-dimensional classical field theory. In this picture, the only nontrivial component of the stress-energy tensor is \(T_{\tau\tau}\), which corresponds to the total energy of the system. For conservative systems, this quantity coincides with the total Hamiltonian. The $\sqrt{T\bar{T}}$ operator can be expressed as
\begin{align}
O_{\sqrt{T \bar{T}}}
= \sqrt{\frac{1}{2} T^{\mu\nu} T_{\mu\nu}
- \frac{1}{4} T_\mu{ }^\mu T_\nu{ }^\nu}.
\end{align}
In a one-dimensional setting, this simplifies considerably:
\begin{align}
O_{\sqrt{T T}}
= \sqrt{\frac{1}{2} T^{00} T_{00}
- \frac{1}{4} (T_0^{\,0})^2}
= \sqrt{\frac{1}{4} H^2}
= \frac{1}{2} H.
\end{align}
Hence, a $\sqrt{T\bar{T}}$-like flow in classical mechanics can be written as
\begin{align}
\frac{\partial L_\lambda}{\partial \lambda}
= \frac{1}{2}\left(
\frac{\partial L_\lambda}{\partial \dot{x}} \dot{x}
- L_\lambda \right),
\qquad
\frac{\partial H_\lambda}{\partial \lambda}
= -\frac{1}{2} H_\lambda.
\label{TTbarflow}
\end{align}
The right-hand side of the Lagrangian flow equation represents the deformed Hamiltonian, obtained as the LT of the deformed Lagrangian. For the initial condition \eqref{freeparticle,Initial.conditions}, the solution spaces have the form
\begin{align}
L_\lambda = \frac{1}{2} \dot{x}^2 e^{\lambda / 2},
\qquad
H_\lambda = \frac{1}{2} p^2 e^{-\lambda / 2}.
\end{align}
One can again see that when $\lambda\neq 0$, the canonical momentum $p \neq \dot x$ like our previous case.
It is straightforward to verify that these expressions are related through the LT which of course depends on $\lambda$. In this case, the operator equivalence~\eqref{operatorequiv} directly yields the required LT from \eqref{TTbarflow} as 
\begin{align}
H_\lambda = \frac{\partial L_\lambda}{\partial \dot{x}}\, \dot{x} - L_\lambda \, .
\end{align}
These examples demonstrate that the connection between the Lagrangian and Hamiltonian formulations remains valid for all values of the deformation parameter, which is, in fact, a direct consequence of the underlying flow equations and the operator equivalence.

\section{Details of few calculations}
\subsection{LT from operator equivalence}\label{opeqv}
Starting with the phase space definition:
\begin{align}
2 \sqrt{T_\alpha \bar{T}_\alpha} &= K_\alpha = -\mathcal{H} \sinh \alpha + \sqrt{\mathcal{H}^2 - \mathcal{J}^2} \cosh \alpha,
\end{align}
where $A,B$ are given as in \eqref{ABdef}, we can show:
\begin{align}
2 \sqrt{T_\alpha \bar{T}_\alpha} &= A \sqrt{\mathcal{H}^2 - \mathcal{J}^2} = \cosh \alpha \sqrt{\mathcal{H}^2 - \mathcal{J}^2} - \sinh \alpha \, \mathcal{H}.
\end{align}
In the configuration space, on the other hand, we have
\begin{align}
2 \sqrt{\mathcal{T}_\alpha \overline{\mathcal{T}}_\alpha} &= \mathcal{K}_\alpha = \sqrt{\sigma^2 - \rho^2} \cosh \alpha + \mathcal{L} \sinh \alpha,
\end{align}
with similar definitions of $C_{1,2}$ as in \eqref{C_1C_2}, this promptly leads to
\begin{align}
2 \sqrt{\mathcal{T}_\alpha \overline{\mathcal{T}}_\alpha} &= C_1 \sqrt{\sigma^2 - \rho^2} - \phi^{\prime 2} \sinh \alpha.
\end{align}
Therefore, the operator equivalence \( K_\alpha = \mathcal{K}_\alpha \), defined at all finite values of $\alpha$, leads to the following via comparison:
\begin{align}
\sqrt{\mathcal{H}^2-\mathcal{J}^2}=\frac{\sinh \alpha \mathcal{J}}{B},  \qquad \sqrt{\sigma^2-\rho^2}=-\frac{\sinh \alpha\rho}{C_2}.\end{align}
Substituting back gives
\begin{align}
A  \frac{\phi_i^\prime\pi_i\sinh \alpha }{B}  =-C_1\frac{\sinh \alpha \rho}{C_2}-\phi^{\prime 2} \sinh \alpha.\label{op115}
\end{align}
From the Hamiltonian EOM we have
\begin{align}
\dot{\phi}_i(x, t)=A(\alpha) \pi_i+B(\alpha) \phi_i^{\prime}\Rightarrow \pi_i=\frac{1}{A} \dot{\phi}_i-\frac{B}{A} \phi_i^{\prime},    
\end{align}
Putting this back in \eqref{op115} gives
\begin{align}
A \frac{\phi_i^{\prime}\sinh \alpha}{B}\left( \frac{1}{A} \dot{\phi}_i-\frac{B}{A} \phi_i^{\prime}\right)=  -C_1 \frac{\sinh \alpha \rho}{C_2}-\phi^{\prime 2} \sinh \alpha. 
\end{align}
This gives
\begin{align}
\frac{1}{B} \rho\sinh \alpha-\phi_i^{\prime}\sinh \alpha&=  -C_1 \frac{\sinh \alpha \rho}{C_2}-\phi^{\prime 2} \sinh \alpha \nonumber\\
\Rightarrow\frac{1}{B} \rho\sinh \alpha&=  -C_1 \frac{\sinh \alpha \rho}{C_2}~~~
\Rightarrow
\frac{1}{B}=-\frac{C_1}{C_2}.\label{OPL.T}    
\end{align}
%We also have:
%\begin{align}
%A \frac{\phi_i^{\prime} \pi_i \sinh \alpha}{B}=-C_1 \frac{\sinh \alpha \phi_i^{\prime} \dot\phi_i}{C_2}-\phi^{\prime 2} \sinh \alpha    
%\label{OPER125}\end{align}
Recall the canonical deformed momenta
\begin{align}
\pi_i=\frac{\partial \mathcal{L}_\alpha}{\partial \dot{\phi}_i}=C_1 \dot{\phi}_i+C_2 \phi_i^{\prime}\Longrightarrow \dot{\phi}_i=\frac{1}{C_1}\pi_i- \frac{C_2}{C_1} \phi_i^{\prime}. 
\end{align}
Substituting this back again, 
\begin{align}
A \frac{\phi_i^{\prime} \pi_i \sinh \alpha}{B}&=-C_1 \frac{\sinh \alpha \phi_i^{\prime} }{C_2}\left(\frac{1}{C_1} \pi_i-\frac{C_2}{C_1} \phi_i^{\prime}\right)-\phi^{\prime 2} \sinh \alpha, \nonumber\\ 
\Longrightarrow A \frac{ \sinh \alpha}{B}\mathcal{J}&=-C_1 \frac{\sinh \alpha }{C_1C_2}\mathcal{J}\\
\Longrightarrow \frac{A}{B}=-\frac{1}{C_2}. 
\end{align}
So, just from the operator equivalence, we can deduce two conditions on $A$ and $B$
\begin{align}
 \frac{1}{B}=-\frac{C_1}{C_2}\qquad \frac{A}{B}=-\frac{1}{C_2}.   
\end{align}
This is actually the same LT equations that we derived before in \eqref{L.T}.
\subsection{Proof of identity \eqref{ID1} \label{APPB}}

The \eqref{ID1} can be proved as follows. We start with definitions of $A,B$ in \eqref{ABdef}.
Corresponding derivatives w.r.t $\alpha$ are
\begin{align}
\frac{\partial A}{\partial \alpha}=
%\frac{\sqrt{\mathcal{H}^2-\mathcal{J}^2} \sinh \alpha-\cosh \alpha \mathcal{J}}{\sqrt{\mathcal{H}^2-\mathcal{J}^2}}=
\frac{-\mathcal{H}_\alpha}{\sqrt{\mathcal{H}^2-\mathcal{J}^2}},~~~
\frac{\partial B}{\partial \alpha}=\frac{\cosh \alpha \mathcal{J}}{\sqrt{\mathcal{H}^2-\mathcal{J}^2}}, \label{A.17}
\end{align}
and also
\begin{align}
 \frac{B}{A}=\frac{\mathcal{J}\sinh \alpha }{\sqrt{\mathcal{H}^2-\mathcal{J}^2} \cosh \alpha-\sinh \alpha \mathcal{H}}=\frac{\mathcal{J}\sinh \alpha }{\sqrt{T_\alpha \bar{T}_\alpha}},   
\label{BbyA}
~~~~A^2 &
%= \frac{\left( \sqrt{\mathcal{H}^2 - \mathcal{J}^2} \cosh \alpha - \mathcal{H}\sinh \alpha \right)^2}{\mathcal{H}^2 - \mathcal{J}^2}
= \frac{T_\alpha\bar T_\alpha}{\mathcal{H}^2-\mathcal{J}^2},
\\
B^2 - 1 &
%= \frac{\mathcal{J}^2\sinh^2 \alpha + \mathcal{J}^2 - \mathcal{H}^2}{\mathcal{H}^2 - \mathcal{J}^2}
= \frac{\mathcal{J}^2 \cosh^2 \alpha - \mathcal{H}^2}{\mathcal{H}^2 - \mathcal{J}^2}.
\end{align}
Therefore, we can write:
\begin{align}
\frac{B^2-1}{2 A^2}+\frac{1}{2}&=\frac{\mathcal{J}^2 \cosh^2 \alpha - \mathcal{H}^2 + T_\alpha \bar{T}_\alpha}{2 T_\alpha \bar{T}_\alpha} \nonumber\\
%&=\frac{\mathcal{J}^2 \cosh^2 \alpha - \mathcal{H}^2 + (\mathcal{H}^2 - \mathcal{J}^2) \cosh^2 \alpha + \mathcal{H}^2 \sinh^2 \alpha - 2\sqrt{\mathcal{H}^2 - \mathcal{J}^2} \mathcal{H} \cosh \alpha \sinh \alpha}{2 T_\alpha \bar{T}_\alpha} \nonumber\\
&=\frac{2 \mathcal{H}^2 \sinh^2 \alpha - 2 \mathcal{H} \sqrt{\mathcal{H}^2 - \mathcal{J}^2} \cosh \alpha \sinh \alpha}{2 T_\alpha \bar{T}_\alpha} \nonumber\\
&=\frac{\mathcal{H} \sinh \alpha \left( \mathcal{H} \sinh \alpha - \sqrt{\mathcal{H}^2 - \mathcal{J}^2} \cosh \alpha \right)}{T_\alpha \bar{T}_\alpha} \nonumber\\
%&=-\frac{\mathcal{H} \sinh \alpha \sqrt{T_\alpha \bar{T}_\alpha}}{T_\alpha \bar{T}_\alpha} \nonumber\\
&=-\frac{\mathcal{H} \sinh \alpha}{\sqrt{T_\alpha \bar{T}_\alpha}}. \label{A.19}
\end{align}
Further using equations (\ref{A.17}, \ref{BbyA}, \ref{A.19})
\begin{align}
 \frac{\partial A}{\partial \alpha}\left(\frac{B^2 - 1}{2 A^2} + \frac{1}{2}\right) - \frac{B}{A} \frac{\partial B}{\partial \alpha}  = \frac{\sinh \alpha \left( \mathcal{H} \mathcal{H}_\alpha - \mathcal{J}^2 \cosh \alpha \right)}{\sqrt{\mathcal{H}^2 - \mathcal{J}^2} \sqrt{T_\alpha \bar{T}_\alpha}}, \label{A.20}
\end{align}
where
\begin{align}
\mathcal{H} \mathcal{H}_\alpha - \mathcal{J}^2 \cosh \alpha &= \mathcal{H}^2 \cosh \alpha - \mathcal{H} \sqrt{\mathcal{H}^2 - \mathcal{J}^2} \sinh \alpha - \mathcal{J}^2 \cosh \alpha \nonumber\\
&= \cosh \alpha (\mathcal{H}^2 - \mathcal{J}^2) - \mathcal{H} \sqrt{\mathcal{H}^2 - \mathcal{J}^2} \sinh \alpha \nonumber\\
%&= \sqrt{\mathcal{H}^2 - \mathcal{J}^2} \left( \cosh \alpha \sqrt{\mathcal{H}^2 - \mathcal{J}^2} - \mathcal{H} \sinh \alpha \right) \nonumber\\
&= \sqrt{\mathcal{H}^2 - \mathcal{J}^2} \sqrt{T_\alpha \bar{T}_\alpha}. \label{A.21}
\end{align}
Substituting \eqref{A.21} in \eqref{A.20} explicitly gives the identity we desire:
\begin{align}
 \frac{\partial A}{\partial \alpha}\left(\frac{B^2 - 1}{2 A^2} + \frac{1}{2}\right) - \frac{B}{A} \frac{\partial B}{\partial \alpha} = \sinh \alpha.
\end{align}

\section{Carroll from the stress energy tensor}\label{StressTvanish}
In the Hamiltonian formulation,
\begin{align}
\partial_0 T^{00}-\partial_1 T^{10}=0,    
\end{align}
and the EOM are:
\begin{align}
\dot{\phi}^i=\frac{\partial \mathcal{H}}{\partial \pi_i},
 \quad \dot{\pi}_i=-\frac{\partial \mathcal{H}}{\partial \phi^i}+\partial_x\left(\frac{\partial \mathcal{H}}{\partial \phi_i^{\prime}}\right).
\end{align}
Using the above, we can calculate
\begin{align}
\partial_t \mathcal{H}&=\frac{\partial \mathcal{H}}{\partial \phi^i} \dot{\phi}^i+\frac{\partial \mathcal{H}}{\partial \pi_i} \dot{\pi}_i+\frac{\partial \mathcal{H}}{\partial \phi_i^{\prime}} \partial_t \phi_i^{\prime}\\ 
&=\frac{\partial \mathcal{H}}{\partial \phi^i}\frac{\partial \mathcal{H}}{\partial \pi_i}-\frac{\partial \mathcal{H}}{\partial \pi_i}\frac{\partial \mathcal{H}}{\partial \phi^i}+\frac{\partial \mathcal{H}}{\partial \pi_i}\partial_x\left(\frac{\partial \mathcal{H}}{\partial \phi_i^{\prime}}\right)+\frac{\partial \mathcal{H}}{\partial \phi_i^{\prime}}\partial_x\dot{\phi}^i\\
&=\frac{\partial \mathcal{H}}{\partial \phi^i}\frac{\partial \mathcal{H}}{\partial \pi_i}-\frac{\partial \mathcal{H}}{\partial \pi_i}\frac{\partial \mathcal{H}}{\partial \phi^i}+\frac{\partial \mathcal{H}}{\partial \pi_i}\partial_x\left(\frac{\partial \mathcal{H}}{\partial \phi_i^{\prime}}\right)+\frac{\partial \mathcal{H}}{\partial \phi_i^{\prime}}\partial_x\frac{\partial \mathcal{H}}{\partial \pi_i}\\
&=\partial_x\left(\frac{\partial \mathcal{H}}{\partial \pi_i}\frac{\partial \mathcal{H}}{\partial \phi_i^{\prime}}\right).
\end{align}
This enables us to write the component of the stress tensor
\begin{align}
T^{10}=\frac{\partial \mathcal{H}}{\partial \phi_i^{\prime}} \cdot \frac{\partial \mathcal{H}}{\partial \pi_i}.    
\end{align}
Then, the full stress energy tensor in the 2d theory can be written in a matrix form:
\begin{align}
\left(\begin{array}{ll}
T^{00} & T^{01} \\
T^{10} & T^{11}
\end{array}\right)=   \left(\begin{array}{cc}
H & \pi_i\phi_i \\
\frac{\partial \mathcal{H}}{\partial \phi_i^{\prime}} \cdot \frac{\partial \mathcal{H}}{\partial \pi_i} & H
\end{array}\right). 
\end{align}
Now we consider our scaled Hamiltonian (using \eqref{HamilAandB} and \eqref{scaledHTTD})
\begin{align}
\hat{\mathcal{H}}_\alpha=\frac{1}{\cosh\alpha}\left(A \mathcal{H}+B \mathcal{J}   \right), \end{align}
which gives
\begin{align}
\begin{aligned}
 \frac{\partial \hat{\mathcal{H}}_\alpha}{\partial \tilde\pi_i}=\frac{1}{\cosh \alpha}\left(A \tilde\pi_i+B \phi_i^{\prime}\right), ~~~~
 \frac{\partial\hat{\mathcal{H}}_\alpha}{\partial \phi_i^{\prime}}=\frac{1}{\cosh \alpha}\left(A \phi_i^{\prime}+B \tilde\pi_i\right).
\end{aligned}    
\end{align}
Then the energy flux is
\begin{align}
T^{10}= &\frac{1}{\cosh ^2\alpha}\left(A \tilde\pi_i+B \phi_i^{\prime}\right)\left(A \phi_i^{\prime}+B \tilde\pi_i\right)\nonumber\\
=&\frac{1}{\cosh ^2 \alpha}\left((A^2+B^2)\tilde\pi_i\phi_i^\prime+AB(\tilde\pi_i^2+\phi_i^{\prime 2})\right)\nonumber\\
=&\frac{1}{\cosh ^2 \alpha}\left((A^2+B^2)\mathcal{J}+2AB\mathcal{H}\right).\label{scaledenergyflux}
\end{align}
To evaluate the quantity within brackets, we note 
\begin{align}
A^2+B^2&=\left(\cosh \alpha-\frac{\sinh \alpha \mathcal{H}}{\sqrt{\mathcal{H}^2-\mathcal{J}^2}}\right)^2+\left(\frac{\sinh \alpha \mathcal{J}}{\sqrt{\mathcal{H}^2-\mathcal{J}^2}}\right)^2 \nonumber
\\
&=\cosh ^2 \alpha-2 \cosh \alpha \sinh \alpha \frac{\mathcal{H}}{\sqrt{\mathcal{H}^2-\mathcal{J}^2}}+\sinh ^2 \alpha \frac{\mathcal{H}^2}{\mathcal{H}^2-\mathcal{J}^2}+\sinh ^2 \alpha \frac{\mathcal{J}^2}{\mathcal{H}^2-\mathcal{J}^2},
\label{F9}\end{align}
and similarly,
\begin{align}
A B&=\left(\cosh \alpha-\frac{\sinh \alpha \mathcal{H}}{\sqrt{\mathcal{H}^2-\mathcal{J}^2}}\right)\left(\frac{\sinh \alpha \mathcal{J}}{\sqrt{\mathcal{H}^2-\mathcal{J}^2}}\right)\nonumber\\&= \cosh \alpha \frac{\sinh \alpha \mathcal{J}}{\sqrt{\mathcal{H}^2-\mathcal{J}^2}}-\sinh ^2 \alpha \frac{\mathcal{H} \mathcal{J}}{\mathcal{H}^2-\mathcal{J}^2}.   
\end{align}
From this, we get the following:
\begin{align}
2 A B \frac{\mathcal{H}}{\mathcal{J}}=2 \cosh \alpha \sinh \alpha \frac{\mathcal{H}}{\sqrt{\mathcal{H}^2-\mathcal{J}^2}}-2 \sinh ^2 \alpha \frac{\mathcal{H}^2}{\mathcal{H}^2-\mathcal{J}^2}. \label{F.11}   
\end{align}
Using \eqref{F9} and \eqref{F.11} we can write,
\begin{align}
&A^2+B^2+2 A B \frac{\mathcal{H}}{\mathcal{J}}\nonumber\\
&=\cosh ^2 \alpha+\sinh ^2 \alpha \frac{\mathcal{H}^2+\mathcal{J}^2-2 \mathcal{H}^2}{\mathcal{H}^2-\mathcal{J}^2}=1.\label{IDE.20}
\end{align}
This identity is mentioned as a consequence of Lorentz invariance in \cite{Bandos:2020jsw}.
Using this identity in \eqref{scaledenergyflux} gives
\begin{align}
T^{10}= \frac{1}{\cosh ^2 \alpha}\left[\left(1-2 A B \frac{\mathcal{H}}{\mathcal{J}}\right) \mathcal{J}+2AB\mathcal{H}\right]\nonumber
=\frac{1}{\cosh ^2 \alpha}\mathcal{J}.
\end{align}
When $\alpha\to\infty$
this becomes $T^{10}=0$,  
which implies the emergence of Carroll symmetries.
\medskip

One can also verify that the leading-order contribution to 
$T^{10}$ for the Lagrangian~\eqref{7.9} vanishes. 
For the Lagrangian~\eqref{7.10}, it can be computed explicitly as
\begin{align}
\hat{\mathcal{T}}^{10}
&= -\frac{\partial \hat{\mathcal{L}}^-_\alpha}{\partial \phi_i'}\,\dot{\phi}_i \nonumber\\
&= -2\,\phi_i'\dot{\phi}_i
   + 2\,\rho\,\dot{\phi}_i\dot{\phi}_i\,
     \frac{1}{\dot{\phi}_j^{\,2}}
= 0 \, .
\end{align}
This again points to the fact that the Lagrangians~\eqref{7.9} and~\eqref{7.10} possess Carroll symmetries.

\bibliographystyle{JHEP}
\bibliography{ref}

\end{document}